%% file: Resub2.tex
\documentclass[twocolumn,english,aps,pra,floatfix,tightenlines,superscriptaddress,nofootinbib,article]{revtex4-2}
\usepackage[T1]{fontenc}
\usepackage[utf8]{inputenc}
\usepackage{color}
\usepackage[english]{babel}
\usepackage{amsmath}
\usepackage{amsthm}
\usepackage{amssymb}
\usepackage{graphicx,subfigure}
\usepackage{esint}
\usepackage{bbold}
\usepackage{amsfonts}
\usepackage{mathtools}
\usepackage{commath}
\usepackage[unicode=true,
 bookmarks=false,
 breaklinks=false,pdfborder={0 0 1},backref=false,colorlinks=true]
 {hyperref}
\hypersetup{
 linkcolor=cyan,citecolor=magenta,filecolor=magenta,urlcolor=cyan,runcolor=cyan}
 \usepackage[normalem]{ulem}
 \usepackage{braket}

\makeatletter

\begin{document}

\date{\today}

\title{
Invalidation of the Bloch-Redfield Equation in Sub-Ohmic Regime via a Practical Time-Convolutionless Fourth-Order Master Equation}
\author{Elyana Crowder}
\affiliation{Georgia Institute of Technology, Atlanta, Georgia, United States}
\author{Lance Lampert}
\affiliation{Georgia Institute of Technology, Atlanta, Georgia, United States}
\author{Grihith Manchanda}
\affiliation{Georgia Institute of Technology, Atlanta, Georgia, United States}
\author{Brian Shoffeitt}
\affiliation{Georgia Institute of Technology, Atlanta, Georgia, United States}
\author{Srikar Gadamsetty}
\affiliation{Georgia Institute of Technology, Atlanta, Georgia, United States}
\author{Yiting Pei}
\affiliation{Georgia Institute of Technology, Atlanta, Georgia, United States}
\author{Shantanu Chaudhary}
\affiliation{Georgia Institute of Technology, Atlanta, Georgia, United States}
\author{Dragomir Davidovi\'c}
\affiliation{Georgia Institute of Technology, Atlanta, Georgia, United States}
\email{dragomir.davidovic@physics.gatech.edu}
\begin{abstract}

Despite recent advances in quantum sciences, a quantum master equation that accurately and simply characterizes open quantum dynamics across extremely long timescales and in dispersive environments is still needed.  In this study, we optimize the computation of the fourth-order time-convolutionless master equation to meet this need.
Early versions of this master equation required computing a multidimensional integral, limiting its use.
Our master equation accounts for simultaneous relaxation and dephasing, resulting in coefficients proportional to the system's spectral density over frequency derivative. In sub-Ohmic environments, this derivative induces infrared divergence in the master equation, invalidating the second-order Bloch-Redfield master equation findings.
We analyze the approach to a ground state in a generic open quantum system and demonstrate that it is not reliably computed by the Bloch-Redfield equation alone. The optimized fourth-order equation shows that the ground-state approach is accurate to second order in bath coupling regardless of the dispersion, even though it can diverge in the fourth order.

\end{abstract}
\maketitle

\section{Introduction}

Understanding the quantum dynamics of small systems coupled to dispersive environments over long timescales is crucial in various scientific fields
~\cite{frohlich2016preparation,zurek2009quantum,shandera2018open,engel2007evidence,collini2009coherent,carleo2012localization}.
Exact state propagation and variational approaches to open quantum systems, such as Feynman–Vernon influence functional~\cite{FEYNMAN1963118} and the multiconfigurational time-dependent Hartree method~\cite{meyer1990multi} scale exponentially versus time in terms of computational cost.
To make these methods work over the long timescales, further approximations are necessary, including singular-value decomposition~\cite{Strathearn2018}, small matrix disentanglement of the path integral~\cite{makri2020small}, and the hierarchical equations of motion~\cite{tanimura1989time}. These further approximations work well only when the correlations in the bath decrease exponentially over time.

The exact time-convolutionless master equation (TCL$2n$) method~\cite{BreuerHeinz-Peter1961-2007TToO}, on the other hand, saturates in computational cost at the bath correlation timescale, which is polynomial at fixed $n$. Here $2n$ is the perturbative order in the expansion of time-ordered cumulants.
Despite this favorable characteristic, little progress has been made beyond the widely used TCL2, which is known as the Bloch-Redfield equation~\cite{bloch1946nuclear,REDFIELD19651}. 
Due to its multiple time integrals, the exact TCL4 generator has been considered ``cumbersome''~\cite{liu2018exact,becker2022canonically} without additional approximations. Our approach simplifies this multidimensional integral into a series of independent one-dimensional integrals, which can be computed efficiently and with relatively low overhead.

A well-known issue with the TCL2$n$ master equation is that its generator is accurate to the level of $O(\lambda^{2n})$, while the solutions it provides are accurate only up to $O(\lambda^{2n-2})$~\cite{Fleming,tupkary2021fundamental}. Here $\lambda$ denotes a dimensionless coupling constant linearly scaling with the interaction Hamiltonian between the system and the bath. This issue is especially significant in the TCL2 master equation, as the precision of the resulting state is only of the magnitude $O(1)$. Thus, the Bloch-Redfield equation precisely represents only the local thermal states as steady states~\cite{davies1974,tupkary2021fundamental}.   Bath interactions cause modifications in these states with the same error as the modifications.  
The current state of the field is fraught with significant problems since the utilization of second-order master equations is widespread while the precise calculation of the most basic modifications remains out of reach.

Many proposals have been made to solve this deficiency by fixing specific issues within the TCL2 master equation. This includes the coarse-graining approach~\cite{Schaller,Benatti,Benatti2,Majenz,Cresser,Giovannetti,Hartmann,mozgunov}; the partial secular approximation~\cite{Vogt,Tscherbul}; the unified master equation~\cite{Trushechkin,gerry2023full};
the geometric-arithmetic approach~\cite{Benoit,perlind,Massimiliano,Kleinherbers,Davidovic2020,Nathan};
the dynamical resonance theory~\cite{merkli2020quantum,merkli2022dynamics,merkli2022dynamics2}; and the adiabatic approximation~\cite{Albash_2012}.
There is also a truncated version of the Bloch-Redfield equation~\cite{Becker}; an improved weak coupling master equation~\cite{Rivas,winczewski}; the Kossakowski matrix approximation approach~\cite{DAbbruzzo}; and master equations that enforce thermodynamic consistency~\cite{potts2021thermodynamically,tupkary2023searching,uchiyama2023dynamics,winczewski2021renormalization}. As stated by Tupkary {\it et al.}~\cite{tupkary2021fundamental}, making adjustments to the Bloch-Redfield equation to ensure complete positivity unavoidably leads to increased errors in the coherences and breaches conservation laws. Therefore, it becomes crucial to consider the fourth-order terms in order to address this issue~\cite{tupkary2021fundamental}. 

 Tokuyama and Mori were among the pioneers who recognized the potential of a time-convolutionless master equation to accurately describe the quantum dynamical map back in 1976~\cite{tokuyama1976statistical}. However,  the utilization of this equation in the fourth-order has not been extensively adopted thus far.
Our goal in this paper is to address the issue of accuracy in the Bloch-Redfield equation by finding a way to simplify the TCL4 master equation so that it can be easily implemented in practice. However, during this process, we realize that there is still a lot of work ahead of us. In the sub-Ohmic regime, the TCL4 exhibits an infrared divergence, invalidating the second-order master equation results from Refs.~\cite{REDFIELD19651,davies1974}.

{\it Brief history:} There are many ways to describe the TCL4 generators in terms of Liouvillians that do not include the bath correlation function over time~\cite{Yoon,Mukamel,shibata1977generalized,shibata1980expansion,laird1991quantum,Reichman,Breuer_1999}.
In the weak coupling and asymptotic limit, i.e., $\lambda\ll 1$ and $t\to\infty$, where $t$ is the time of the dynamics, Silbey and colleagues developed the first easy-to-use formulation for the kernels of the TCL4 generator~\cite{JangS}.  However, their methods included additional assumptions that the extremely singular Dirac delta functions may approximate the bath correlation functions.

Trushechkin~\cite{trushechkin2019higher} has recently published a simplified TCL4 and TCL6 generator with bath correlation functions that decay exponentially versus time. However, we cannot use these to study the asymptotic dynamics when the bath correlations decay as a power-law versus time.
The same author has also put up a
simplified fourth-order generator for the asymptotic dynamics in arbitrary baths, based on
Bogoliubov's approach to deriving the Boltzmann equation~\cite{trushechkin2021derivation}. 
It would be interesting to study the differences between Trushechkin's Bogoliubov approach and the TCL4 generator derived here to better understand their domains of validity.

References~\cite{nestmann2019timeconvolutionless,karasev2023timeconvolutionless}  recently reformulated the perturbative TCL$2n$ expansion and evaluated all integrals over time in terms of the expansion. Nestmann and Timm, in particular, pinpoint sequences of matrix products
that alternate between common matrix products and Hadamard products in each term of the expansion within Liouiville space~\cite{nestmann2019timeconvolutionless}. In our study, we utilized a comparable technique, which we refer to as the ``Hadamard trick''.  None of the preceding methodologies provide the same level of simplicity in implementation for general applications as the one offered in the present study.

{\it Main result.}  We will show that the TCL4 generator exhibits {\it infrared divergence} in sub-Ohmic baths. The underlying cause is that a small system emits soft bosons faster than the sub-Ohmic bath can remove them, leading to the diverging boson numbers~\cite{de2013approach}. The infrared divergence is well studied in gauge quantum field theories, where the scattering cross-sections involving a finite number of asymptotic bosons result in integrals that diverge in the infrared region~\cite{derezinski2004scattering}. In the usual open quantum system setup, states that have an unlimited number of bosons are located outside the Fock space~\cite{araki1963representations,merkli2006ideal}.
Then, if the ground state or the asymptotic state is computed using perturbation theory restricted within the Fock space, then the energy diverges
even if lower bounds on the ground-state energy are demonstrated~\cite{spohn1989ground}. Infrared divergence supports the presence of the {\it infrared triangle}~\cite{strominger2018lectures}, which has not yet been explored in the open quantum system setup. Consequently, we anticipate that this result will prompt additional investigation into this triangle. 

{\it Application.} To demonstrate the usefulness of the TCL4 master equation, we will look at how it models the {\it approach to a ground-state} property of open quantum systems~\cite{jakvsic1996model,bach2000return}.
We say that an open quantum system approaches a ground state if and only if the dynamics' reduced asymptotic state at absolute zero is the reduced ground state of the combined system. We prove the identity between asymptotic and ground states in generic open quantum systems with bounded Hamiltonians for the first time. However, this identity is valid only in the second-order perturbation theory and breaks down in the fourth-order theory, with strong and weak corrections depending on whether the bath is sub-Ohmic. When the bath dispersion crosses Ohmic damping, a first-order phase transition occurs in the asymptotic states, while a continuous transition occurs in the ground states.

The paper is organized as follows. We present the terminology and assumptions in Sec.~\ref{Sec:Ham}. The TCL4 generator, the asymptotic limit, and convergence criteria are developed in Sec.~\ref{Sec:generators}. The approach to a ground state will be discussed in Sec.~\ref{Sec:Approach}.
We discuss conclusions and future prospects in Sec.~\ref{Sec:conclusion}.

\section{\label{Sec:Ham} Notation, Assumptions, and Generic Open Quantum Systems.}

Our open quantum system has the total system-bath Hamiltonian
\begin{equation}
\label{Eq:Htotal}
    H_T = H_S + H_B + H_I.
\end{equation}
$H_S$ is the isolated system Hamiltonian on the $N$-dimensional system Hilbert space $\mathcal{H}_S$,
\begin{equation}
    H_S=\sum_{n=1}^N E_n\vert n\rangle\langle n\vert
\end{equation}
with nondegenerate energy levels $E_1<E_2<\cdot\cdot\cdot<E_N$. The system Bohr frequencies are defined as $\omega_{nm}=E_n-E_m$.
$H_B$ is the isolated Hamiltonian of the baths, likewise on the bath Fock space $\mathcal{H}_F$.
The baths will be considered to be contained in cubes with volumes $L^d$ and periodic boundary conditions, with delocalized discrete normal modes of linear harmonic oscillators
\begin{equation}
\label{Eq:Hbath}
    H_B=\sum_{\alpha,k}\omega_k^\alpha b_{k}^{\alpha\dagger}b_{k}^{\alpha},
\end{equation}
where $\alpha=1,2,.\, .\, .\, , N_b$ label different baths, $d$ is the spatial dimension of the bath, 
$k$ labels oscillators with frequencies $\omega_k^\alpha> 0$ in bath $\alpha$ (with $\hbar =1$), and
$b_{k}^{\alpha\dagger}$ ($b_{k}^{\alpha}$) are the creation (annihilation) operators satisfying the canonical commutation relations
\begin{equation}  
\label{Eq:CCR}[b_{k}^{\alpha},b_{q}^{\gamma\dagger}]=\delta_{k,q}\delta_{\alpha,\gamma}~\text{(Kronecker deltas)}.
\end{equation}
The energy spectrum of Eq.~\ref{Eq:Hbath} is discrete, because the bath has a finite size.

$H_I$ is a Hermitian operator describing the system bath interaction
proportional to a dimensionless (weak) coupling constant $\lambda$.
The bath is much larger than the system and the interaction Hamiltonian has bilinear form~\cite{ulrich},
\begin{equation}
    H_I = \sum_{\alpha=1}^{N_b}A^\alpha\otimes F^{\alpha},
\end{equation}
where $A^\alpha$ are Hermitian operators on the system Hilbert space (system coupling operators) and $F^{\alpha}$ are operators on corresponding bath Fock space (bath coupling operators). The system coupling operators are normalized on the Frobenious norm, i.e., $\Vert A^\alpha \Vert_2=1$, while $\lambda$ is absorbed in $F^{\alpha}$, which has the physical unit of frequency. 
We shall assume that the bath coupling operators are local oscillator displacements, which can be expanded in terms of the normal modes as
\begin{equation}
\label{Eq:BathCoupling}
     F^{\alpha} = \sum_{k}g_{k}^{\alpha}(b_{k}^{\alpha}+b_{k}^{\alpha\dagger}),
\end{equation}
where $g_{k}^{\alpha}$ are the (real) form factors, which have the physical units of frequency and scale with the bath size as $1/L^{d/2}$~\cite{nemati2022coupling}.

Let us further use $H_0 = H_S + H_B$, for
the free Hamiltonian.
Consequently, the master equation appears in the interaction picture
\begin{equation}\label{eqn:ME_start}
    \frac{d \varrho(t)}{dt} = -i [H_I(t),\varrho(t)] \equiv \mathcal{L}_I(t)\varrho(t),
\end{equation}
where the interaction picture operators are $H_I(t) = e^{iH_0 t} H_I e^{-iH_0 t}$ and $\varrho(t)=e^{iH_0 t} \rho(t) e^{-iH_0 t}$ and the Liouvillian is
$\mathcal{L}_I(t)= -i [H_I(t),\,]$.
{\it Throughout this paper, the system and baths are initialized into a factorized state at time $t=0$.}

The operators $F^\alpha$ are fully described by two-point time correlations computed in the bath's initial state ($\varrho_B$), which will serve as the reference state for deriving the TCL4 master equation utilizing the projector-operator method~\cite{BreuerHeinz-Peter1961-2007TToO}. 
We shall label the bath correlation functions (BCF) as 
\begin{equation}
    C_{\alpha}(t) = \langle F^\alpha(t) F^{\alpha}(0)\rangle=\text{Tr}[\varrho_B F^\alpha(t) F^{\alpha}(0)],
\end{equation}
which has the physical unit of frequency squared. The baths themselves are not correlated, that is,
\begin{equation}
    \langle F^\alpha(t_1) F^{\gamma}(t_2)\rangle = \text{Tr}_{B}[\varrho_B F^\alpha(t_1) F^{\gamma}(t_2)] = 0,
\end{equation}
if $\alpha \not = \gamma$, which also follows from Eq.~\ref{Eq:CCR}. 

These BCFs will be determined from a characteristic spectral density at zero temperature, which has the physical unit of frequency and is defined as
\begin{equation}
\label{Eq:J_sum}
\tilde{\mathcal{J}}_{\omega}^{\alpha}=\pi\sum_{k}  (g_{k}^{\alpha})^2\delta(\omega-\omega_k^\alpha),
\end{equation}
in terms of which the BCF at any temperature can be expressed as
\begin{equation}
\label{Eq:BCF}
C_\alpha(t)=\frac{1}{\pi}\int_0^\infty d\omega\,\tilde{\mathcal{J}}_{\omega}^{\alpha}(\cos\omega t\coth\frac{1}{2}\beta_\alpha\omega-i\sin \omega t),
\end{equation}
where $\beta_\alpha=1/(k_BT_\alpha)$. {\it The labels with (without) a tilde will indicate quantities at zero (any) temperature.}

The time-dependent or timed spectral density  in bath $\alpha$ is defined as
\begin{equation}
\label{Eq:timedSD}
\Gamma_\omega^\alpha(t)=\int_0^t d\tau C_\alpha(\tau)e^{i\omega \tau}.
\end{equation}
The spectral density  and the principal density  at temperature $T_\alpha$ are the real and imaginary parts of the half-sided Fourier transform of the BCF, e.g.,
\begin{equation}
\label{Eq:deftsd}
\Gamma_\omega^\alpha = \lim_{t\to\infty}\Gamma_\omega^\alpha(t)= \int_0^\infty d\tau  C_\alpha(\tau)e^{i\omega \tau},
\end{equation}
namely,
$\mathcal{J}_{\omega}^{\alpha}=\text{Re}(\Gamma_\omega^\alpha)$ and
$\mathcal{S}_{\omega}^{\alpha}=\text{Im}(\Gamma_\omega^\alpha$).
The frequency-derivatives of the spectral and principal densities at zero frequency, will be shown to be related to the soft-boson numbers in asymptotic and ground states, respectively. They can be obtained by taking the derivative of Eq.~\ref{Eq:deftsd}, substituting $\omega=0$, and taking the real and imaginary parts. The result is
\begin{eqnarray}
\label{Eq:genderivativeJ}
\frac{\partial \mathcal{J}_{\omega}^\alpha}{\partial\omega}\bigg\vert_{\omega=0}&=&-\int_0^\infty t\text{Im}\,C_\alpha(t)dt,\\
\label{Eq:genderivativeS}
\frac{\partial \mathcal{S}_{\omega}^\alpha}{\partial\omega}\bigg\vert_{\omega=0}&=&\int_0^\infty t\text{Re}\,C_\alpha(t)dt.
\end{eqnarray} 
Because of Eq.~\ref{Eq:BCF}, Eq.~\ref{Eq:genderivativeJ} is temperature independent.

The spectral density at nonzero temperature is related to that at zero temperature as
\begin{equation}
\label{Eq:KMSplus}
    \mathcal{J}_\omega^\alpha=\frac{\tilde{\mathcal{J}_\omega^\alpha}}{1-e^{-\beta_\alpha\omega}}, \,\omega>0 
\end{equation}
and satisfies the Kubo-Martin-Schwinger (KMS) condition 
\begin{equation}
\label{Eq:KMS}
\mathcal{J}^\alpha_{-\omega}=\exp(-\beta_\alpha\omega)\mathcal{J}^\alpha_{\omega}, \forall \omega.
\end{equation} The principal and the spectral densities are related via
the Kramers-Kronig relation
\begin{equation}\label{Eq:KramersK}
\mathcal{S}_{\omega}^{\alpha}=\frac{1}{\pi}\mathcal{P}\int_{-\infty}^{\infty}d\omega'\frac{\mathcal{J}_{\omega'}^{\alpha}}{\omega-\omega'},
\end{equation}
where $\mathcal{P}$ indicates the principal value.  See Appendix~\ref{Sec:KMS} for derivation of Eqs.~\ref{Eq:KMSplus}-\ref{Eq:KramersK}.

\subsection{Assumptions}

We make the following assumptions on the system.
\begin{enumerate}
    \item System spectral density condition. We assume that in the the thermodynamic limit, i.e., $L\to\infty$, 
\begin{equation}
\tilde{\mathcal{J}}_{\omega}^{\alpha}\mapsto2\pi\lambda^2\omega_c\left(\frac{\omega}{\omega_c}\right)^se^{-\frac{\omega}{\omega_c}}\Theta(\omega); s>0.
\label{Eq:SDexp}
\end{equation}
Here and in the rest of the main text, the symbol $\mapsto$ implies that the thermodynamic limit is taken.
The cutoff frequency $\omega_c$, the power-law exponent $s$, and $\lambda$  can implicitly vary between the baths.
$\Theta(\omega)$ is the Heaviside step-function.
    \item Fermi golden rule condition. We assume that  in the thermodynamic limit
    \begin{equation}
    \label{Eq:FGR}
    \sum_{\alpha=1}^{N_b}\vert A^\alpha_{nm}\vert^2\mathcal{J}_{\omega_{nm}}^\alpha>0; n,m=1,...,N, \vert n-m\vert>0,
    \end{equation}
        \item Dephasing condition. We assume that
    \begin{equation}
    \label{Eq:MDF}
\sum_{\alpha=1}^{N_b}\sum_{n,m=1}^N\vert A^\alpha_{nn}-A^\alpha_{mm}\vert >0.\\
    \end{equation}
\end{enumerate}

Equation~\ref{Eq:SDexp} is the regularity condition on the spectral density (SD). 
The Heaviside step function in Eq.~\ref{Eq:SDexp} follows from the definition in Eq.~\ref{Eq:J_sum} and the fact that the frequencies of the bath modes are positive. It complies with the second law of thermodynamics, Clausius's statement, that heat does not spontaneously flow from a colder to a hotter body. Namely, when the bath is at zero temperature, only energy quanta at a positive frequency can be emitted from the system into the bath.

The bath is defined as sub-Ohmic for $s<1$, Ohmic for $s=1$, and super-Ohmic for  $s>1$. 
In terms of physical considerations, it has been observed that if the small quantum system is connected to a local displacement of the bath's oscillator, then $s=d-2$, with $d$ representing the dimension of the bath~\cite{nemati2022coupling}.
Consequently, a cube shaped bath should be Ohmic.
We specifically require that $s>0$ so that the ground-state energy of the system is lower bounded~\cite{spohn1989ground}. This also ensures that the Davies generator of the dynamics and its asymptotic state are bounded~\cite{davies1974}. 
The asymptotic and ground states, however, need not lie in the Fock space of the infinite bath because the number of bosons can diverge, if $s<1$~\cite{de2013approach,spohn1989ground}, invalidating the results of the second-order master equations at a sufficiently long timescale.
The sufficient condition for  $H_T$ to admit asymptotic~\cite{faupin2014rayleigh,de2015asymptotic} and ground states~\cite{arai1997existence,bach1999spectral,gerard2000existence,griesemer2001ground,hasler2011ground,abdesselam2011ground} is $s>1$, which is significantly stricter than the existence of the lower energy bound and the Davies generator.

The weak coupling constant, $\lambda>0$, is defined such that an unbiased spin boson model with energy splitting $E_2-E_1=1$ has a Berezinskii–Kosterlitz–Thouless transition at  $\lambda=1$ when $s=1$ and $\omega_c\gg 1$~\cite{Leggett}. 
Throughout the paper, we refer to the dynamics as being {\it infrared-divergent}, if and only if it diverges at $t\to\infty$ for arbitrarily small but nonzero $\lambda$. 

Consider infrared regular dynamics, where the asymptotic generator lacks eigenvalues with positive real parts when $\lambda <\lambda_0$,
for some $\lambda_0>0$.
We call such $\lambda$ as {\it sufficiently small}. 
We will assume throughout the paper that $\lambda$ is sufficiently small whenever the dynamics is regular. Thus, we do not consider phase transitions driven by the coupling strength, such as the Berezinskii–Kosterlitz–Thouless transition~\cite{Leggett}.

Inserting $\beta_\alpha=\infty$ and Eq.~\ref{Eq:SDexp} into Eq.~\ref{Eq:BCF}, we obtain
\begin{equation}
\label{Eq:Cexplicit}
\tilde{C}_\alpha(t)\mapsto 2\Gamma(s+1)\lambda^2\omega_c^2\frac{1}{(1+i\omega_ct)^{s+1}},
\end{equation}
which has the real part
\begin{equation}
\label{Eq:BCFreal}
    \text{Re}[\tilde{C}_\alpha(t)]=2\Gamma(s+1)\lambda^2\omega_c^2\frac{\cos[(s+1)\arctan(\omega_ct)]}{[1+\omega_c^2t^2]^\frac{s+1}{2}}
\end{equation}
and the imaginary part
\begin{equation}
\label{Eq:BCFimag}
    \text{Im}[\tilde{C}_\alpha(t)]=-2\Gamma(s+1)\lambda^2\omega_c^2\frac{\sin[(s+1)\arctan(\omega_ct)]}{[1+\omega_c^2t^2]^\frac{s+1}{2}}.
\end{equation}
Due to the imaginary part's temperature independence in  Eq.~\ref{Eq:BCF}, the bath correlations decay as power-law at any temperature. At $T_\alpha=0$, we see that the
the real part of the BCF decays 
as $t^{-s-1}$ in the open interval $0<s<2$.
The
imaginary part of the BCF in Eq.~\ref{Eq:BCFimag} also decays 
as $t^{-s-1}$, except in the one and only one case,  $s=1$, where it decays as
$t^{-s-2}$.
The different decaying power-laws in Eqs.~\ref{Eq:BCFreal} and~\ref{Eq:BCFimag} will ultimately result in distinct order phase-transitions in the ground and asymptotic states at $s=1$ (Theorem 1, Sec.~\ref{Sec:ergodicity}).

 If a theorem is valid under conditions ~\ref{Eq:FGR} and~\ref{Eq:MDF}, then it may not be true for all open quantum systems. However, since the open quantum systems satisfying ~\ref{Eq:FGR} and~\ref{Eq:MDF} form a dense set, all slight perturbations of them will also satisfy the conditions, making these open quantum systems the most important to address in theorems. These open quantum systems are referred to as generic.

The condition~\ref{Eq:FGR} implies that for each Bohr frequency, there exists at least one system coupling operator with a nonzero matrix element and nonzero spectral density. This condition is the Fermi golden rule condition from Refs.~\cite{merkli2022dynamics,merkli2022dynamics2} extended for multiple independent baths. It ensures that the energy exchange between the system and the baths has a unique asymptotic state, in the absence of infrared divergence.

At least one $A^\alpha$ must have inhomogeneous diagonal matrix elements in order to satisfy the requirement~\ref{Eq:MDF}. 
The system then displays inhomogeneous dephasing, where the bath states corresponding to the various system energy eigenstates shift over time and gradually become more orthogonal at irregular rates (\cite[Sec.~4.2]{BreuerHeinz-Peter1961-2007TToO}). 
The conditions~\ref{Eq:FGR} and~\ref{Eq:MDF} are realistic because relaxation and dephasing are prevalent in real world.

The time-dependent spectral density, its derivative at zero frequency, and the principal density, all at zero temperature, are
\begin{widetext}
\begin{equation}
\label{Eq:Aomegat}
\tilde{\Gamma}_\omega^\alpha (t)=\left\{
  \begin{array}{ll}
2i\lambda^2\Gamma(s+1)\omega_c\left(-\frac{\omega}{\omega_c}\right)^s e^{-\frac{\omega}{\omega_c}}\left[\Gamma\left(-s,-\frac{\omega}{\omega_c}-i\omega t\right)-\Gamma\left(-s,-\frac{\omega}{\omega_c}\right)\right], & \hbox{$\omega\neq 0$;} \\
2i\lambda^2\Gamma(s)\omega_c\left[(1+i\omega_ct)^{-s}-1\right],& \hbox{$\omega= 0$,}
  \end{array}
\right.
\end{equation}
\begin{equation}
\label{Eq:powerlaw}
\frac{\partial \tilde{\Gamma}_{\omega}^\alpha(t) }{\partial\omega}\bigg\vert_{\omega=0}=
\left\{
  \begin{array}{ll}
-2i\lambda^2\Gamma(s+1)\bigg[\frac{1}{s(s-1)}-\frac{(1+i\omega_ct)^{1-s}}{s-1}
+\frac{(1+i\omega_ct)^{-s}}{s}\bigg],  & \hbox{$s\neq 1$;}\\
-2i\lambda^2\Gamma(s+1)\bigg[
\frac{1}{2}\ln[1+(\omega_ct)^2]+i\arctan \omega_ct-\frac{i\omega_ct}{1+i\omega_ct}\bigg],
& \hbox{$s= 1$,}  \end{array}
\right.
\end{equation}
\begin{equation}
\tilde{\mathcal{S}}_{\omega}^{\alpha}=\left\{
  \begin{array}{ll}
-2\lambda^2\Gamma(s+1)\omega_c\text{Re}\left[\left(-\frac{\omega}{\omega_c}\right)^s\right]e^{-\frac{\omega}{\omega_c}}\Gamma\left(-s,-\frac{\omega}{\omega_c}\right), & \hbox{$\omega\neq 0$;}\\
    -2\lambda^2\Gamma(s)\omega_c, & \hbox{$\omega= 0$.}
\end{array}
\right.
\label{Eq:PDexp}
\end{equation}
\end{widetext}
$\Gamma(z)$ and $\Gamma(\mu,z)$ within Eqs.~\ref{Eq:Aomegat}-\ref{Eq:PDexp} are the gamma and the upper incomplete gamma functions respectively. 
\begin{figure}[h]
  \includegraphics[width=0.47\textwidth]{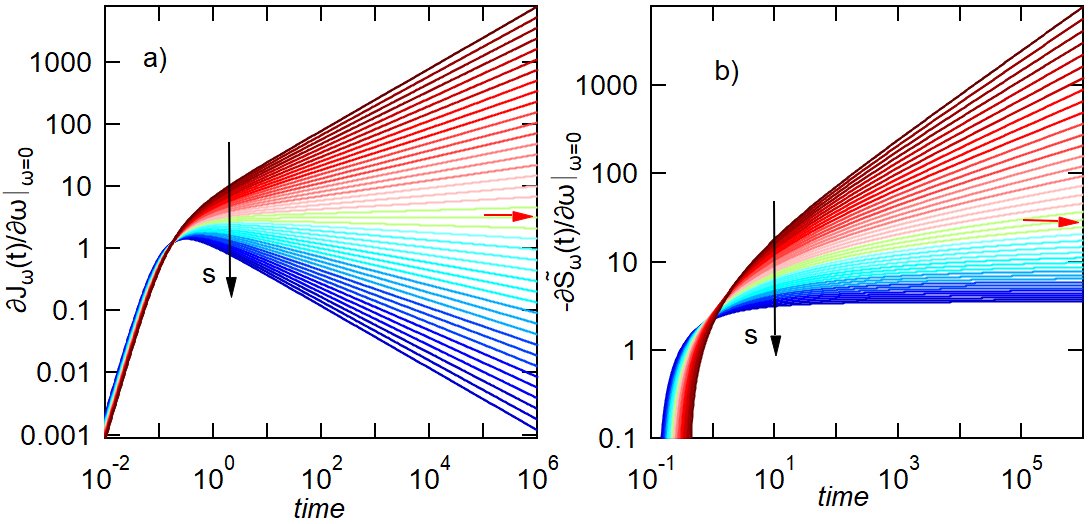}
\caption{\label{Fig:rainbow}The derivatives of the spectral and principal density divided by $\lambda^2$ versus time. The derivatives are interpreted as the number of soft bosons emitted into the bath in the asymptotic and ground states, respectively.
 $s$ varies the interval
$0.5\leq s\leq 1.5$ in increments of $0.025$ from top to bottom.  Red arrow points to the Ohmic bath values.} 
\end{figure}


Figures~\ref{Fig:rainbow}~(a) and (b) display the zero frequency derivatives of the time-dependent spectral and principal densities versus time of the dynamics, as a function of $s$.  At $t\to\infty$, $\partial{\mathcal{J}}/\partial\omega\vert_{\omega=0}$ approaches zero, a constant, and infinity,
if $s>1$, $s=1$, and $s<1$, respectively. In contrast, $\partial\tilde{\mathcal{S}}/\partial\omega\vert_{\omega=0}$ approaches a constant
dependent on $s$ when $s>1$, and infinity when $s\leq 1$.

Taking the limit $t\to\infty$  in Eq.~\ref{Eq:powerlaw}, we obtain the explicit derivatives of the spectral and principal densities at $\omega=0$:
\begin{eqnarray}
\label{Eq:dJdw}
\frac{\partial \tilde{\mathcal{J}}_{\omega}^\alpha}{\partial\omega}\bigg\vert_{\omega=0}&=&\left\{
\begin{array}{ll}
0, & \hbox{$s>1$;}\\
\pi\lambda^2, & \hbox{$s=1$;} \\
\infty, & \hbox{$s<1$.}
\end{array}
\right.\\
\label{Eq:dSdw}
\frac{\partial\tilde{\mathcal{S}}_{\omega}^\alpha}{\partial\omega}\bigg\vert_{\omega=0}&=&\left\{
\begin{array}{ll}
-2\lambda^2\Gamma(s-1), & \hbox{$s>1$;}\\
\infty, & \hbox{$s\leq 1$.} 
\end{array}
\right.
\end{eqnarray}
Recall from the discussion below Eq.~\ref{Eq:genderivativeS} that the derivative in Eq.~\ref{Eq:dJdw} is valid at arbitrary temperature. 

\subsection{Number of emitted soft bosons\label{Sec:ergodicity}}

Let us establish a connection between the number of emitted soft bosons in the bath at zero temperature and the spectral density.
In quantum field theory, 
the number of emitted bosons is closely related to the problem of asymptotic completeness~\cite{de2013approach}. 
In order to prove that the asymptotic states are spanned by the Fock space basis, it is necessary to show that the number of bosons released into the bath is limited~\cite{de2013approach}.  

Consider the small system initially in an excited eigenstate with a Bohr frequency $\epsilon>0$. During the time interval $t$ that fulfills $1/\epsilon<t<1/(\vert A_{21}\vert^2 \tilde{\mathcal{J}}_\epsilon)$, the population of the excited state is decreased by $\vert A_{21}\vert^2 \tilde{\mathcal{J}}_\epsilon t$ according to the Fermi golden rule.  $\epsilon \vert A_{21}\vert^2 \tilde{\mathcal{J}}_\epsilon t$ is the resulting energy emitted into the bath. The number of emitted bosons is determined by dividing by $\epsilon$ using $\vert A_{21}\vert\sim 1$ in our units, i.e., $\langle N_{sb}(t)\rangle\sim \tilde{\mathcal{J}}_\epsilon t\sim t\epsilon \tilde{\mathcal{J}}_\epsilon /\epsilon$.  By taking the limits $\epsilon\to 0_+$, $t\to +\infty$, and $\epsilon t=\text{const}$, we find the soft-boson number $\langle N_{sb}(\infty)\rangle\approx\partial \tilde{\mathcal{J}}_\omega /\partial\omega\vert_{\omega=0}$.
On the other hand, Eq.~\ref{Eq:dSdw} gives the soft boson number in the combined system-bath ground state. See Appendix~\ref{Sec:vanHove} for further details.

Roeck and Kupiainen computed rigorous boson number bounds in generic scattering systems  at zero temperature using a polymer expansion in statistical physics~\cite{de2013approach}. At time $t>0$, they find
 \begin{equation}
 \label{Eq:roeck}
 \langle e^{\kappa N(t)}\rangle\leq \left\{
 \begin{array}{ll}
\breve{C}\exp(C), & \hbox{$s\geq 1$};\\
\breve{C}\exp[Ct^{1-s}], & \hbox{$s< 1$},
\end{array}
\right.
 \end{equation}
 where $N(t)$ is the number of emitted bosons.  $\breve{C}$ and $C$ are independent of $\lambda$, $\kappa$, and $t$, if  $0<\lambda\leq \lambda_0$ and $\vert \kappa\vert \leq \kappa_0$, for some $\lambda_0>0$ and $\kappa_0>0$.  

On the other hand, we have the asymptotic behavior from Eq.~\ref{Eq:powerlaw} at time $t\gg 1/\omega_c$, 
 \begin{eqnarray}
\nonumber
\frac{\partial\tilde{\mathcal{J}}_{\omega}^\alpha(t)}{\partial\omega}\bigg\vert_{\omega=0}&\mapsto&\left\{
\begin{array}{ll}
\frac{2\lambda^2\Gamma(s+1)\sin \frac{(1-s)\pi}{2}}{1-s} (\omega_ct)^{1-s}, & \hbox{$s\neq 1$;}\\
\pi\lambda^2(1-\frac{2}{\pi\omega_ct}), & \hbox{$s=1$}
\end{array}.
\right.
\end{eqnarray}
Comparing this with Eq.~\ref{Eq:roeck}, 
along with the previous paragraph showing that
$\partial\tilde{\mathcal{J}}_{\omega}^\alpha/\partial\omega\vert_{\omega=0}$ is the soft boson number at $t\to+\infty$,
suggests that $\partial\tilde{\mathcal{J}}_{\omega}^\alpha(t)/\partial\omega\vert_{\omega=0}$ is the number of soft bosons that have been emitted up to time $t$.
We shall see that the TCL4 master equation's generator is linear with $\partial\tilde{\mathcal{J}}_{\omega}^\alpha(t)/\partial\omega\vert_{\omega=0}$.
Consequently, the boson number determines the existence of the generator of the TCL master equation.

The primary outcome of the work is the following theorem.
If we assume that the open quantum system is generic, the states are computed up to the fourth order of the perturbation theory, the interaction $\lambda$ is arbitrarily small but not zero, and the thermodynamic limit is conventional, then, as a function of the dispersion $s$, we have the following theorem:

{\it\underline{Theorem 1}: The asymptotic states undergo a discontinuous phase
transition, whereas the ground
states undergo a continuous phase transition at $s=1$.}

Outline of the proof:
We will demonstrate in Sec.~\ref{Sec:CPT}
that the reduced asymptotic states at $T=0$K are linear with $\partial \tilde{\mathcal{J}}/\partial\omega$ and independent of $\partial \tilde{\mathcal{S}}/\partial\omega$. 
Consequently, Eq.~\ref{Eq:dJdw} 
represents a discontinuous transition from a bounded to an unbounded reduced asymptotic state,
which takes place when $s$ changes from $1$ to $1_-$.
The transition is consistent with the discontinuous transition in the boson number bound~\ref{Eq:roeck}.
The norm of the asymptotic state is the order parameter representing the discontinuous transition of the asymptotic state out of the Fock space. 

In Sec.~\ref{Sec:RSPT} we will show that the reduced ground states are linear with $\partial \tilde{\mathcal{S}}/\partial\omega$. Consequently, Eq.~\ref{Eq:dSdw} 
represents a continuous transition 
from a bounded to an unbounded reduced ground state, computed perturbatively,
which takes place when $s$ approaches $1_+$. 

\section{\label{Sec:generators} TCL4 Generator}

The system and the bath are initialized into a factorized state at time $t=0$. If the resulting quantum-dynamical map is invertible, then it may be represented by the exact time-convolutionless master equation as follows~\cite{BreuerHeinz-Peter1961-2007TToO}:
\begin{equation}
\label{Eq:TCL}
    \frac{d\rho}{dt}=\mathcal{R}^{0}\rho+\mathcal{R}^{2}(t)\rho+\mathcal{R}^{4}(t)\rho+...
\end{equation} The generator of the dynamics is thus expressed in terms of tensors $\mathcal{R}^{2n}(t)$, each proportional to $\lambda^{2n}$.
This will be the TCL2n master equation when it is truncated to the first $2n$ terms. Due to the hermiticity preservation of the master equation, the generator's contributions can be written as 

\begin{equation}
\label{Eq:deltaRHC}
\mathcal{R}_{nm,ij}^{2k}(t)=\delta\mathcal{R}_{nm,ij}^{2k}(t)+\delta\mathcal{R}_{mn,ji}^{2k\star}(t), k=0,1,2,...
\end{equation}

The matrix elements of the generator components are as follows:
\begin{eqnarray}
\label{Eq:R0th}
\delta\mathcal{R}_{nm,ij}^{0}&=&-iE_n\delta_{ni}\delta_{mj}\\
\nonumber
\delta\mathcal{R}_{nm,ij}^{2}(t)&=&\sum_{\alpha=1}^{N_b}\big[A_{ni}^\alpha A_{
jm}^\alpha\Gamma_{in}^\alpha(t)\\
\label{Eq:R2nd}
&-&\delta_{jm}\sum_{k=1}^{N}A_{nk}^\alpha A_{ki}^\alpha\Gamma_{ik}^\alpha(t)\big].
\end{eqnarray}
\begin{widetext}
\begin{eqnarray}
\label{Eq:ExplicitLiouv}
       \delta\mathcal{R}_{nm,ij}^{4}(t) &=\delta_{jm} &\sum_{\alpha,\beta=1}^{N_b}\sum_{a,b,c=1}^N\Big\{     
        A_{na}^\alpha A_{ab}^\beta A_{bc}^\alpha A_{ci}^\beta [\mathsf{F}_{cb,ci,ac}^{\alpha\beta}(t) - \mathsf{R}_{cb,ab,bi}^{\alpha\beta}(t)]
        + A_{na}^\alpha A_{ab}^\beta A_{bc}^\beta A_{ci}^\alpha \mathsf{R}_{ic,ab,bi}^{\alpha\beta}(t)
        \\
        \label{Eq:Expl1}
        &-& A_{na}^\alpha A_{ab}^\alpha A_{bc}^\beta A_{ci}^\beta \mathsf{F}_{ba,ci,ac}^{\alpha\beta}(t)\Big\}
        +\sum_{\alpha,\beta=1}^{N_b}\sum_{a,b=1}^N\Big\{  A_{na}^\beta A_{ab}^\alpha A_{bi}^\beta A_{jm}^\alpha  [\mathsf{R}_{ba,na,ai}^{\alpha\beta}(t)-\mathsf{F}_{ba,bi,nb}^{\alpha\beta}(t)]
        \\
        \label{Eq:Expl2}
        &+& A_{na}^\alpha A_{ab}^\beta A_{bi}^\beta A_{jm}^\alpha \mathsf{F}_{an,bi,nb}^{\alpha\beta}(t)
        - A_{na}^\beta A_{ab}^\beta A_{bi}^\alpha  A_{jm}^\alpha \mathsf{R}_{ib,na,ai}^{\alpha\beta}(t)
        \\
        \label{Eq:Expl3}
        &+& A_{na}^\alpha A_{ab}^\alpha A_{bi}^\beta A_{jm}^\beta [\mathsf{C}_{ba,jm,ai}^{\alpha\beta}(t)+\mathsf{R}_{ba,jm,ai}^{\alpha\beta}(t)]
        - A_{na}^\alpha A_{ab}^\beta A_{bi}^\alpha A_{jm}^\beta [\mathsf{C}_{ib,jm,ai}^{\alpha\beta}(t)+\mathsf{R}_{ib,jm,ai}^{\alpha\beta}(t)]
        \\
        \label{Eq:Expl4}
        &-& A_{na}^\alpha A_{ai}^\beta A_{jb}^\beta A_{bm}^\alpha[\mathsf{C}_{an,jb,ni}^{\alpha\beta}(t)+\mathsf{R}_{an,jb,ni}^{\alpha\beta}(t)]
        + A_{na}^\beta A_{ai}^\alpha A_{jb}^\beta A_{bm}^\alpha [\mathsf{C}_{ia,jb,ni}^{\alpha\beta}(t)+\mathsf{R}_{ia,jb,ni}^{\alpha\beta}(t)]\Big\},\\
    \label{Eq:preferredF}
    \mathsf{F}_{\omega_1\omega_2\omega_3}^{\alpha\beta}(t) &=&-\int_0^{t} d\tau \Delta \Gamma_{\omega_1}^\alpha(t,\tau) \Delta \Gamma_{\omega_2}^{\beta T}(t,t-\tau) e^{-i(\omega_1 + \omega_2 + \omega_3)\tau}+i\Gamma_{\omega_2}^{\beta T}(t)\frac{\Gamma_{-\omega_2-\omega_3}^\alpha(t)-\Gamma_{\omega_1}^\alpha(t)}{\omega_1+\omega_2+\omega_3}\\
    \label{Eq:preferredC}
    \mathsf{C}_{\omega_1\omega_2\omega_3}^{\alpha\beta}(t) &=&-\int_0^{t} d\tau \Delta \Gamma_{\omega_1}^\alpha(t,\tau) \Delta\Gamma_{\omega_2}^{\beta\star}(t,t-\tau) e^{-i(\omega_1
     + \omega_2 + \omega_3)\tau}
+i\Gamma_{\omega_2}^{\beta\star}(t)\frac{\Gamma_{-\omega_2-\omega_3}^\alpha(t)-\Gamma_{\omega_1}^\alpha(t)}{\omega_1+\omega_2+\omega_3}\\
\label{Eq:preferredR}
   \mathsf{R}_{\omega_1\omega_2\omega_3}^{\alpha\beta}(t) &=&-\int_0^{t} d\tau \Delta \Gamma_{\omega_1}^\alpha(t,\tau) \Delta \Gamma_{\omega_2}^\beta(t,\tau) e^{-i(\omega_1 + \omega_2 + \omega_3)\tau}
   +i\Gamma_{\omega_2}^\beta(t)\frac{\Gamma_{-\omega_2-\omega_3}^\alpha(t)-\Gamma_{\omega_1}^\alpha(t)}{\omega_1+\omega_2+\omega_3},
\end{eqnarray}
\end{widetext}
where
\begin{equation}
\label{Eq:DeltaGamma}
\Delta \Gamma_{\omega}^\alpha(t,\tau) = \Gamma_{\omega}^\alpha(t) - \Gamma_{\omega}^\alpha(\tau).
\end{equation}

The derivation of the TCL4 generator is given in Appendix~\ref{Sec:appendixA}.
Equations~\ref{Eq:preferredF}-\ref{Eq:preferredR} define the system's three-dimensional (3D) spectral densities.
In Equations~\ref{Eq:ExplicitLiouv}-\ref{Eq:Expl4}, $\mathsf{F}_{(ab)(cd)(ef)}^{\alpha\beta}(t)=\mathsf{F}_{\omega_1\omega_2\omega_3}^{\alpha\beta}(t)$,
where $\omega_1=\omega_{ab}$, $\omega_2=\omega_{cd}$, and $\omega_3=\omega_{ef}$; {\it mutadis mutandis} $\mathsf{C}$ and $\mathsf{R}$.
The superscript $T$ in Eq.~\ref{Eq:preferredF} represents transposition in the energy basis, e.g., $\Delta \Gamma_{\omega_2}^{\beta T}=\Delta \Gamma_{-\omega_2}^{\beta}$. The star in
the superscript in Eq.~\ref{Eq:preferredC} indicates complex conjugation. The right-hand side of Eqs.~\ref{Eq:preferredF}-\ref{Eq:preferredR} should be interpreted as a limit in the case of $\omega_1+\omega_2+\omega_3= 0$, i.e.,
\begin{equation}
\label{Eq:fraction}
\lim_{\Sigma_i\omega_i\to 0}\frac{\Gamma_{-\omega_2-\omega_3}^\alpha(t)-\Gamma_{\omega_1}^\alpha(t)}{\omega_1+\omega_2+\omega_3}=-\frac{\partial \Gamma_{\omega_1}^\alpha(t)}{\partial\omega_1}.
\end{equation}

Once we have the time-dependent spectral density, we can express both the second- and fourth-order generator contributions in terms of that spectral density. Therefore, the TCL4 master equations, in this particular form, has the same level of universality as the TCL2 master equation. The TCL4 master equation is trickier to use at this time because we have to integrate over the products of the time-dependent spectral densities in Eqs.~\ref{Eq:preferredF}-\ref{Eq:preferredR}. Nevertheless, this may be simply managed through numerical computations.
The integrals may be computed with a moderate amount of diligence, approaching double-precision accuracy, as we shall elaborate on in Sec.~\ref{Sec:Return}.
After calculating the integrals, we may modify the system operators without the need to recalculate the 3D-spectral densities, resulting in faster computing. 

We take note of the claim made by Karasev and  Teretenkov that the TCL expansion was expressed without the use of any integrals in time~\cite{karasev2023timeconvolutionless}, which suggests that the remaining integrals in~\ref{Eq:preferredF}-\ref{Eq:preferredR} can be further reduced. We will defer this for future investigation.

\subsection{Asymptotic TCL4 generator}

In the thermodynamic limit, we substitute the smooth function from Eq.~\ref{Eq:powerlaw} into the spectral density.
Then, the analytical characteristics of the TCL4 generator can be different from those of the TCL2 generator, because they depend on the derivative of the SD. In particular, Eqs.~\ref{Eq:dJdw} and~\ref{Eq:dSdw} are infrared divergent in a sub-Ohmic bath leading to a diverging generator and diverging asymptotic state.

Let us consider the baths in the thermodynamic limit. According to the definition in Eq.~\ref{Eq:DeltaGamma}, $\Delta\Gamma_{\omega}^\alpha(t,\tau)$ and $\Delta\Gamma_\omega^{\beta T}(t,t-\tau)$ 
vary from $\Gamma_{\omega}^\alpha(t)$ to $0$ and from $0$ to $\Gamma_\omega^{\beta T}(t)$, respectively, when $\tau$ changes from $0$ to $t$.
If now we take the asymptotic limit, e.g., $t\to\infty$, then $\Delta\Gamma_{\omega}^\alpha(t,\tau)\to \Delta\Gamma_{\omega}^\alpha(\tau)=\Gamma_{\omega}^\alpha-\Gamma_{\omega}^\alpha(\tau)$, which approaches zero when $\tau$ is larger than the characteristic timescale of the bath. In contrast, $\Delta\Gamma_\omega^{\beta T}(t,t-\tau)$ approaches zero when 
$\tau$ is smaller than $t$ minus the characteristic timescale of the bath, when $t$ is very large.
As a result, we have
\begin{equation}
\label{Eq:Overlap}
\lim_{t\to\infty}\Delta\Gamma_{\omega_1}^\alpha(t,\tau)\Delta\Gamma_{\omega_2}^{\beta T}(t,t-\tau)=0,\forall \tau\in (0,t).
\end{equation}

The same applies after replacing $\Delta\Gamma_{\omega_2}^{\beta T}$ with $\Delta\Gamma_{\omega_2}^{\beta \star}$ in the above.
Furthermore, we propose a stronger condition, 
\begin{equation}
\lim_{t\to\infty}\int_0^td\tau\,\Delta\Gamma_{\omega_1}^\alpha(t,\tau)\Delta\Gamma_{\omega_2}^{\beta\, T,\star}(t,t-\tau)e^{-i(\omega_1+\omega_2+\omega_3)\tau}=0.
\label{Eq:OverlapIntegral}
\end{equation}
See Appendix~\ref{Appendix:TCL4integrals} for the verification of~\ref{Eq:Overlap} and~\ref{Eq:OverlapIntegral} for any $s>0$. 

Hence, we obtain the asymptotic TCL4 generator by utilizing the asymptotic 3D spectral densities, defined as
\begin{eqnarray}
\label{Eq:preferredFasymptotic}    \mathsf{F}_{\omega_1\omega_2\omega_3}^{\alpha\beta}& = &i\Gamma_{\omega_2}^{\beta T}\frac{\Gamma_{-\omega_2-\omega_3}^\alpha-\Gamma_{\omega_1}^\alpha}{\omega_1+\omega_2+\omega_3}\\
\label{Eq:preferredCasymptotic}
    \mathsf{C}_{\omega_1\omega_2\omega_3}^{\alpha\beta}& = &i\Gamma_{\omega_2}^{\beta\star}\frac{\Gamma_{-\omega_2-\omega_3}^\alpha-\Gamma_{\omega_1}^\alpha}{\omega_1+\omega_2+\omega_3}\\
\label{Eq:preferredRasymptotic}
    \mathsf{R}_{\omega_1\omega_2\omega_3}^{\alpha\beta}& = &i\Gamma_{\omega_2}^\beta\frac{\Gamma_{-\omega_2-\omega_3}^\alpha-\Gamma_{\omega_1}^\alpha}{\omega_1+\omega_2+\omega_3} -\\
\label{Eq:3DSD}
    \int_0^{\infty} &d\tau&\Delta\Gamma_{\omega_1}^\alpha(\tau)\Delta\Gamma_{\omega_2}^\beta(\tau) e^{-i(\omega_1 + \omega_2 +\omega_3)\tau}.
\end{eqnarray}
These formulas are applicable for any $s > 0$, temperature, and frequencies $\vert\omega_1\vert+\vert\omega_2\vert+\vert\omega_3\vert>0$, but not at $\omega_1=\omega_2=\omega_3=0$ and $s<1/2$, as shown in Appendix~\ref{Appendix:TCL4integrals}. However, the Appendix also shows that the TCL4 generator is independent of the 3D spectral densities at $\omega_1=\omega_2=\omega_3=0$.
Thus, the asymptotic TCL4 generator is computed accurately by replacing the time-dependent 3D spectral densities in Eqs.~\ref{Eq:ExplicitLiouv}-\ref{Eq:Expl4} with the respective asymptotic limits in Eqs.~\ref{Eq:preferredFasymptotic}-\ref{Eq:3DSD}.

\subsection{TCL4 generator's convergence at $T=0$ K.\label{Sec:convergence}}

At zero temperature, according to Eqs.~\ref{Eq:SDexp} and~\ref{Eq:PDexp}, $\tilde{\Gamma}_\omega^\alpha=\tilde{\mathcal{J}}_\omega^\alpha+i\tilde{\mathcal{S}}_\omega^\alpha$ is continuous when $s>0$.
However, at a positive temperature, 
the spectral density diverges at zero frequency if $s<1$~\cite{ulrich},
as $1/\omega^{1-s}$. 
To keep the paper simple, we will narrow the scope of the convergence analysis to $T=0$\,K only, and defer the analysis of the convergence at a positive temperature
to a future publication. 

We first determine the 3D spectral density existence condition.  
At $T=0$\,K, we only need to find the convergence condition for the three ratios in Eqs.~\ref{Eq:preferredFasymptotic}-\ref{Eq:preferredRasymptotic}. That is, we can ignore the integral in~\ref{Eq:3DSD} for the time being because it will converge at any $s>0$ and $\vert\omega_1\vert+\vert\omega_2\vert+\vert\omega_3\vert>0$.
This claim is justified in Appendix~\ref{Appendix:TCL4integrals}.

The three ratios diverge if and only if $\omega_1+\omega_2+\omega_3=0$. After taking the limit $t\to\infty$ in Eq.~\ref{Eq:fraction},
\begin{equation}
\label{Eq:Rest1}
\frac{\tilde{\Gamma}_{-\omega_2-\omega_3}^\alpha-\tilde{\Gamma}_{\omega_1}^\alpha}{\omega_1+\omega_2+\omega_3}=-\frac{\partial\tilde{\Gamma}_{\omega_1}^\alpha}{\partial\omega_1}=-i\int_0^\infty dt\,t \tilde{C}_{\alpha}(t)e^{i\omega_1 t}.
\end{equation}
For the BCF that decays as $t^{-s-1}$, the integrand alternates in sign versus time at $\omega_1\neq 0$. In this case the integral exists under the original premise that $s>0$.
However, if $\omega_1=0$, then in addition to $\omega_1+\omega_2+\omega_3=0$,
the integral will diverge if $s\leq 1$. 
The infrared divergence at $T=0$\,K is due to the frequency subset $\omega_1=0$ and 
$\omega_2=-\omega_3\neq 0$. 

To assess the impact of the divergence in the 3D spectral densities on the generator, we will isolate the contributions from this frequency subset in Eqs.~\ref{Eq:ExplicitLiouv}-\ref{Eq:Expl4}. 
After adding those terms up, we determined that e Eqs.~\ref{Eq:ExplicitLiouv}-\ref{Eq:Expl2} result in zero. 
In the remaining equation lines~\ref{Eq:Expl3}-\ref{Eq:Expl4},  the requirement $\omega_{2}+\omega_{3}=0$, for $\omega_2=\omega_{ab}$ and $\omega_3=\omega_{cd}$ splits into
two possibilities: $(a=b \wedge c=d) \vee (a=d \wedge b=c)$. 
The scenario $a=b \wedge c=d$ has all frequencies equal to zero, which do not contribute to the generator, as previously stated.
However, the divergences corresponding to $a=d \wedge b=c$ do not cancel and may involve nonzero frequencies.

After adding the Hermitian conjugate to Eq.~\ref{Eq:ExplicitLiouv} and some
algebra, the sum of all the terms that can diverge for $0<s\leq 1$ is
\begin{widetext}
\begin{equation}
\label{Eq:Divergent}
\mathcal{R}_{nm,ij}^{4d}(t)=2\delta_{ij}\sum_{\alpha\beta=1}^{N_b}\big\{i(A_{mm}^\alpha-A_{jj}^\alpha)A_{nm}^\alpha\vert A_{mj}^\beta\vert^2\text{Im}[\mathsf{C}(t)+\mathsf{R}(t)]_{0,jm,mj}^{\alpha\beta}-i(A_{nn}^\alpha-A_{ii}^\alpha)A_{nm}^\alpha\vert A_{ni}^\beta\vert^2\text{Im}[\mathsf{C}(t)+\mathsf{R}(t)]_{0,in,ni}^{\alpha\beta}\big\}
.\end{equation}
\end{widetext}
The second term on the right-hand side of Eq.~\ref{Eq:Divergent} is obtained from the first term through complex conjugation and realignment $n  \leftrightarrows m$ and $i  \leftrightarrows j$, ensuring the preservation of hermiticity defined in Eq.~\ref{Eq:deltaRHC}. 

It can be easily demonstrated through index substitutions that $\mathcal{R}_{nn,ii}^{4d}(t)=0$ and $\mathcal{R}_{ni,ni}^{4d}(t)=0$. Therefore, $\mathcal{R}_{nn,ii}^{4}$ and $\mathcal{R}_{ni,ni}^{4}$ are bounded.
These rates are the ones that survive time averaging in the interaction picture. As a result, the secular approximation of the TCL4 (e.g., the fourth-order Davies master equation) is infrared regular.

Due to the $\delta_{ij}$ on the right-hand side of Eq.~\ref{Eq:Divergent}, only the population-to-coherence transfer matrix elements $\mathcal{R}_{nm,ii}^{4d}$, $n\neq m$, can diverge.  
When the ratios in Eqs.~\ref{Eq:preferredC} and~\ref{Eq:preferredR} are inserted into the equation for $\mathsf{C}+\mathsf{R}$, the real part in $\Gamma_{\omega}^\beta$ is isolated. In more precise terms, $\Gamma_\omega^\beta+\Gamma_{\omega}^{\beta\star}=2Re(\Gamma_{\omega}^\beta)$.
The result is
\begin{eqnarray}
\nonumber
\text{Im}[\mathsf{C}(t)+\mathsf{R}(t)]_{0,in,ni}^{\alpha\beta}&=&\text{Im}\left\{-2i\text{Re}[\Gamma^\beta_{in}(t)]\frac{\partial\Gamma_\omega^\alpha(t)}{\partial\omega} \bigg\vert_{\omega=0}\right\}\\
&=&-2\mathcal{J}_{in}^\beta(t)\frac{\partial \mathcal{J}_{\omega}^\alpha(t) }{\partial\omega}\bigg\vert_{\omega=0}
\end{eqnarray}
Substituting into Eq.~\ref{Eq:Divergent} for $i=j$, we isolate the diverging term into its final form:
\begin{eqnarray}
\label{Eq:DivergentA}
\mathcal{R}_{nm,ii}^{4d}(t)&=&-4i\sum_{\alpha\beta=1}^{N_b}\frac{\partial \mathcal{J}_{\omega}^\alpha(t) }{\partial\omega}\bigg\vert_{\omega=0}\bigg[\\
\label{Eq:DivergentB}
&&(A_{mm}^\alpha-A_{ii}^\alpha)A_{nm}^\alpha\mathcal{J}_{im}^\beta(t)\vert A_{im}^\beta\vert^2\\
\label{Eq:DivergentC}
&-&(A_{nn}^\alpha-A_{ii}^\alpha)A_{nm}^\alpha\mathcal{J}_{in}^\beta(t)\vert A_{in}^\beta\vert^2\bigg].
\end{eqnarray}

The terms on lines~\ref{Eq:DivergentB} and~\ref{Eq:DivergentC} converge at $t\to \infty$ for any $s>0$. It follows that
$\mathcal{R}_{nm,ii}^{4d}(t)$ diverges if and only if the derivative of the {\it real part} of the spectral density diverges
at $\omega=0$ [see line~\ref{Eq:DivergentA}]. Thus, the sufficient and necessary condition
for the existence of the fourth-order asymptotic generator is 
\begin{equation}
\label{Eq:genexists}
\frac{\partial \mathcal{J}_{\omega}^\alpha}{\partial\omega}\bigg\vert_{\omega=0}=-\int_0^\infty t\text{Im}\, C(t) dt<\infty.
\end{equation}

\begin{figure}[h]
\includegraphics[width=0.40\textwidth]{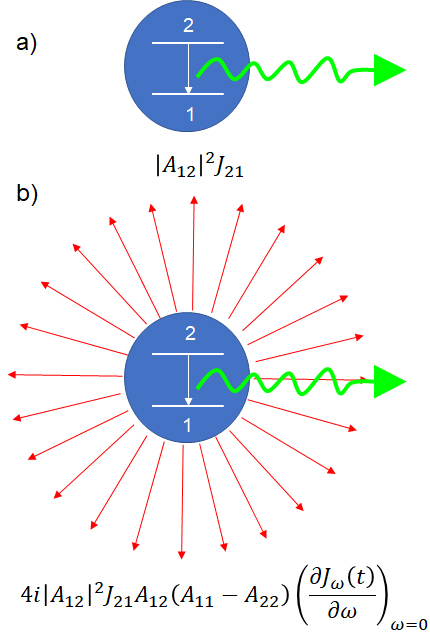}
\caption{\label{Fig1:diagram} Soft boson emission in a two-level system. Picture (a) represents the Fermi golden rule for the relaxing system. The fourth-order relaxation-dephasing hybrid process is depicted in picture (b). The red arrows in Fig.~2(b) represent the
production of infra-red bosons in addition to the population decay
(the green arrow).}
\end{figure}

 Figure~\ref{Fig1:diagram} illustrates the Fermi golden rule and the relaxation-dephasing hybrid processes (Eqs.~\ref{Eq:DivergentA}-\ref{Eq:DivergentC}) in a two-level system, which is initially in the excited state. 
The system relaxes into a quasistatic equilibrium state because the Fermi golden rule rate, sketched in Fig.~\ref{Fig1:diagram}(a), is initially much higher than the rate of the fourth-order hybrid process. But, the latter grows in time as the soft bosons are emitted into the bath, and will out-compete the Fermi golden rule over a long time if $s<1$.

The dephasing subprocess has a probability proportional to $A_{12}(A_{11}-A_{22})$ and to the derivative of the spectral density at zero frequency, from Eqs.~\ref{Eq:DivergentA}-\ref{Eq:DivergentC}.  As previously discussed,  the derivative represents the number of soft bosons emitted until time $t$.
The system dynamics described by the TCL4 master equation initially reaches a steady state at the Fermi-golden rule rate and for any $s>0$. 
However, when the boson number diverges at $s<1$ and $t\to\infty$, the global quantum state leaves the Fock space and the perturbative approach diverges. In the sub-Ohmic regime, the second-order Bloch-Redfield and Davies master equations lack this divergence, rendering them invalid.

The asymptotic TCL4 generator is uniformly bounded in the interval $s\geq 1$, since the left-hand side of Eq.~\ref{Eq:genexists} is bounded according to Eq.~\ref{Eq:dJdw}. 
The condition~\ref{Eq:genexists} is temperature independent (recall Eq.~\ref{Eq:BCF}) and less stringent than the existence of  Eq.~\ref{Eq:Rest1}, where the derivatives of the real and imaginary parts must {\it both} be bounded to guarantee that the 3D spectral functions are bounded.

\section{Approach to a ground state~\label{Sec:Approach}}

As an application, we examine if and how the open quantum system dynamics approaches the combined system and bath ground state. Our method for calculating the asymptotic and ground states will be detailed in the next two parts.

\subsection{\label{Sec:CPT} Asymptotic states}

Here we use the perturbation theory of the asymptotic TCL4 generator to compute the corrections to the system's asymptotic dynamics. The working hypothesis will be that the asymptotic limit exists for $s\geq 1$. 

Let us begin with the master Eq.~\ref{Eq:TCL}, which can be rewritten in the asymptotic limit as
\begin{equation}
\label{Eq:TCLas}
    \frac{d\rho(t)}{dt}=\mathcal{R}^{0}\rho(t)+\lambda^2 \mathcal{R}^{2}\rho(t)+\lambda^4 \mathcal{R}^{4}\rho(t)+.\,.\,.\,.
\end{equation}
For clarity, the generator's explicit dependency on $\lambda$ has been excised as a prefactor in each term of the expansion.
Since the coefficients of this equation are time independent, we seek the solution $\rho(t)=e^{\nu t}\rho$, which leads to
the linear system 
\begin{equation}
\label{Eq:TCLalg}
    (\mathcal{R}^{0}+\lambda^2 \mathcal{R}^{2}+\lambda^4 \mathcal{R}^{4}+.\,.\,.)\rho=\nu\rho,
\end{equation}
with the unknowns $\nu$ and $\rho$. We will solve this eigenvalue equation using the perturbation theory, which is sometimes referred to as the canonical perturbation theory. This method has been applied to explore the asymptotic states in Refs.~\cite{Fleming,tupkary2021fundamental}. 

$\nu$ and $\rho$ are expanded in the Taylor series as $\nu=\nu_{0}+\lambda^2\nu_{2}+\lambda^4\nu_{4}+\cdot\,\cdot\,\cdot$ and $\rho=\rho^{0}+\lambda^2\rho^{2}+\lambda^4\rho^{4}+\cdot\,\cdot\,\cdot$.
Inserting in the eigenvalue equation~\ref{Eq:TCLalg}, we find
\begin{widetext}
\begin{equation}
\label{Eq:TCLeig}
    (\mathcal{R}^{0}+\lambda^2 \mathcal{R}^{2}+\lambda^4 \mathcal{R}^{4}+\cdot\,\cdot\,\cdot\,)(\rho^{0}+\lambda^2\rho^{2}+\lambda^4\rho^{4}+\cdot\,\cdot\,\cdot\,)\\
    =(\nu_{0}+\lambda^2\nu_{2}+\lambda^4\nu_{4}+\cdot\,\cdot\,\cdot\,)(\rho^{0}+\lambda^2\rho^{2}+\lambda^4\rho^{4}+\cdot\,\cdot\,\cdot\,).
\end{equation}
\end{widetext}
Expanding the requirement $\text{tr}\rho=1$ in Taylor series in
$\lambda$, we find 
\begin{equation}
\label{Eq:TCLeigb}
    \text{tr}\rho^{2k}=\delta_{k0}.
\end{equation}    

Next we equalize the coefficients of the expansion on the left-hand side and the right-hand side of Eq.~\ref{Eq:TCLeig}. In the zeroth order and using Eq.~\ref{Eq:R0th},
we find
\begin{equation}
    -i\omega_{nm}\rho_{nm}^{0}=\nu_0\rho_{nm}^{0},
\end{equation}
where $\omega_{nm}=E_n-E_m$.
Since we assume that the eigenenergies of the isolated system are nondegenerate, it follows that there are $N^2-N$ nondegenerate 
eigenvalues $-i\omega_{nm}, n\neq m$, with the eigenvectors equal to the coherences $\rho=\vert n\rangle\langle m\vert$. In addition, there is a 
zero eigenvalue with a degenerate manifold of populations $\rho=\vert n\rangle\langle n\vert, n=1,2,\,.\,.\,.\,,N$. 

The application of the Fermi golden rule on the Liouvillian operating in the representation of the algebra of local operators shows that $N^2-1$ of these $N^2$ eigenvalues disintegrate in the continuous spectrum when the interaction is turned on in Ref.~\cite{bach2000return}, which leads to return to equilibrium. 
In our case, the reduced system's Liouvillian retains a fully discrete spectrum.  Proceeding with the perturbation theory, 
in the second  order, we find
\begin{equation}
\label{Eq:CPT2nd}
\mathcal{R}^{2}\rho^{0}+\mathcal{R}^{0}\rho^{2}=
\nu_2\rho^{0}+\nu_0\rho^{2}.
\end{equation}

The examination of the dephasing modes and their rates, while intriguing on its own, does not offer any data directly pertinent to this section and will be skipped. 
The relaxation modes and their rates are computed using degenerate perturbation theory. We have $\nu_0=0$
and consider the superpositions $\rho^{0}=\sum_{k=1}^N\rho_{kk}^{0}\vert k\rangle\langle k\vert$.
Then let us take the matrix element $\langle n\vert\,\cdot\,\cdot\,\cdot\,\vert m\rangle$ of Eq.~\ref{Eq:CPT2nd}. The result is
\begin{equation}
\label{Eq:CPT2ndEig}
    \sum_{k=1}^N \mathcal{R}^{2}_{nm,kk}\rho_{kk}^{0}-i\omega_{nm}\rho_{nm}^{2}=\nu_2\rho_{nn}^{0}\delta_{nm}.
\end{equation}
For $n=m$, we find 
\begin{equation}
\label{Eq:CPT2ndEigB}
    \sum_{k=1}^N \mathcal{R}^{2}_{nn,kk}\rho_{kk}^{0}=\nu_2\rho_{nn}^{0},
\end{equation}
which is the usual eigenvalue problem in degenerate perturbation theory. 

Let us make a few remarks.  First, it should be noted that~\ref{Eq:CPT2ndEigB} determines both the eigenvalues $\nu_2$ and the populations $\rho_{nn}^{0}$ to precision $O(1)$. However, after multiplying $\nu_2$
with $\lambda^2$, the eigenvalues of the generator will be known to precision $O(\lambda^2)$.
Second, trace preservation of the master equation requires that $\sum_k \mathcal{R}_{kk,ij}^{2q}=0$. When we take the sum over $n$ in Eq.~\ref{Eq:CPT2ndEigB} we see that $\nu_2\sum_{k}\rho_{kk}^{0}=0$. If $\nu_2=0$, then $\sum_k\rho_{kk}^{0}$ can be nonzero. This eigenvector represents the reduced system's asymptotic state in the zeroth order, which is unique by the condition~\ref{Eq:FGR}. 
If $\nu_2\neq 0$, then $\sum_k\rho_{kk}^{0}=0$. These eigenvectors turn out to be the same as the decaying population modes in the Davies master equation~\cite{davies1974}. There are $N-1$ such modes. 

The nonsecular matrix elements of the generator will generate coherences in these modes.
Let $n\neq m$ in Eq.~\ref{Eq:CPT2ndEig}. Then
\begin{equation}
    \rho_{nm}^{2}=\frac{-i}{\omega_{nm}}\sum_{k=1}^N\mathcal{R}_{nm,kk}^{2}\rho_{kk}^{0}.
\end{equation}
Thus from the $O(\lambda^2)$ generator, the  coherences are known to precision $O(\lambda^2)$.

At this time, the coherences are more precise than the populations, which is one of the flaws of the Bloch-Redfield equation. 
To make the precision of the populations consistent with the coherences, we must look at the populations' fourth-order corrections, where we have
\begin{equation}
\label{Eq:CPT4th}
\mathcal{R}^{0}\rho^{4}+\mathcal{R}^{2}\rho^{2}+\mathcal{R}^{4}\rho^{0}=
\nu_4\rho^{0}+\nu_2\rho^{2}.
\end{equation}
Taking the matrix element $\langle n\vert\,\cdot\,\cdot\,\cdot\,\vert m\rangle$, this equation transforms into
\begin{eqnarray}
\nonumber
-i\omega_{nm}\rho_{nm}^{4}&+&\sum_{i,j=1}^N\mathcal{R}^{2}_{nm,ij}\rho_{ij}^{2}
    +\sum_{k=1}^N\mathcal{R}^{4}_{nm,kk}\rho_{kk}^{0}\\
    &=&\nu_4\delta_{nm}\rho_{nn}^{0}+\nu_2\rho_{nm}^{2}.
\end{eqnarray}
Take $n = m$ to obtain the next in-order population correction. We have the following linear system
\begin{eqnarray}
\nonumber
&&
\sum_{k=1}^N\left(\mathcal{R}^{2}_{nn,kk}-\nu_2\delta_{nk}
\right)\rho_{kk}^{2}=\nu_4\rho_{nn}^{0}\\
\label{Eq:ranks}
&-&\sum_{i,j=1}^{N}{}^{'} \mathcal{R}^{2}_{nn,ij}\rho_{ij}^{2}-\sum_{k=1}^N\mathcal{R}^{4}_{nn,kk}\rho_{kk}^{0}.
\end{eqnarray}
The primed sum indicates summations over unequal indices $i$ and $j$.
Both the fourth-order corrections of the relaxation rates and the second-order corrections of the populations are generated by this equation.

The key finding is that the second-order populations depend on $\mathcal{R}^{4}_{nn,kk}$.
So, in order to compute the population dynamics to precision $O(\lambda^2)$ of the TCL2-generator, we must compute the population-to-population matrix elements of the TCL4 generator. This discovery extends the findings of Refs.~\cite{Fleming,tupkary2021fundamental} from asymptotic states to asymptotic dynamics. 
As an example, see Appendix~\ref{Sec:inaccuracy} for the Bloch-Redfield equation's dynamics inaccuracy at times longer than the relaxation time.

Let us summarize the corrections to the asymptotic state, which can be obtained by substitution
$\nu_0=\nu_2=\,\cdot\,\cdot\,\cdot\,=0$ in all of the above equations.
For the coherences, $i\neq j$, we find
\begin{eqnarray}
\label{Eq:CPT1}
\rho^{0}_{ij}&=&0,\\
\label{Eq:CPT2}
\rho^{2}_{ij}&=&\frac{-i}{\omega_{ij}}\sum_{k=1}^N\mathcal{R}^{2}_{ij,kk}\rho^{0}_{kk},\\
\nonumber
\rho^{4}_{ij}&=&\frac{-i}{\omega_{ij}}\Big(\sum_{k=1}^N\mathcal{R}^{2}_{ij,kk}\rho^{2}_{kk}+\mathcal{R}^{4}_{ij,kk}\rho^{0}_{kk}\\
\label{Eq:CPT3}
&+&\sum_{l,m=1}^{N}{}^{'} \mathcal{R}^{2}_{ij,lm}\rho^{2}_{lm}\Big),\\
\nonumber
\rho^{6}_{ij}&=&\frac{-i}{\omega_{ij}}\Big(\sum_{k=1}^N\mathcal{R}^{2}_{ij,kk}\rho^{4}_{kk}+
\mathcal{R}^{4}_{ij,kk}\rho^{2}_{kk}+\mathcal{R}^{6}_{ij,kk}\rho^{0}_{kk}\\
\label{Eq:CPT3b}
&+&\sum_{l,m=1}^{N}{}^{'}\mathcal{R}^{4}_{ij,lm}\rho^{2}_{lm}+
\mathcal{R}^{2}_{ij,lm}\rho^{4}_{lm}\Big),
\end{eqnarray}
while for the populations,  we have
\begin{eqnarray}
\label{Eq:CPT4}
\sum_{k=1}^N \mathcal{R}^{2}_{nn,kk}\rho^{0}_{kk}&=&0\\
\nonumber
\sum_{k=1}^N \mathcal{R}^{2}_{nn,kk}\rho^{2}_{kk}&=&-\sum_{k=1}^N\mathcal{R}^{4}_{nn,kk}\rho^{0}_{kk}\\
\label{Eq:CPT5}
&-&\sum_{l,m=1}^{N}{}^{'}\mathcal{R}^{2}_{nn,lm}\rho^{2}_{lm}\\
\nonumber
\sum_{k=1}^N \mathcal{R}^{2}_{nn,kk}\rho^{4}_{kk}&=&-\sum_{k=1}^N\mathcal{R}^{6}_{nn,kk}\rho^{0}_{kk}+\mathcal{R}^{4}_{nn,kk}\rho^{2}_{kk}\\
\label{Eq:CPT6}
&-&\sum_{k,l=1}^{N}{}^{'}\mathcal{R}^{4}_{nn,kl}\rho^{2}_{kl}+\mathcal{R}^{2}_{nn,kl}\rho^{4}_{kl}.
\end{eqnarray}

Applying Eq.~\ref{Eq:R2nd} to Eq.~\ref{Eq:CPT4} and applying Eq.~\ref{Eq:TCLeigb} results in $\rho_{kk}^{0}=\delta_{k1}$, showing that the isolated system's ground state is the zeroth-order asymptotic reduced state. The right-hand side of the first line in Eq.~\ref{Eq:CPT5} demonstrates that we need to compute the diagonal elements of the fourth-order asymptotic generator in order to compute the second-order corrections of the populations, as already demonstrated in Refs.~\cite{Fleming,tupkary2023searching}. 

When $s<1$, the TCL4 generator's infrared divergences will propagate via these equations, leading to a diverging asymptotic density matrix. 
First, let us look at Eq.~\ref{Eq:CPT3}, which gives us the fourth order coherence correction. The infrared-divergent terms make a contribution of
$(-i/\omega_{ij})\mathcal{R}^{4d}_{ij,11}$, given that $\rho_{kk}^{0}=\delta_{k1}$. In this case, it can be demonstrated from Eqs.~\ref{Eq:DivergentA}-\ref{Eq:DivergentC} that as time approaches infinity, 
$\mathcal{R}^{4d}_{ij,11}(t)$ tends to zero given that $\tilde{ \mathcal{J}}_{1k} = 0$ for all $k$. That is, in the relaxation-dephasing hybrids, a nonzero relaxation rate requires that the initial population's energy be higher than an energy of the two states involved in the final coherence. With the diverging terms going to zero, the asymptotic coherences in the fourth order are finite. 

By inserting the corrections of the excited-state populations, e.g., $\rho_{kk}^{2}$, into Eq.~\ref{Eq:CPT3b}, we can see that they induce infrared divergence through the fourth-order population to coherence transfers. As a result, the density matrix is divergent in the sixth order of coupling to the bath, but not before that. {\it This completes the proof of the first claim stated in Theorem 1 that the function
$\partial\tilde{\mathcal{J}}/\partial\omega\vert_{\omega=0}$ determines the infrared divergence of the asymptotic states of the TCL4 master equation.}

\subsection{\label{Sec:RSPT}Ground states}

To make the discussion easier, let us proceed with the assumption that there is only one bath.  Nondegenerate Rayleigh-Schr{\"o}dinger perturbation theory will be used to determine the ground state of the whole Hamiltonian, which is nondegenerate for the free Hamiltonian. After that, we will take the thermodynamic limit, to be consistent with how we determine the generators. Recently, Cresser and Andrés adapted this method to the second-order Rayleigh-Schr\"dinger perturbation theory~\cite{cresser2021weak}, which we now extend to the fourth-order. 

We are attempting to calculate the ground state's reduced density matrix. That is, we are assuming that the perturbed (coupled) Hamiltonian $\tilde H = H_0 + \lambda V$ has a ground state $\vert\tilde 0\rangle$  that can be expanded in powers of $\lambda$, and $\vert\tilde 0\rangle= \vert 0\rangle + O(\lambda^1)$. 
  From this total ground-state approximation, we will then calculate the reduced density matrix for just the system part, that is,
\[\tilde \varrho_S = \text{Tr}_B\vert\tilde 0\rangle\langle \tilde 0\vert.\]

In the zeroth order, the ground state of the free Hamiltonian 
is the tensor product of the ground states of the system and the bath.
See Appendix~\ref{Appendix:TIPT} for the derivation of the second- and fourth-order corrections of the ground state.
Reducing the ground state of the total system to the density matrix of the system, we find the second-order corrections
\begin{eqnarray}
\label{Eq:popu2TIPT}
\rho_{\text{mfgs},nn}^{2}&=&
    -\vert A_{1n}\vert^2\partial\tilde{\mathcal{S}}_{1n}, \hbox{$n=2,3,.\,.\,.,N$;}
\\
\nonumber
\rho_{\text{mfgs},ij}^{2}&=&\frac{A_{i1}A_{1j}(\tilde{\mathcal{S}}_{1i}-\tilde{\mathcal{S}}_{1j})}{E_i-E_j}\\
\label{Eq:CoherencesCPT2}
&+&\sum_k\frac{(\delta_{1i}A_{1k}A_{kj}-\delta_{1j}A_{ik}A_{k1})\tilde{\mathcal{S}}_{1k}}{E_i-E_j},
\end{eqnarray}
where $\tilde{\mathcal{S}}_{1k}=\tilde{\mathcal{S}}_{\omega_{1k}}$ and $\partial\tilde{\mathcal{S}}_{ab}=\partial\tilde{\mathcal{S}}(\omega)/\partial\omega$ at $\omega=\omega_{ab}$. Here mfgs refers to the mean-force ground state, which is equivalent to the mean-force Gibbs state at zero temperature. Alternatively,  the mean-force ground state is the reduced ground state of the combined system and bath.

These findings are identical to Ref.~\cite{cresser2021weak}  after applying the Kramers-Kronig transform $\tilde{\mathcal{S}}\mapsto\tilde{\mathcal{J}}$.
There is a bounded perturbative ground state in the second order of the coupling at any $s>0$.

The perturbative corrections of the reduced ground-state coherences in the fourth order of coupling are
\begin{widetext}
\begin{eqnarray}
\nonumber
&&\rho_{\text{mfgs},ij}^{4}=\sum_{a,b=1}^N\frac{A_{ib}A_{b1}A_{1a}A_{aj}}{\omega_{1i}\omega_{1j}}\tilde{\mathcal{S}}_{1a}\tilde{\mathcal{S}}_{1b}-\sum_{a=1}^N\Big[
A_{1a}A_{a1}A_{i1}A_{1j}\frac{\tilde{\mathcal{S}}_{1a}(\partial\tilde{\mathcal{S}}_{1i}-\partial\tilde{\mathcal{S}}_{1j})+\partial\tilde{\mathcal{S}}_{1a}(\tilde{\mathcal{S}}_{1i}-\tilde{\mathcal{S}}_{1j})}{\omega_{ij}}\Big]\\
\nonumber
&+&\sum_{a=1}^N\sum_{b=2}^N(A_{i1}A_{1a}A_{ab}A_{bj}+A_{ib}A_{ba}A_{a1}A_{1j})
\frac{\tilde{\mathcal{S}}_{1a}(\tilde{\mathcal{S}}_{1j}-\tilde{\mathcal{S}}_{1i})}{\omega_{b1}\omega_{ij}}
\\&+&\sum_{a=1}^N\sum_{b=2}^N(A_{i1}A_{1a}A_{ab}A_{bj}+A_{ib}A_{ba}A_{a1}A_{1j})\frac{\tilde{\mathcal{S}}_{{1a}}}{\pi\omega_{ij}} 
\int_{-\infty}^\infty d\omega' \frac{\tilde{\mathcal{J}}_{\omega'}}{\omega_{ab} -\omega'}\left(\frac{1}{\omega_{1j} - \omega'} - \frac{1}{\omega_{1i} - \omega'} \right)\nonumber
\\
&+& \sum_{a=1}^N\sum_{b=1}^N A_{ib}A_{b1}A_{aj}A_{1a}
\frac{ \tilde{\mathcal{S}}_{{1a}}}{\pi\omega_{ij}}\int_{-\infty}^\infty d\omega'\frac{\tilde{\mathcal{J}}_{\omega'}}{\omega_{1b} - \omega'}\left(\frac{1}{\omega_{aj} - \omega'} - \frac{1}{\omega_{ai} - \omega'} \right)\nonumber
\\
&-&\sum_{a=1}^N\sum_{b=1}^NA_{ib}A_{b1}A_{aj}A_{1a}
\frac{1}{\pi\omega_{ij}}\int_{-\infty}^\infty d\omega'\left[\frac{\tilde{\mathcal{J}}_{\omega'}\tilde{\mathcal{S}}_{{1j}-\omega'}}{(\omega_{1b} - \omega')(\omega_{1a} - \omega')} -
\frac{\tilde{\mathcal{J}}_{\omega'}\tilde{\mathcal{S}}_{{1j}-\omega'}}{(\omega_{1b} - \omega')(\omega_{aj} - \omega')}\right]+h.c.
\nonumber
\\
&-&\sum_{a=1}^N\sum_{b=1}^N(A_{i1}A_{1a}A_{ab}A_{bj}+A_{ib}A_{ba}A_{a1}A_{1j})
\frac{1}{\pi\omega_{ij}}\int_{-\infty}^\infty d\omega'\left[ \frac{\tilde{\mathcal{J}}_{\omega'}\tilde{\mathcal{S}}_{\omega_{1b}-\omega'}}{(\omega_{1j} - \omega')(\omega_{ab} - \omega')}-\frac{\tilde{\mathcal{J}}_{\omega'}\tilde{\mathcal{S}}_{\omega_{1b}-\omega'}}{(\omega_{1j} + \omega')(\omega_{1a} + \omega')}\right] +h.c. 
\nonumber\\
&& \hbox{$i,j=2,3,.\,.\,.,N$;$i\neq j$}.\label{Eq:coherences4}
\end{eqnarray}
\end{widetext}
The derivative 
$\partial\tilde{\mathcal{S}}$ is present on the first line, as well as certain cases of both the third and fourth lines, demonstrating that both $\partial\tilde{\mathcal{S}}$ as well as $\tilde{\mathcal{S}}$ govern the ground states. 
According to Eq.~\ref{Eq:dSdw}, $\partial\tilde
{\mathcal{S}}$  at zero frequency diverges in the limit $s\to 1_+$. Isolating the terms in the summations in ~\ref{Eq:coherences4} that contain this derivative, we find
\begin{eqnarray}
\label{Eq:divergingTerms4} \rho_{\text{mfgs},ij}^{4}&=&\cdot\,\cdot\,\cdot + A_{11}\left(A_{i1}A_{1j}A_{11}-|A_{i1}|^2A_{ij}\right)\nonumber\\
&\times&\left(
\frac{\tilde{\mathcal{S}}_{\omega_{1i}}\partial \tilde{\mathcal{S}}_{0}}{\omega_{ij}}+ \int_{-\infty}^\infty d\omega'\frac{\tilde{\mathcal{J}}_{\omega'}\tilde{\mathcal{S}}_{\omega_{1i}-\omega'}}{\pi\omega_{ij} \omega'^2}\right)
\nonumber\\
\label{Eq:gsDivergence}
&-&A_{i1}A_{1j}A_{11}\left(
A_{11}-A_{jj}\right)\nonumber\\&\times&\left(\frac{  \tilde{\mathcal{S}}_{\omega_{1j}}\partial\tilde{\mathcal{S}}_{0}}{\omega_{ij}} + \int_{-\infty}^\infty d\omega'\frac{\tilde{\mathcal{J}}_{\omega'}\tilde{\mathcal{S}}_{\omega_{1j}-\omega'}}{\pi\omega_{ij}\omega'^2}\right)
\end{eqnarray}
As a result, the fourth-order correction to the coherence $\rho_{\text{mfgs},ij}^{4}$ diverges continuously in the limit $s\to 1_+$ due to the term $\partial\tilde{\mathcal{S}}_0$ given by Eq.~\ref{Eq:dSdw}.
See Appendix~\ref{Appendix:TIPT} for the terms $\rho_{\text{mfgs},1j}^{4}$, which also diverge due to the term $\partial\tilde{\mathcal{S}}_0$. The distinction between asymptotic and ground states lies not only in their continuity at $s=1$, but also in their perturbative order, which exhibits divergence.
{\it This completes the proof of the second claim stated in Theorem 1 that the function
$\partial\tilde{\mathcal{S}}/\partial\omega\vert_{\omega=0}$ controls the infrared divergence of the reduced ground state.}

\subsection{\label{Sec:Return}Approach to ground state to precision $O(\lambda^2)$}

In the traditional thermodynamic limit method, it is relatively easy to show and widely known that the ground-state coherences provided by Eq.~\ref{Eq:CoherencesCPT2} and the coherences of the asymptotic solutions of the TCL2 master equation are identical~\cite{Fleming,Thingna,tupkary2021fundamental,lee2022perturbative,lobejko2022towards}. 
However, establishing the identity of the second-order populations is substantially more challenging.
Recently, there has been some progress in finding a solution to this problem~\cite{lobejko2022towards}.
According to Eqs.~\ref{Eq:popu2TIPT} and~\ref{Eq:CoherencesCPT2}, the ground state in the second order in $\lambda$ is bounded at any $s>0$. Since it solely depends on the secular matrix elements of the TCL4 generator, which are infrared regular, the asymptotic-state populations determined using Eq.~\ref{Eq:CPT6} are bounded, too.  
Still, it is not evident that the asymptotic and ground-state populations are identical. 

Hänggi and coworkers hypothesized that the asymptotic coherences can be analytically continued into the asymptotic populations~\cite{Thingna}. In that situation, they discovered agreement between the asymptotic and equilibrium populations to precision $O(\lambda^2)$. This is due to the fact that such analytical continuation holds true for the equilibrium states. In our case, as an example, when $i$ and $j$ are equal to $n$, it may be said that Eq.~\ref{Eq:popu2TIPT} is the limit of Eq.~\ref{Eq:CoherencesCPT2}.
Given that the asymptotic and the ground-state coherences are identical, the analytical continuations of their coherences are clearly also identical. Therefore, the analytic continuation cannot provide new information until it can be demonstrated that it is applicable to the asymptotic states. Here we numerically computed the second-order population corrections from the TCL4 generator, and compared them to the second-order corrections in the ground states using the following example.

\begin{figure}[h]
  \includegraphics[width=0.45\textwidth]{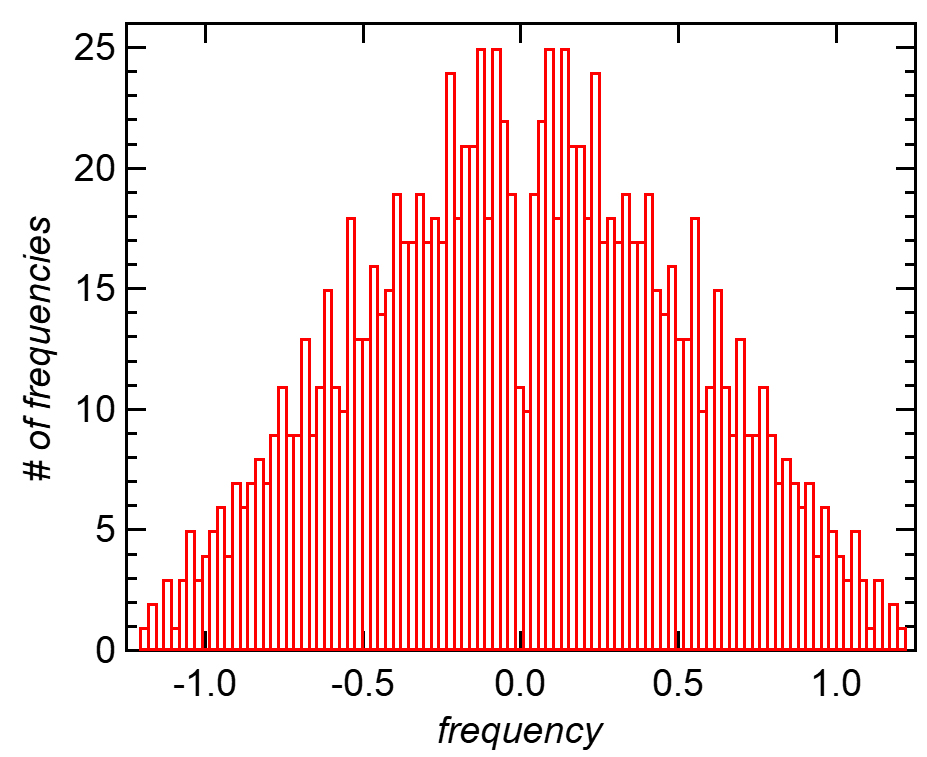}
\caption{\label{Fig1:histo}Histogram of Bohr frequency of the random-matrix example. $N=35$ In these units, the bath cutoff frequency is $\omega_c=10$}
\end{figure}

The system Hamiltonian has $N=35$ eigenenergies taken from the eigenvalue distribution of a Gaussian unitary ensemble of big matrices at the spectrum's center (GUE). 
The histogram in Fig.~\ref{Fig1:histo} displays the Bohr frequency distribution. There are 1191 distinct Bohr frequencies, with 0.0114, 0.4194, and 1.2108 being the lowest, average, and greatest absolute oscillation frequency, respectively.
The energy level repulsion, which prevents energy levels from being too close to one another, is what causes the minimum around 0 frequency. The bath cutoff frequency is $\omega_C=10$ the BCF-decay is Ohmic ($s=1$), which guarantees the existence of the TCL4 generator.
The vast number of frequencies is significant since it indicates our TCL4 generator's capacity to account for a large plethora of frequencies. This will be useful when extending the generator to time-dependent system Hamiltonians. 

For the system coupling operator, we randomly select an $N\times N$ Hermitian matrix from the GUE. All levels of the system are connected by this matrix in a way that allows for direct relaxation between them. It also exhibits significant diagonal element fluctuations, indicating the existence of dephasing. We separated the integration region into sections before computing the integral in~\ref{Eq:3DSD}. Smaller time steps are employed when the integrand is big. We may accept larger time steps without compromising accuracy if the integrand is small.

\begin{figure}[h]
\includegraphics[width=0.45\textwidth]{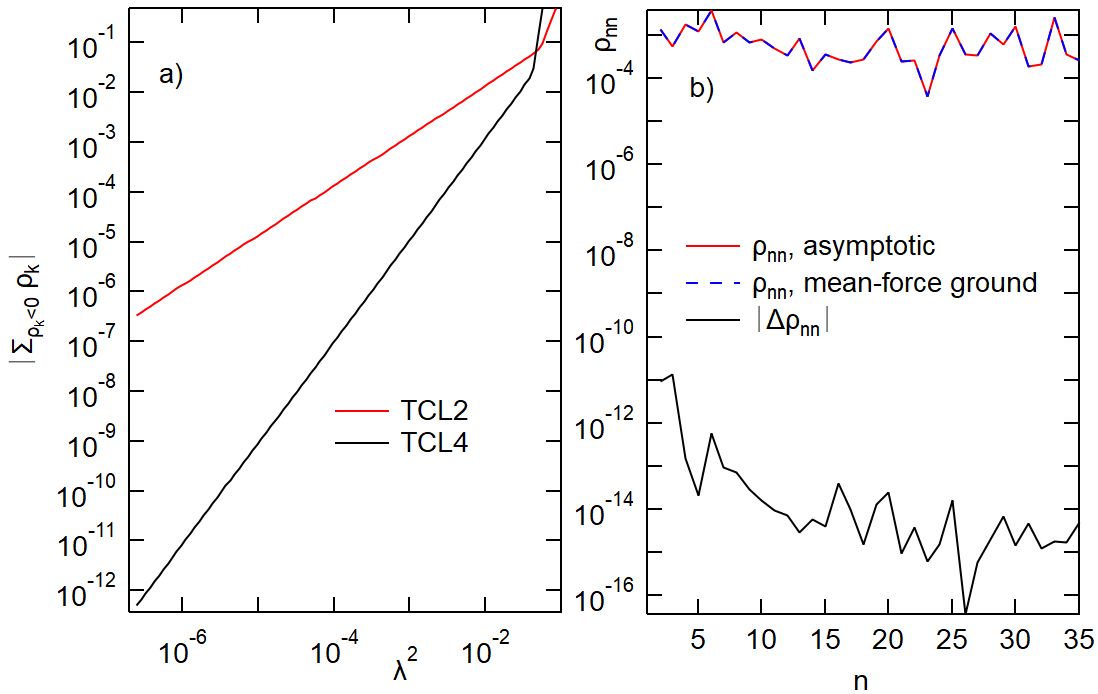}
\caption{\label{Fig1:mfg}Asymptotic-state properties in a 35-level system. a) Absolute negativity of the asymptotic density matrix versus coupling constant. b) Populations of the excited states $\rho_{nn}^{2}$ and the absolute value of the difference between the populations of the asymptotic and the reduced ground state versus $n$; $\omega_c=10$, $s=1$, and $T=0$ K.}
\end{figure}

The results for the asymptotic states in the Ohmic bath at $T=0$ K are shown in Fig.~\ref{Fig1:mfg}. In (a), we compute the absolute negativity, or the absolute sum of the negative eigenvalues of the asymptotic states of the TCL2 and TCL4 generators. The figure shows that the negativity scales as $\lambda^2$ and $\lambda^4$ in the TCL2 and TCL4 asymptotic states, respectively. The negativity also measures the state inaccuracy because the precise quantum state's negativity must be zero. Therefore, the inaccuracy of the asymptotic states scales as $\lambda^2$ and $\lambda^4$, which is in agreement with the downgrade in accuracy of quantum master equations discussed by Fleming and Cummings~\cite{Fleming}. 
The decrease in the negativity of the TCL4 asymptotic states relative to those of the TCL2 shows that all inaccuracies owing to the second order have been properly canceled out by the fourth-order terms. So the negativity may be used as a test to confirm that there are no algebraic errors in the TCL2n generators' computation.
The asymptotic states turn positive in the limit $n\to\infty$ since the time-convolutionless master equation is exact. This is implied here by the observation that the asymptotic-state negativity seems to scale with $\lambda^{2n}$, based on $n=1$ and $2$. 

The second-order corrections of the asymptotic and ground-state populations computed using  Eqs.~\ref{Eq:CPT5} and~\ref{Eq:popu2TIPT} are shown in Fig.~\ref{Fig1:mfg}(b).
{\it The difference between
the corrections is approximately ten orders of magnitude smaller than the corrections.} 
The similarity is unlikely to be coincidental and provides an additional proof that our calculation of the TCL4 generator is error-free.
As a function of $N$, we have reached the limit of computing precision at $N=4$, where we discovered that the asymptotic and ground state populations agree in 13 significant digits (not shown). 

Within the second-order perturbation theory, we can relax the working hypothesis $s\geq 1$, since the asymptotic states converge at any $s>0$. For $s=0.9,1.1$, and $1.2$, the 13-digit agreement was valid. In the open range $0<s<5$, we have likewise discovered numerical agreement; however, assuming the same computational bandwidth, the number of significant digits decreases dramatically when $s$ is much lower than 1.
This is due to the longer integration time required to achieve the necessary accuracy at lower $s$. 

In summary, the reduced asymptotic state of a small open quantum is almost certainly identical to the reduced ground state computed in order of $O(\lambda^2)$,
in the entire physical range of baths ($s>0$).  We believe that algebraic proof of this statement is within reach. 

\subsection{\label{Sec:WeakCorrections} Approach to a ground state to precision $O(\lambda^4)$}

Because the canonical perturbation theory requires the TCL6 generator to generate the asymptotic populations in the fourth order~\cite{Fleming}, we can only directly compare the coherences between the asymptotic and ground states in the fourth order, based on the work done so far. 
According to Theorem 1, the asymptotic states of the TCL4 master equation have a first-order phase transition at $s=1$, represented by Eq.~\ref{Eq:dJdw}.
On the other hand, the ground-state coherences diverge continuously in the limit $s\to 1_+$ as $1/(s-1)$ according to Eqs.~\ref{Eq:dSdw}. 
Therefore, the asymptotic and ground states cannot be the same.
Their distance is larger the closer $s$ approaches $1$ and diverges in the limit $s\to 1_+$. 

The following asymptotic limits capture the approach to ground state in the zeroth and second order in $\lambda$ and its breakdown in the fourth order:
\begin{eqnarray}
\label{Eq:DDbbound}
    \lim_{t\to\infty} \Vert \rho_{\text{TCL4}}(t) - \rho_{\text{mfgs,0}}\Vert &=& O(\lambda^2); s\geq 1,\\ 
\label{Eq:DDabound}
    \lim_{t\to\infty} \Vert \rho_{\text{TCL4}}(t) - \rho_{\text{mfgs,2}}\Vert &=& O(\lambda^4); s\geq 1,\\
\label{Eq:DDbound}
    \lim_{t\to\infty} \Vert \rho_{\text{TCL4}c}(t) - \rho_{\text{mfgs,4c}}\Vert &=& O\left(\frac{\lambda^4}{s-1}\right); s>1,
\end{eqnarray}
where $\rho_\text{mfgs,2k}=\rho_\text{mfgs}^{0}+\lambda^2\rho_\text{mfgs}^{2}+\,\cdot\,\cdot\,\cdot\,\lambda^{2k}\rho_\text{mfgs}^{2k}$, for $k=0,1,$ and $2$.
The inequality~\ref{Eq:DDbound} applies to the ground and asymptotic state coherences only, which are both computed with errors of order $O(\lambda^6)$. Since their distance 
is of order $O(\lambda^4)$ while the error of this distance is 
of order $O(\lambda^6)$, the dynamics does not approach the ground state to precision  $O(\lambda^4)$.

Proof of Eqs.~\ref{Eq:DDbbound}-\ref{Eq:DDbound}: Applying the perturbation theory, we have $\lim_{t\to\infty}\rho_\text{TCL4}(t)\equiv\tilde{\rho}=\tilde{\rho}^{0}+\lambda^2\tilde{\rho}^{2} +\lambda^4\tilde{\rho}^{4}+\,\cdot\,\cdot\,\cdot\,$, which we substitute in the steady-state equation for the TCL4 generator. Equations~\ref{Eq:CPT1}–\ref{Eq:CPT6} result from this; however, they lack the sixth- and higher-order terms.  Since the sixth-order terms are absent from Eqs.~\ref{Eq:CPT1} and~\ref{Eq:CPT2} as well as ~\ref{Eq:CPT4} and~\ref{Eq:CPT5}, the solutions for $\tilde{\rho}^{0}$ and $\tilde{\rho}^{2}$ will match those found with the canonical perturbation theory, which also match the mean-force ground states based on the previous section. It follows that $\tilde{\rho}=\rho_{\text{mfgs}}^0+\lambda^2\rho_{\text{mfgs}}^2+\lambda^4 \tilde{\rho}^{4}+\,\cdot\,\cdot\,\cdot\,$.

$\tilde{\rho}^{4}$ is determined by Eqs.~\ref{Eq:CPT3} and~\ref{Eq:CPT6} after excluding the sixth- and higher-order terms.
 Since the TCL4 generator is bounded uniformly in $s$ when $s\geq 1$, then $\tilde{\rho}^{4}$ will also be bounded uniformly in $s$.
Inserting $\tilde{\rho}$ in the left-hand sides of Eqs.~\ref{Eq:DDbbound} and~\ref{Eq:DDabound}, these equations are proven. The left-hand side of Eq.~\ref{Eq:DDbound} yields $\lambda^4\Vert \tilde{\rho}^{4}_c- \rho_\text{mfgs,c}^{4}\Vert$ when $\tilde{\rho}$ is substituted, where subscript ``c'' indicates the coherence part of the density matrix. For $s\geq 1$, $\rho_\text{mfgs,c}^{4}$ is linear with $1/(s-1)$, as shown in Sec:~\ref{Sec:RSPT}, while $\tilde{\rho}^{4}_c$ is bounded uniformly in $s$. As a result, the bound~\ref{Eq:DDbound} is proportional to $O(\lambda^4)$ and is not uniform in $s$. QED.

Prior investigations of the return to equilibrium demonstrated agreement between asymptotic and equilibrium states in all orders of $\lambda$~\cite{trushechkin2022open}. The difference between our technique and others is how we perform the thermodynamic limit. In the usual return-to-equilibrium computations, the thermodynamic limit is assumed from the start of the problem, e.g., the bath has an infinite size, and the computation is done in the algebra of local observables (i.e., the $C^\star$ algebra)~\cite{jakvsic1996model,bach2000return}. In our example, we perform the usual thermodynamic limit by substituting a sum over discrete bath modes of a finite-sized bath with an integral over the wavevectors of bath excitations.
It is well-known that the selection of the thermodynamic limit can influence the outcome of the computation~\cite{trushechkin2022open}.

\subsubsection*{~\label{Sec:} Counterexamples}

We briefly discuss two nongeneric open quantum systems. 
If $(A^\alpha_{ii}-A^\alpha_{nn})A^\alpha_{nm}\mathcal{J}_{in}^\beta\vert A_{in}^\beta\vert^2=0$, $\forall i,n,\alpha,\beta$, then
equation lines~\ref{Eq:DivergentB} and~\ref{Eq:DivergentC} demonstrate that the TCL4 generator's divergence at $s<1$ is suppressed. 
Consider for example the unbiased spin boson model where $A=\sigma_z/2$ and $H_S=\sigma_x/2$, with $\sigma_x$ and $\sigma_z$ the Pauli sigma matrices.
$A$ has no diagonal matrix elements in the energy eigenbasis, and hence the TCL4 generator exhibits no infrared divergence for any $s>0$. In Ohmic and sub-Ohmic baths, ground states are known to exist for the unbiased spin boson model Hamiltonian~\cite{spohn1989ground,hasler2011ground,bulla2003numerical,PhysRevLett.107.160601}. Theorem 1 does not apply. 

Another example is the exactly solvable reduced dynamics of pure dephasing Hamiltonians where $[A,H_S]=0$. 
This dynamics converges at any $s>0$. 
There are no off-diagonal matrix elements in the energy eigenbasis of $A$. As a consequence, the TCL4 generator converges for any $s>0$. Again, Theorem 1 does not apply.
It is reassuring to know that in these two well proven examples of infrared regular dynamics and ground states, the TCL4 generator is also infrared regular at zero temperature.

\section{\label{Sec:conclusion}Conclusion}

For the fourth-order perturbative time-convolutionless quantum master equation, we provided a concise yet exact generator given by Eqs.~\ref{Eq:ExplicitLiouv}-\ref{Eq:preferredR}. To set up the generator, only the bath's time-dependent spectral density and one integral over time need to be determined. As a result, this equation can now be used to study a variety of complex dynamical systems, without worrying about the inaccuracies that build up over timescales exceeding the relaxation time. These systems may include glasses, conjugated polymers, light-harvesting biomolecular complexes, black holes, as well as quantum-state preparation and gate fidelity in noisy quantum computing. 

As an application, we investigated the approach to ground-state characteristics in rather large open quantum systems. 
The approach is numerically exact only in the second-order perturbation theory. The situation is considerably more complex in the fourth order, though. The system asymptotic and ground states are both present in super-Ohmic baths, but they can differ significantly.
The system's asymptotic state is still present in Ohmic baths, but the ground state is not perturbatively present. In contrast, the system initially stabilizes near a Fock space ground state in sub-Ohmic baths, which is still present in the order of $O(\lambda^2)$, but after an escape time, the generator approaches divergence and the the reduced state starts to diverge. 

We expect the reformulated fourth-order time convolutionless master equation to pave the way for new approaches to solid-state quantum computing noise comprehension and reduction. The presence of $1/f$ noise in these systems at very low temperatures~\cite{muller2015interacting,burnett2019decoherence,jock2022silicon,carroll2022dynamics} indicates sub-Ohmic spectral density~\cite{shnirman2003dephasing}. As demonstrated Here this necessitates a strategy for open quantum systems that considers the emission of soft bosons and renders second-order master equations invalid. The loss of phase coherence in this situation is caused by the entanglement with the soft bosons emitted by relaxation. Carney {\it et al.}~\cite{carney2017infrared} created the term ''{\it infrared quantum information}'' to describe this phenomenon.
The emission of infrared quantum information could potentially have consequences that are still unexplored in the field of quantum computing. In quantum gauge field theories, the presence of infrared bosons leads to intricate memory effects, soft theorems, and asymptotic symmetries~\cite{strominger2018lectures}. We believe that the TCL4 master equation will let us explore some of these effects in mesoscopic devices and in today's noisy quantum computers.

We greatly appreciate the insightful comments that Anton Trushechkin, Marco Merkli, Michael Loss, Brian Kennedy, Itamar Kimchi, Sami Hakani, and Oliver Dial provided. We acknowledge support from the School of Physics at the Georgia Institute of Technology.

\appendix
\section{KMS AND THE KRAMERS-KRONIG RELATIONS\label{Sec:KMS}}
In this section we rewrite the spectral functions of the finite-sized system bath in terms of the sums over discrete bath modes, as opposed to the integrals over frequency. In the main text, we introduced a highly irregular spectral density function Eq.~\ref{Eq:J_sum} and expressed the generators in terms of that function, so that the thermodynamic limit can be performed simply by replacing the irregular spectral density with the smooth function~\ref{Eq:SDexp}. 
Here we convert the integrals back to the sums over discrete bath modes and demonstrate that the key relationships in Sec.~\ref{Sec:Ham} hold without the assumption~\ref{Eq:SDexp}. 

The BCF of the finite-sized combined system, corresponding to Eq.~\ref{Eq:BCF}  translates to
\begin{equation}
\label{Eq:BCFsum}
C_\alpha(t)=\sum_k (g_k^\alpha)^2(\cos \omega_k^\alpha t \coth\frac{1}{2}\beta_\alpha\omega_k^\alpha-i\sin \omega_k^\alpha t).
\end{equation}
The time-dependent spectral density given by Eq.~\ref{Eq:timedSD} recasts as
\begin{eqnarray}
\label{Eq:BCFsum}
\Gamma_\omega^\alpha(t)&=&\int_0^t d\tau C_\alpha(\tau)e^{i\omega\tau}\\
\nonumber
&=&
\sum_k (g_k^\alpha)^2\int_0^td\tau e^{i\omega\tau}[e^{i\omega_k^\alpha\tau}f_k^\alpha+e^{-i\omega_k^\alpha\tau}(1+f_k^\alpha)].
\end{eqnarray}
where $f_k^\alpha=1/(e^{\beta_\alpha\omega_k^\alpha}-1)$. Performing the integral over time, and taking the real and imaginary parts, we obtain
\begin{multline}
\label{Eq:JS}
\mathcal{J}_\omega^\alpha(t)=\sum_k(g_k^\alpha)^2\Big[\frac{\sin(\omega+\omega_k^\alpha)t}{\omega+\omega_k^\alpha}f_k^\alpha\\
+\frac{\sin(\omega-\omega_k^\alpha)t}{\omega-\omega_k^\alpha}(1+f_k^\alpha)\Big]\\
\mathcal{S}_\omega^\alpha(t)=\sum_k(g_k^\alpha)^2\big[\frac{1-\cos(\omega+\omega_k^\alpha)t}{\omega+\omega_k^\alpha}f_k^\alpha\\
+\frac{1-\cos(\omega-\omega_k^\alpha)t}{\omega-\omega_k^\alpha}(1+f_k^\alpha)\big].
\end{multline}
These are the spectral functions on which the TCL2n generators depends.  Neither the asymptotic ($t\to\infty$) nor the thermodynamic ($L\to\infty$) limits have been applied.

Let us now take the limit $t\to\infty$ in Eq.~\ref{Eq:JS} before the thermodynamic limit. Using the  representation of the delta function, e.g.,  $\delta(y)=\lim_{x\to\infty}\sin (xy)/(\pi y)$, we get the asymptotic spectral density 
\begin{equation}
\label{Eq:Jirreg}
\mathcal{J}_\omega^\alpha=\pi\sum_k(g_k^\alpha)^2\Big[\delta(\omega+\omega_k^\alpha)f_k^\alpha
+\delta(\omega-\omega_k^\alpha)(1+f_k^\alpha)\Big]
\end{equation}
At $\omega>0$, we can drop the left delta-function since $\omega_k^\alpha\geq 0$. Using the general property $\delta(x-y)h(x)=\delta(x-y)h(y)$, we can pull $(1+f_k^\alpha)$ as $1+f(\omega)$ in front of the sum, where $f(\omega)=1/[\exp(\beta_\alpha\omega)-1]$. This results in the relation~\ref{Eq:KMSplus}. Using similar manipulations, at $\omega<0$, this equation reproduces the KMS condition~\ref{Eq:KMS}. 

We can rewrite the principal density at time $t$ in Eq.~\ref{Eq:JS}
as
\begin{multline}
   \mathcal{S}_\omega^\alpha(t)=\frac{1}{\pi}\int_{0}^{\infty} d\omega'~\tilde{\mathcal{J}}_{\omega'}^\alpha
    \big[
    \frac{1-\cos(\omega+\omega')t}{\omega+\omega'}f(\omega')\\+
\frac{1-\cos(\omega-\omega')t}{\omega-\omega'}(1+f(\omega')\big].
\end{multline}
Changing the variable on the first line $\omega'\to  -\omega'$, and utilizing the relations~\ref{Eq:KMSplus} and~\ref{Eq:KMS}, we find
\begin{equation}
    \mathcal{S}_\omega^\alpha(t)=\frac{1}{\pi}\int_{-\infty}^{\infty}d\omega'\mathcal{J}_{\omega'}^\alpha \frac{1-\cos(\omega-\omega')t}{\omega-\omega'}.
\end{equation}
This equation is an extension of the Kramers-Kronig relation to the time dependent spectral density. The ratio in the integrand is regular at $\omega=\omega'$. We can thus pull a principal value in front of the integral, with the understanding that the integration region excludes an infinitesimal interval $[\omega-\epsilon,\omega+\epsilon]$ only, e.g., the singularities of the delta functions within $\mathcal{J}_\omega^\alpha$ do not count under the principal value. Then, we take the limit $t\to\infty$ and apply the Rieman-Lebesgue lemma. This results in the Kramers-Kronig relation~\ref{Eq:KramersK}.

\section{\label{Sec:vanHove} BOSON NUMBERS IN THE VAN HOVE AND PURE DEPHASING HAMILTONIANS}

Let us take a look at the boson numbers in the ground and asymptotic states of the van Hove~\cite{derezinski2003van} and
pure dephasing Hamiltonians~\cite{BreuerHeinz-Peter1961-2007TToO}, where these can be calculated exactly. 

For the van Hove  Hamiltonian, we take $H_S=0$, $N_b=1$, and $A=1$. We rewrite the Hamiltonian in Eq.~\ref{Eq:Htotal} as
\begin{equation}
    H_T=\sum_k\omega_k(b_k+\frac{g_k}{\omega_k})^\dagger(b_k+\frac{g_k}{\omega_k})-\sum_k\frac{g_k^2}{\omega_k},
\end{equation}
by ``completing the square''.
The ground state is a displaced vacuum
with the oscillator displacements $\alpha_k=-g_k/\omega_k$.
The ground-state energy is
\begin{equation}
    E_g=\frac{1}{\pi}\int_0^\infty d\omega\frac{\tilde{\mathcal{J}}}{\omega}\mapsto 2\pi\lambda^2\Gamma(s)\omega_c,
\end{equation}
where, we note that $\mapsto$ means taking the thermodynamic limit.
It is bounded if and only if $s>0$. 
The average boson number is 
\begin{equation}
\label{Eq:bosonnumber}
    \langle N\rangle =\sum_k\frac{g_k^2}{\omega_k^2}=\frac{1}{\pi}\int_0^\infty d\omega\frac{\tilde{\mathcal{J}}(\omega)}{\omega^2}\mapsto-\frac{\partial\tilde{\mathcal{S}}}{\partial\omega}\bigg\vert_{\omega=0}.
\end{equation}
The last equality can be shown by integrating by parts (e.g., $\int udv=uv-\int vdu$, where $u=\tilde{\mathcal{J}},dv=d\omega/\omega^2$), and replacing $\tilde{\mathcal{J}}$ and $\tilde{\mathcal{S}}$ with $\partial\tilde{\mathcal{J}}/\partial\omega$ and
$\partial\tilde{\mathcal{S}}/\partial\omega$, respectively,  in the Kramers-Kronig equation~\ref{Eq:KramersK}.
This assumes that 
$\tilde{\mathcal{J}}(\omega)/\omega=0$ at $\omega = 0$,
$\partial\tilde{\mathcal{J}}/\partial\omega<\infty$,
and $\partial\tilde{\mathcal{S}}/\partial\omega<\infty$, all of which holds true for the spectral density  in Eq.~\ref{Eq:SDexp} if $s>1$.

Next, we compute the dynamics at $T=0K$. The annihilation operators in the Heisenberg picture satisfy the equation of motion,
\begin{equation}
    \frac{db_k}{dt}=i[H_T,b_k]=-i\omega_kb_k-ig_k,
\end{equation}
which has the solution
\begin{equation}
    b_k(t)=b_k(0)e^{-i\omega_kt}-\frac{g_k}{\omega_k}(1-e^{-i\omega_kt}).
\end{equation}
In the Heisenberg picture the bath is in the vacuum state $\vert 0\rangle$. Thus, $b_k(t)\vert 0\rangle=
\alpha_k(t)\vert 0\rangle $, where $\alpha_k(t)=-(g_k/\omega_k)(1-e^{-i\omega_kt})$. After rotating back to the Schr\"odinger picture, the bath state, $\exp(-iH_Tt)\vert 0\rangle$, is the displaced vacuum with the oscillator displacements $\alpha_k(t)$. 
The state of the small system does not change
while the bath is being displaced because the Hilbert space of the system is one dimensional.

The average boson number is
\begin{eqnarray}
\label{Eq:RLebesgue}
    &&\langle N(t)\rangle=\sum_k\vert\alpha_k(t)\vert^2=
    \nonumber
    \frac{2}{\pi}\int_0^\infty d\omega \frac{\tilde{\mathcal{J}}(\omega)}{\omega^2}(1-\cos \omega t)\\
    &\mapsto&4\lambda^2\Gamma(s-1)\left\{1-\frac{\cos[(s-1)\arctan \omega_ct ]}{(1+\omega_c^2t^2)^{\frac{s-1}{2}}}\right\}.
\end{eqnarray}
The number of bosons grows as $\vert t\vert^{1-s}$ for $s<1$ and as $\log(t)$ at $s=1$.
For $s>1$, the boson number is bounded uniformly in time. 

Taking the limits $t\to\infty$ and $L\to\infty$ on the first line of Eq.~\ref{Eq:RLebesgue}, (in arbitrary order), applying the Riemann–Lebesgue lemma,  and integrating by parts as described below Eq.~\ref{Eq:bosonnumber}, we find
\begin{equation}
\label{Eq:bosonnumber1}
    \lim_{t\to\infty}\langle N(t)\rangle=-2\frac{\partial\tilde{\mathcal{S}}}{\partial\omega}\bigg\vert_{\omega=0}; s>1.
\end{equation}
By Eq.~\ref{Eq:dSdw}, the ground and asymptotic-state boson numbers both diverge as $1/(s-1)$ in the limit $s\to 1_+$.

Equations~\ref{Eq:dSdw},~\ref{Eq:bosonnumber}, and~\ref{Eq:bosonnumber1} show that the Hamiltonian admits both asymptotic and ground states, if and only if $s>1$. 
No infrared divergence is present in the system's reduced dynamics since there is no dynamics.
Extension to pure dephasing Hamiltonians yields similar boson numbers~\cite[Sec. 4.2]{BreuerHeinz-Peter1961-2007TToO}. In this case, we can take $H_s=0$ and $A=\sigma_z$ (Pauli matrix). Once again, the diverging boson numbers at $s\leq 1$ have no impact on the dynamics. Pure dephasing Hamiltonians 
do not satisfy the Fermi golden rule condition 2 in Sec.~\ref{Sec:Ham}.

\section{DERIVATION OF THE TCL4 GENERATOR\label{Sec:appendixA}}

Let us begin with Ref.~\cite[Eq.~(29)]{Breuer_1999} as the expression of the TCL4 generator in the interaction picture, 
\begin{widetext}
\begin{multline}
    \label{Eq:TCL4_3D_com}
    \Big{(}\frac{\partial \varrho}{\partial t}\Big{)}_4=
\sum_{\alpha,\beta=1}^{N_b}\int_0^t d t_1 \int_0^{t_1} d t_2 \int_0^{t_2} d t_3
\big{\{}\langle02\rangle_{\alpha}\langle13\rangle_{\beta}[\hat 0^{\alpha},[\hat 1^{\beta}, \hat 2^{\alpha}] \hat 3^{\beta}\varrho]\\
-\langle02\rangle_{\alpha}\langle31\rangle_{\beta}[\hat 0^{\alpha},[\hat 1^{\beta}, \hat 2^{\alpha}] \varrho \hat 3^{\beta}]
+\langle03\rangle_{\alpha}\langle12\rangle_{\beta}([\hat0^{\alpha},[\hat 3^{\alpha}, \hat 2^{\beta}] \varrho \hat 1^{\beta}]+[\hat 0^{\alpha},[\hat 1^{\beta} \hat 2^{\beta}, \hat 3^{\alpha}] \varrho])
\\-\langle03\rangle_{\alpha}\langle21\rangle_{\beta}[\hat 0^{\alpha},[\hat1^{\beta}, \hat3^{\alpha}] \varrho \hat2^{\beta}]\big{\}} + h.c.
\end{multline}
\end{widetext}
Here we have expanded the authors' original short-hand notation to account for our to our multiple uncorrelated baths, defining $\hat i^\alpha = A^\alpha(t_i)$ and $\langle ij\rangle_\alpha = C_\alpha(t_i - t_j)$ with $i=0,\ldots,3$ where $t_0 = t$. We have omitted the explicit time dependence of the reduced density matrix as well, defining $\varrho = \varrho(t)$, the interaction picture operator.   
\subsection{Integral simplification}
For finite $t$, the integration in Eq.~\eqref{Eq:TCL4_3D_com} is over a bounded region in $D \subset \mathbb{R}^3$.

Let us consider three general operators, 
\begin{align}
    K_{ijk}^{\alpha\beta} &= \langle 0 i\rangle_\alpha \langle jk \rangle_{\beta} [\hat 0^{\alpha},[\hat i^\alpha, \hat k^\beta] \varrho \hat j^\beta] \\
    H_{ijk}^{\alpha\beta}&= \langle 0 i\rangle_\alpha \langle jk \rangle_{\beta} [\hat 0^{\alpha},[\hat j^\beta\hat k^\beta, \hat i^\alpha] \varrho ]\\
    M_{ijk}^{\alpha\beta} &= \langle 0 i\rangle_\alpha \langle jk \rangle_{\beta} [\hat 0^{\alpha},\hat j^\beta[ \hat i^\alpha,\hat k^\beta] \varrho ]
\end{align}
with which we can rewrite Eq.~\eqref{Eq:TCL4_3D_com} as

\begin{multline}   \label{Eq:TCL4_3D}
     \Big{(}\frac{\partial \varrho}{\partial t}\Big{)}_4=
\sum_{\alpha,\beta=1}^{N_b}\int_0^t d t_1 \int_0^{t_1} d t_2 \int_0^{t_2} d t_3 \big{(}K_{231}^{\alpha\beta} +H_{213}^{\alpha\beta} \\+ M_{213}^{\alpha\beta} + K_{321}^{\alpha \beta} + K_{312}^{\alpha \beta} + H_{312}^{\alpha \beta}\big{)} + h.c.
\end{multline}

Note that the variables $t_1,t_2,t_3$ are independent and that, for any two successive integrals, we can visualize the region of integration as a triangle in $\mathbb{R}^2$. Assuming the integral is bounded, we can change the order of integration as 
\begin{equation}
    \int_{0}^{t_{n-1}} dt_{n} \int_{0}^{t_{n}}dt_{n+1} \mapsto \int_{0}^{t_{n-1}} dt_{n+1} \int_{t_{n+1}}^{t_{n-1}}dt_{n} 
\end{equation}

We will use this to transform the integral over  $D$. Our choice of transformation will depend only on the order of the time arguments within the bath correlation functions, which we separate into two cases, $i=2$ and $i=3$.
Let us also employ the short-hand notation 
\begin{equation}
    \int_0^i d_j = \int_{0}^{t_i} dt_j
\end{equation}
\paragraph{Case 1 ($i=2$):}
\begin{multline}
    \iiint_D  = \int_{0}^t d_1 \int_{0}^{1} d_{2} \int_{0}^{1}d_3  \\+\int_{0}^t d_3 \int_0^{3} d_2\int_{0}^{3} d_1 - \int_{0}^t d_3 \int_0^{3} d_2\int_{0}^{t} d_1 
\end{multline}
and we have
\paragraph{Case 2 ($i=3$):}
\begin{equation}
    \iiint_D  = \int_0^t d_2 \int_{0}^{t} d_1 \int_{0}^{2} d_3 -\int_0^t d_2 \int_{0}^{2} d_1 \int_{0}^{2} d_3 
\end{equation}

While the overall ordering of the time arguments is not arbitrary, as long as the relation between the scalar correlation functions and the and operator product ordering is preserved with the $\alpha, \beta$ bath index labels, we are free to swap the arbitrary $t_1,t_2,t_3$ labels. Thus, we will permute the labels such that the bounds only depend on time $t_1$ as the outermost integration variable. Applying, we arrive at a simplified form of Eq.~\eqref{Eq:TCL4_3D}, 

\begin{equation}
\label{Eq:TCL4_3D_bounds}
\begin{split}
     \Big{(}\frac{\partial \varrho}{\partial t}\Big{)}_4=
\sum_{\alpha,\beta=1}^{N_b}\Big{[}\int_0^t d t_1 \int_0^{t_1} d t_3 \int_0^{t_1} d t_2 \left(Q_{213}^{\alpha\beta} +M_{231}^{\alpha \beta}\right)
\\+ \int_0^t dt_1 \int_0^t dt_3 \int_0^{t_1} dt_2
P_{231}^{\alpha \beta} 
\Big{]} + h.c., 
\end{split}
\tag{\ref*{Eq:TCL4_3D}'}
\end{equation}
where we have defined two new functions,
\begin{align}
Q_{ijk}^{\alpha,\beta} &=H_{ijk}^{\alpha,\beta} + M_{ijk}^{\alpha,\beta}= \langle 0 i\rangle_\alpha \langle jk \rangle_{\beta} [\hat 0^{\alpha},[ \hat j^\beta,\hat i^\alpha]\hat k^\beta \varrho] 
\\
P_{ijk}^{\alpha,\beta} &=K_{ijk}^{\alpha,\beta} - M_{ijk}^{\alpha,\beta}= \langle 0 i\rangle_\alpha \langle jk \rangle_{\beta} [\hat 0^{\alpha},[ \hat j^\beta,[\hat k^\beta,\hat i^\alpha] \varrho ]]
 \end{align}
We will also define the double integrals of our remaining functions as,  
\begin{subequations}
\label{Eq:TCL4_2D_Terms}
\begin{align}    \mathcal{Q}^{\alpha \beta}(t,t_1) &= \int_0^{t_1}dt_2 \int_0^{t_1} dt_3 Q^{\alpha \beta}_{213}\\
\mathcal{P}^{\alpha \beta}(t,t_1) &= \int_0^{t_1}dt_2 \int_0^{t} dt_3 P^{\alpha \beta}_{231}\\
\mathcal{M}^{\alpha \beta}(t,t_1) &= \int_0^{t_1}dt_2 \int_{0}^{t_1} dt_3 M^{\alpha \beta}_{231}
\end{align}
\end{subequations}

Our goal now is to write Eq.~\eqref{Eq:TCL4_3D_bounds} as single quadrature,
\begin{equation} \label{Eq:TCL4_1D}
\begin{split}
         \Big{(}\frac{\partial \varrho}{\partial t}\Big{)}_4=
\sum_{\alpha,\beta=1}^{N_b}\int_0^t d t_1[ \mathcal{Q}^{\alpha \beta}(t,t_1) + \mathcal{P}^{\alpha\beta}(t,t_1)\\+  \mathcal{M}^{\alpha\beta}(t,t_1)] + h.c.
\end{split}
\end{equation}
by evaluating the integrals in Eq.~\eqref{Eq:TCL4_2D_Terms}, which can be expressed in terms of products of system operators and the time-dependent spectral density for each bath, $\Gamma^\alpha_\omega(t)$, defined in Eq.~\ref{Eq:timedSD}.

\subsection{Time-dependent spectral density}
As mentioned in the main text, we assume that for each bath the BCF has the property $C_{\alpha}(-s) = C_\alpha^*(s)$, from which we can define the complex conjugate of $\Gamma^\alpha_\omega(t)$ as
\begin{equation}
    \Gamma^{\alpha*}_\omega(t) = \int_{0}^{t} ds C_\alpha^*(s) e^{-i\omega s} = -\Gamma^\alpha_\omega(-t)
\end{equation}

We define transposition as 
    $\Gamma^{\alpha T}_\omega(t) = \Gamma^\alpha_{-\omega}(t)$, since we will be evaluating the function only for the Bohr frequencies, which can be expressed in terms of the system energies as $\omega_{nm} = E_n - E_m$, leading to an  antisymmetric matrix representation in the $H_S$ eigenbasis, $\omega_{mn}^T = -\omega_{mn}$. With these two operations, we have a Hermitian conjugate,
$\Gamma^{\alpha\dagger}_\omega(t) = -\Gamma^\alpha_{-\omega}(-t)$.

We will define a Hadamard product in the system energy basis which is equivalent to a point wise product in the frequency domain, 
\begin{equation}
    f \circ g = f(\omega_{nm})g(\omega_{nm}) = f_\omega g_{\omega}.
\end{equation}
With this product we can express the interaction picture operators as 
\begin{equation}
    \hat i^\alpha =  A^\alpha(t_i) = U^\dagger(t_i)A^\alpha U(t_i) = A^\alpha \circ e^{i \omega t_i},
\end{equation}
where $A^\alpha$ is a time-independent Hermitian system coupling operator, and $U(t)$ is the unitary time evolution operator 
$U(t) = e^{-i \hat H_S t}$.

To evaluate  \eqref{Eq:TCL4_2D_Terms}, we could rotate each expression to the Schrödinger picture and expand each coefficient in the system energy basis, in which can we  would find that each bath index contributes an integral of one of the following forms,
\begin{subequations}
\label{Eq:generalIntegrals}
\begin{align}
    \int_{t_a}^{t_b}dsC(t - s)e^{i \omega s} &=
    e^{i \omega t}\Delta\Gamma^T_\omega(t-t_a,t-t_b)\\
    \int_{t_a}^{t_b}ds C(s - t)e^{i \omega s} & = e^{i \omega t}\Delta \Gamma_\omega^*( t - t_a, t - t_b),
\end{align}
\end{subequations}
where we have used the Eq.~\ref{Eq:DeltaGamma} definition of $\Delta \Gamma_\omega(t,\tau)$. 

However, examining the operators $P^{\alpha \beta}_{213},Q^{\alpha \beta}_{213},M^{\alpha \beta}_{231}$, we note that since only a single system operator will depend on $t_2$ and $t_3$, we can use the linearity of integration and the linearity of operators themselves to see that the results of \ref{Eq:generalIntegrals} can be accounted for while still in the interaction picture using  
\begin{subequations}
\label{Eq:HadamardTrick}
\begin{align}
    \int_{t_a}^{t_b} dt_j \langle ij \rangle_\alpha \hat j^\alpha &=  \hat i^\alpha \circ \Delta\Gamma^{\alpha T}(t_i-t_a,t_i-t_b)\\
    \int_{t_a}^{t_b} dt_j \langle ji \rangle_\alpha \hat j^\alpha &= \hat i^\alpha \circ\Delta\Gamma^{\alpha*}(t_i-t_a,t_i-t_b),
\end{align}
\end{subequations}
where we have removed frequency subscript, as it is determined by the product. Equations \ref{Eq:HadamardTrick} define our ``Hadamard trick'', which enables fast evaluation of the TCL4 generator kernels over the different integration regions. 

Explicitly evaluating \ref{Eq:TCL4_2D_Terms}, we find

\begin{widetext}
    \begin{align}
        \mathcal{Q}^{\alpha \beta}(t,t_1) &= \int_0^{t_1}dt_2 \int_0^{t_1} dt_3 \langle 0 2\rangle_\alpha \langle 13 \rangle_{\beta} [\hat 0^{\alpha},[ \hat 1^\beta,\hat 2^\alpha]\hat 3^\beta \varrho ] = [\hat 0^{\alpha},[ \hat 1^\beta,\{\hat 0^\alpha\circ \Delta \Gamma^{\alpha T} (t,t-t_1)\}]\{\hat 1^\beta \circ \Delta \Gamma^{\beta T}(t_1,0) \}\varrho ]
        \\ 
        \mathcal{P}^{\alpha \beta}(t,t_1) &= \int_0^{t_1}dt_2 \int_0^{t} dt_3 \langle 0 2\rangle_\alpha \langle 31 \rangle_{\beta} [\hat 0^{\alpha},[ \hat 3^\beta,[\hat 1^\beta,\hat 2^\alpha] \varrho ]] =[\hat 0^{\alpha},[ \{\hat 1^\beta\circ \Delta \Gamma^{\beta *}(t_1,t_1 -t)\},[\hat 1^\beta,\{\hat 0^\alpha\circ \Delta \Gamma^{\alpha T}(t,t - t_1)\}] \varrho ]] 
        \\
        \mathcal{M}^{\alpha \beta}(t,t_1) &= \int_0^{t_1}dt_2 \int_0^{t_1} dt_3 \langle 0 2\rangle_\alpha \langle 31 \rangle_{\beta} [\hat 0^{\alpha},\hat 3^\beta[ \hat 2^\alpha,\hat 1^\beta] \varrho ] = -[\hat 0^{\alpha},\{\hat 1^\beta \circ \Delta \Gamma^{\beta*}(t_1,0)\}[\hat 1^\beta, \{\hat 0^\alpha\circ \Delta \Gamma^{\alpha T}(t,t-t_1)\}] \varrho ]
    \end{align} 
\end{widetext}
Note that $\mathcal{M}^{\alpha\beta}$ will cancel with part of $\mathcal{P}^{\alpha\beta}$. 
So we can define
\begin{equation}
    \mathcal{P}^{\alpha \beta}(t,t_1) + \mathcal{M}^{\alpha \beta}(t,t_1) = \mathcal{K}^{\alpha\beta}(t,t_1)-\mathcal{Y}^{\alpha \beta}(t,t_1)
\end{equation}
where 
\begin{align}
    &\mathcal{K}^{\alpha\beta}(t,t_1)  \notag\\&  = [\hat 0^{\alpha},[\{\hat 1^\beta\circ  \Gamma^\beta(t-t_1)\},[\hat 1^\beta,\{\hat 0^\alpha\circ \Delta \Gamma^{\alpha T}(t,t - t_1)\}] \varrho] ]\\
    &\mathcal{Y}^{\alpha\beta}(t,t_1) \notag\\& = [\hat 0^{\alpha},[\hat 1^\beta,\{\hat 0^\alpha\circ \Delta \Gamma^{\alpha T}(t,t -t_1)\}] \varrho\{\hat 1^\beta\circ \Gamma^{\beta *}(t_1)\}] 
\end{align}

So we have 
\begin{equation}
\label{Eq:TCL_1D_final}
\begin{split}
    \Big{(}\frac{\partial \varrho}{\partial t}\Big{)}_4=
\sum_{\alpha,\beta=1}^{N_b}\int_0^t d t_1( \mathcal{Q}^{\alpha \beta}(t,t_1) \\+ \mathcal{K}^{\alpha\beta}(t,t_1)- \mathcal{Y}^{\alpha\beta}(t,t_1)) + h.c. 
\end{split}
\tag{\ref*{Eq:TCL4_1D}'}
\end{equation}Let us adjust our notation to use $A^\alpha = A$, $A^\beta = B$. 
Define 
\begin{align}
    \Delta\mathcal{A}(t,t_1) &= A(t) \circ \Delta\Gamma^{\alpha T}(t,t-t_1)\\
    \mathcal{B}^\dagger(t_1) &= B(t_1) \circ \Gamma^{\beta *}(t_1)
\end{align}Translating into this notation we find
\begin{align}
\mathcal{Q}^{\alpha\beta}(t,t_1) &=  [A(t),[ B(t_1),\Delta\mathcal{A}(t,t_1)]\mathcal{B}(t_1)\varrho ]\\
    \mathcal{Y}^{\alpha\beta}(t,t_1) &= [A(t), [B(t_1),\Delta\mathcal{A}(t,t_1)]\varrho\mathcal{B}^\dagger(t_1)]\\
\mathcal{K}^{\alpha\beta}(t,t_1) & \notag\\&=[A(t),[B(t_1)\circ\Gamma^\beta(t - t_1),[B(t_1),\Delta\mathcal{A}(t,t_1)]\varrho]] 
\end{align}

Looking at the common terms,in each case we can factor the unitary operators such that the time dependence is grouped with the 
time-dependent spectral density,
\begin{align}
[B(t_1),\Delta\mathcal{A}(t,t_1)] &= e^{i\omega t_1} \notag \\
     &\circ[B(0), A(t-t_1) \circ\Delta \Gamma^{\alpha T}(t,t-t_1)]  \\
     \varrho \mathcal{B}^\dagger(t_1) &= e^{i\omega t} \circ\{ \rho [B(t_1 - t) \circ \Gamma^{\beta *} (t_1)]\}\\
     \mathcal{B}(t_1)\varrho  &= e^{i\omega t} \circ \{[B(t_1 - t) \circ \Gamma^{\beta T}(t_1) ]\rho\}.
\end{align}
We now rotate each term to the Schrödinger picture, noting that as we transform $\frac{d \varrho}{dt} \mapsto \frac{d\rho}{dt}$ our fourth-order generator will pick up an additional factor of $U(t)$ on the left, and $U^\dagger(t)$ on the right,  such that overall we have the expression
\begin{widetext}
    

\begin{align}
\mathcal{Q}^{\alpha,\beta}(t,t_1) &= 
[A, \{e^{-i\omega(t-t_1)}\circ[ B,A \circ \Delta \Gamma^{\alpha T}(t,t-t_1)e^{i\omega(t-t_1)}]\}\{B \circ\Gamma^{\beta T} (t_1)e^{-i\omega(t-t_1)} \}\rho ]
\\
\mathcal{Y}^{\alpha,\beta}(t,t_1) &= 
[A, \{e^{-i\omega(t-t_1)}\circ[B,A \circ\Delta\Gamma^{\alpha T}(t,t-t_1)e^{i\omega(t-t_1)}]\}\rho\{B\circ \Gamma^{\beta*} (t_1)e^{-i\omega(t-t_1)} \}]
\\
\mathcal{K}^{\alpha,\beta}(t,t_1) &= 
[A,[B\circ\Gamma^\beta(t - t_1)e^{-i\omega(t-t_1)},\{e^{-i\omega(t-t_1)}\circ[B,A \circ \Delta\Gamma^{\alpha T}(t,t-t_1)e^{i\omega(t-t_1)}]\}\rho]]. 
\end{align}

Next, indexing each term in the $H_S$ eigenbasis, using implicit notation for inner contractions (i.e.,
$(AB)_{ij} = A_{ik}B_{kj}$) we will set $i,j$ for the indices of $\rho$ and $n,m$ as outer, open indices of the total product of each term in the summation. With this notation, for $\mathcal{Q}^{\alpha,\beta}(t,t_1) $ we find

\begin{equation}
\label{Eq:Q_coefficients}
    \begin{split}
\mathcal{Q}^{\alpha,\beta}(t,t_1) 
    &\mapsto 
    +A_{na}\rho_{ij}\delta_{jm}
    B_{ab}A_{bc}B_{ci}\Delta\Gamma^{\alpha}_{cb}(t,t-t_1)\Gamma^{\beta T}_{ci}(t_1)e^{-i(\omega_{cb} + \omega_{ci}+\omega_{ac})(t-t_1)}
        \\ 
        &- A_{na}\rho_{ij}\delta_{jm}
        A_{ab}B_{bc}B_{ci}\Delta\Gamma^{\alpha}_{ba}(t,t-t_1) \Gamma^{\beta T}_{ci}(t_1)e^{-i(\omega_{ba}+\omega_{ci}+\omega_{ac})(t-t_1)}
        \\
        &- \rho_{ij}A_{jm}
        B_{na}A_{ab}B_{bi} \Delta\Gamma^{\alpha}_{ba}(t,t-t_1) \Gamma^{\beta T}_{bi}(t_1)e^{-i(\omega_{ba} + \omega_{bi}+\omega_{nb})(t-t_1)}
        \\
        &+ \rho_{ij}A_{jm}
        A_{na}B_{ab}B_{bi}\Delta\Gamma^\alpha_{an}(t,t-t_1) \Gamma^{\beta T}_{bi}(t_1)e^{-i(\omega_{an} + \omega_{bi} + \omega_{ni} )(t-t_1)}.
    \end{split}
\end{equation}
Using $\mapsto$ to indicate the translation of the operator from matrix product form to the indexed form of its $nm$-th coefficient. We likewise compute the other two operators, 
\begin{equation}
\label{Eq:Y_coefficients}
    \begin{split}
    \mathcal{Y}^{\alpha,\beta}(t,t_1)
\mapsto &+ A_{na}\rho_{ij}
A_{ab}B_{bi}B_{jm}\Delta\Gamma^\alpha_{ba}(t,t-t_1)  \Gamma^{\beta *}_{jm}(t_1)e^{-i(\omega_{ba}+\omega_{jm} + \omega_{ai})(t-t_1)}\\
        &- A_{na}\rho_{ij}
        B_{ab}A_{bi}B_{jm}\Delta\Gamma^\alpha_{ib}(t,t-t_1) \Gamma^{\beta *}_{jm}(t_1)e^{-i(\omega_{ib}+\omega_{jm} + \omega_{ai})(t-t_1)}\\
        &- A_{bm}\rho_{ij}
        A_{na}B_{ai}B_{jb} \Delta\Gamma^\alpha_{an}(t,t-t_1) \Gamma^{\beta *}_{jb}(t_1)e^{-i(\omega_{an} + \omega_{jb}+\omega_{ni})(t-t_1)}\\
        &+ A_{bm}\rho_{ij} 
        B_{na}A_{ai}B_{jb}\Delta\Gamma^\alpha_{ia}(t,t-t_1) \Gamma^{\beta *}_{jb}(t_1)e^{-i(\omega_{ia}+\omega_{jb} + \omega_{ni})(t-t_1)}.
    \end{split}
\end{equation}

\begin{equation}
\label{Eq:K_coefficients}
    \begin{split}
\mathcal{K}^{\alpha,\beta}(t,t_1)
\mapsto &+ A_{na}\rho_{ij}\delta_{jm}
A_{bc}B_{ci}B_{ab}\Delta\Gamma^\alpha_{cb}(t,t-t_1)\Gamma^\beta_{ab}(t-t_1)e^{-i(\omega_{cb}+\omega_{ab} + \omega_{bi})(t-t_1)}
\\
&- A_{na}\delta_{jm}\rho_{ij}
B_{bc}A_{ci}B_{ab}\delta_{jm}\Delta\Gamma^\alpha_{ic}(t,t-t_1) \Gamma^\beta_{ab}(t-t_1)e^{-i(\omega_{ic} + \omega_{ab} + \omega_{bi})(t-t_1)}
\\
&- A_{na}\rho_{ij}
A_{ab}B_{bi}B_{jm}\Delta\Gamma^\alpha_{ba}(t,t-t_1) \Gamma^\beta_{jm}(t-t_1)e^{-i(\omega_{ba}+\omega_{jm} + \omega_{ai})(t-t_1)}
        \\
        &+ A_{na}\rho_{ij}
        B_{ab}A_{bi}B_{jm}\Delta\Gamma^\alpha_{ib}(t,t-t_1) \Gamma^\beta_{jm}(t-t_1)e^{-i(\omega_{ib}+\omega_{jm} + \omega_{ai})(t-t_1)}
        \\
        &- \rho_{ij}A_{jm}
        A_{ab}B_{bi}B_{na}\Delta\Gamma^\alpha_{ba}(t,t-t_1)  \Gamma^\beta_{na}(t-t_1)e^{-i(\omega_{ba}+\omega_{na}+\omega{ai})(t-t_1)}
        \\
        &+ \rho_{ij}A_{jm}
        B_{ab}A_{bi}B_{na}\Delta\Gamma^\alpha_{ib}(t,t-t_1) \Gamma^\beta_{na}(t-t_1)e^{-i(\omega_{ib} + \omega_{na} + \omega_{ai})(t-t_1)}
        \\
        &+ A_{bm}\rho_{ij}
        A_{na}B_{ai}B_{jb} \Delta\Gamma^\alpha_{an}(t,t-t_1) \Gamma^\beta_{jb}(t-t_1)e^{-i(\omega_{an}+\omega_{jb} + \omega_{ni})(t-t_1)}
        \\
        &- A_{bm}\rho_{ij}
        B_{na}A_{ai}B_{jb}\Delta\Gamma^\alpha_{ia}(t,t-t_1) \Gamma^\beta_{jb}(t-t_1)e^{-i(\omega_{ia}+\omega_{jb} + \omega_{ni})(t-t_1)}
    \end{split}
\end{equation}

We will introduce the following three-dimensional spectral densities that will encode the integration of the time-dependent component of each of the coefficients contributing to Eq.~\eqref{Eq:TCL_1D_final}.

\begin{eqnarray}
    \label{Eq:Ft}
     \mathsf{F}_{ab,cd,ef}^{\alpha\beta}(t)&=&  \mathsf{F}_{\omega_1\omega_2\omega_3}^{\alpha \beta}(t) =\int_0^{t} ds \Delta \Gamma^{\alpha}_{\omega_1}(t,t-s) \Gamma_{\omega_2}^{\beta T}(s) e^{-i(\omega_1 + \omega_2 + \omega_3)(t - s)}, \\
    \label{Eq:Ct}
     \mathsf{C}_{ab,cd,ef}^{\alpha \beta}(t) &=& \mathsf{C}_{\omega_1\omega_2\omega_3}^{\alpha \beta}(t)=\int_0^{t} ds \Delta \Gamma^{\alpha}_{\omega_1}(t,t-s) \Gamma_{\omega_2}^{\beta *}(s) e^{-i(\omega_1 + \omega_2 + \omega_3)(t - s)},\\
    \label{Eq:Rt}
     \mathsf{R}_{ab,cd,ef}^{\alpha \beta}(t) &=& \mathsf{R}_{\omega_1\omega_2\omega_3}^{\alpha \beta}(t)=\int_0^{t} ds \Delta \Gamma^{\alpha}_{\omega_1}(t,t-s) \Gamma^\beta_{\omega_2}(t - s) e^{-i(\omega_1 + \omega_2 + \omega_3)(t - s)}.
\end{eqnarray}
We will rewrite these expressions by making a trivial insertion of the type $0 = -\Gamma^\beta(t)+\Gamma^\beta(t)$ and making the change of variables $\tau = t-s$ within each of the equations \eqref{Eq:Ft}~--~\eqref{Eq:Rt}, we arrive at

\begin{align}
\label{Eq:Ft2}
\mathsf{F}_{\omega_1\omega_2\omega_3}^{\alpha \beta}(t) &=\int_0^{t} d\tau \Delta \Gamma^{\alpha}_{\omega_1}(t,\tau) \Delta\Gamma_{\omega_2}^{\beta T}(t-\tau,t) e^{-i(\omega_1 + \omega_2 + \omega_3)\tau} + \Gamma_{\omega_2}^{\beta T}(t)\int_0^{t} d\tau \Delta \Gamma^{\alpha}_{\omega_1}(t,\tau) e^{-i(\omega_1 + \omega_2 + \omega_3)\tau} \tag{\ref*{Eq:Ft}'},
\\
\label{Eq:Ct2}
\mathsf{C}_{\omega_1\omega_2\omega_3}^{\alpha \beta}(t)&=\int_0^{t} d\tau \Delta \Gamma^{\alpha}_{\omega_1}(t,\tau) \Delta\Gamma_{\omega_2}^{\beta *}(t-\tau,t) e^{-i(\omega_1 + \omega_2 + \omega_3)\tau}  
+\Gamma_{\omega_2}^{\beta *}(t) \int_0^{t} d\tau \Delta \Gamma^{\alpha}_{\omega_1}(t,\tau) e^{-i(\omega_1 + \omega_2 + \omega_3)\tau}\tag{\ref*{Eq:Ct}'} ,\\
    \label{Eq:Rt2}
 \mathsf{R}_{\omega_1\omega_2\omega_3}^{\alpha \beta}(t)&=\int_0^{t} d\tau \Delta \Gamma^{\alpha}_{\omega_1}(t,\tau) \Delta\Gamma^\beta_{\omega_2}(
\tau,t) e^{-i(\omega_1 + \omega_2 + \omega_3)\tau}+ \Gamma^\beta_{\omega_2}(t)\int_0^{t} d\tau \Delta \Gamma^{\alpha}_{\omega_1}(t,\tau) e^{-i(\omega_1 + \omega_2 + \omega_3)\tau}\tag{\ref*{Eq:Rt}'}.
\end{align}

Evaluating the second integral in each case, we have
\begin{multline}
\label{Eq:FRCt_Singularity}
        \int_0^{t} d\tau \Delta \Gamma^{\alpha}_{\omega_1}(t,\tau) e^{-i(\omega_1 + \omega_2 + \omega_3)\tau}
    =\int_0^{t} d\tau \int_{\tau}^{t} ds C_{\alpha}(s)e^{i\omega_1 s} e^{-i(\omega_1 + \omega_2 + \omega_3)\tau} 
    =\int_0^{t} ds C_{\alpha}(s) \frac{i[e^{-i(\omega_2 + \omega_3)s} - e^{i\omega_1 s}}{\omega_1 + \omega_2 + \omega_3}
   \\ = \frac{i[\Gamma^\alpha_{-\omega_2 - \omega_3}(t) - \Gamma^\alpha_{\omega_1}(t)]}{\omega_1 + \omega_2 + \omega_3}.
\end{multline}
 Plugging this back in to Eqs.~\eqref{Eq:Ft2}~--~\eqref{Eq:Rt2}, we arrive at the densities defined in Eqs.~\ref{Eq:preferredF}-\ref{Eq:preferredR}. 
Using these definitions, and combining \eqref{Eq:Q_coefficients}~--~\eqref{Eq:K_coefficients} back into the generator \eqref{Eq:TCL_1D_final}, using the original $\alpha,\beta$ bath indices, we arrive at the following equation
\begin{subequations}
\label{Eq:TCL4_Indexed}
\begin{eqnarray}
    \left[\Big{(}\frac{\partial \rho}{\partial t}\Big{)}_4\right]_{nm} =
\sum_{\alpha,\beta=1}^{N_b} \sum_{i,j=1}^{N} \rho_{ij}&&\Big{\{}\sum_{a,b,c=1}^{N}A^\alpha_{na}\delta_{jm}A^\beta_{ab}A^\alpha_{bc}A^\beta_{ci}[\mathsf{F}_{cb,ci,ac}^{\alpha \beta}(t)-\mathsf{R}_{cb,ab,bi}^{\alpha \beta}(t)]\\&&-
        \sum_{a,b,c=1}^{N}A^\alpha_{na}\delta_{jm}A^\beta_{bc}[A^\beta_{ci}A^\alpha_{ab}\mathsf{F}_{ba,ci,ac}^{\alpha \beta}(t)-A^\alpha_{ci}A^\beta_{ab}\mathsf{R}_{ic,ab,bi}^{\alpha \beta}(t)]
\\
 && - \sum_{a,b=1}^{N}A^\alpha_{jm}A^\beta_{na}A^\alpha_{ab}A^\beta_{bi}[ \mathsf{F}_{ba,bi,nb}^{\alpha \beta}(t)-\mathsf{R}_{ba,na,ai}^{\alpha \beta}(t)]
 \\&&+ \sum_{a,b=1}^{N}A^\alpha_{jm}A^\beta_{ab}[A^\alpha_{na}A^\beta_{bi}\mathsf{F}_{an,bi,nb}^{\alpha \beta}(t)-A^\beta_{na}A^\alpha_{bi}\mathsf{R}_{ib,na,ai}^{\alpha \beta}(t)
        ] \\
         &&+ \sum_{a,b=1}^{N}A^\alpha_{na}A^\alpha_{ab}A^\beta_{bi}A^\beta_{jm}[\mathsf{C}_{ba,jm,ai}^{\alpha \beta}(t)
         +\mathsf{R}_{ba,jm,ai}^{\alpha \beta}(t)]\\
        &&- \sum_{a,b=1}^{N}A^\alpha_{na}A^\beta_{ab}A^\alpha_{bi}A^\beta_{jm}[\mathsf{C}_{ib,jm,ai}^{\alpha \beta}(t)
        +\mathsf{R}_{ib,jm,ai}^{\alpha \beta}(t)]\\
        &&- \sum_{a,b=1}^{N}A^\alpha_{na}A^\alpha_{bm}A^\beta_{ai}A^\beta_{jb}[\mathsf{C}_{an,jb,ni}^{\alpha \beta}(t)+\mathsf{R}_{an,jb,ni}^{\alpha \beta}(t)
        ]\\
        &&+ \sum_{a,b=1}^{N}A^\beta_{na}A^\beta_{jb}A^\alpha_{bm}A^\alpha_{ai}[\mathsf{C}_{ia,jb,ni}^{\alpha \beta}(t)+\mathsf{R}_{ia,jb,ni}^{\alpha \beta}(t)]\Big{\}} + h.c.
\end{eqnarray}
\end{subequations}
The superoperator $\delta \mathcal{R}^{4}_{nm,ij}$ can immediately be identified from Eq.~\eqref{Eq:TCL4_Indexed} as the expression within the curly brackets, giving us Eqs.~\ref{Eq:ExplicitLiouv}-\ref{Eq:Expl4}.
\end{widetext}

\section{CONVERGENCE OF THE TCL4 INTEGRALS\label{Appendix:TCL4integrals}}
Both the verification of proposition~\ref{Eq:OverlapIntegral} and the existence of  integral~\ref{Eq:3DSD} have yet to be demonstrated.
We can assume the spectral densities of the baths are identical without losing generality because the generator is summing over the baths, i.e., see Eqs.~\ref{Eq:ExplicitLiouv}-\ref{Eq:Expl4}.
If not, then the spectral density with the smallest $s$ will result in the generator's convergence being determined by the same criterion.

The values of the 3D spectral densities at zero frequency, e.g., $\mathsf{F}_{000}(t)$, $\mathsf{R}_{000}(t)$, and $\mathsf{C}_{000}(t)$, have no effect on the TCL4 generator $\delta\mathcal{R}_{nm,ij}^{4}(t)$ given by Eqs.~\ref{Eq:ExplicitLiouv}-\ref{Eq:Expl4}. Thus, the frequency set $\omega_1=\omega_2=\omega_3=0$ is irrelevant to the dynamics governed by the TCL4 master equation. 
To show this, we trim the sum over $a,b$, and $c$ in Eqs.~\ref{Eq:ExplicitLiouv}-\ref{Eq:Expl4} to only include terms with zero frequencies in the subscripts, resulting in zero. 

The convergence analysis is aided by the well-known asymptotic formula for the incomplete gamma function~\cite{dingle1973asymptotic,olver1997asymptotics}:
\begin{equation}
\Gamma(a,x)= x^{a-1}e^{-x}[1+O(1/x)], \vert x\vert\gg 1
\end{equation}
Substituting into Eq.~\ref{Eq:Aomegat}, we find
\begin{equation}
\label{Eq:DGasymptotic}
  \Delta\Gamma_\omega(t)= \Gamma_\omega-\Gamma_\omega(t)\sim \left\{
\begin{array}{ll}
t^{-s-1}e^{i\omega t}, & \hbox{$\omega\neq 0$;}\\
t^{-s}, & \hbox{$\omega=0$.} 
\end{array}
\right.
\end{equation}

The integral~\ref{Eq:3DSD} at large $t$ becomes
\begin{equation}
\label{Eq:Int50}
\sim\left\{
\begin{array}{lll}
\int_a^t d\tau\frac{e^{-i\omega_3\tau}}{\tau^{2s+2}}, & \hbox{$\omega_1\neq 0$ and $\omega_2\neq 0$;}\\
\int_a^t d\tau\frac{e^{-i\omega_3\tau}}{\tau^{2s+1}}, & \hbox{$\omega_1=0,\omega_2\neq 0$ or vice versa} ;\\
\int_a^t d\tau\frac{e^{-i\omega_3\tau}}{\tau^{2s}}, & \hbox{$\omega_1=\omega_2= 0$}.\\
\end{array}
\right.
\end{equation}
where $a>0$. At the relevant frequencies,
e.g., $\vert\omega_{1}\vert+\vert\omega_{2}\vert+\vert\omega_{3}\vert> 0$, 
these integrals converge when $s>0$. 
If $\omega_1=\omega_2=\omega_3=0$ and $s<1/2$, then the third integral  diverges as $t^{1-2s}$, but this is irrelevant because the generator is independent of the 3D-spectral densities when all three frequencies are zero.

Next we verify the proposition
~\ref{Eq:OverlapIntegral}. First, we want to show that
\begin{equation}
\label{Eq:absoverlap}
\lim_{t\to\infty}\Delta\Gamma_{\omega_1}(t,\tau)\Delta\Gamma_{\omega_2}(t,t-\tau)=0,\forall \omega_{1,2} \text{ and } 0<\tau<t.
\end{equation}
This condition is invariant with respect to the exchange $t\leftrightarrow t-\tau$ and $\omega_1\leftrightarrow\omega_2$.  Thus, we can assume $\tau=xt$ where $x\leq 1/2$, without loss in generality. In this range of $x$ we find, applying Eq.~\ref{Eq:DGasymptotic},
\begin{equation}
\label{Eq:ineqx}
    \Delta\Gamma_{\omega}(t,t-\tau)=\Delta\Gamma_\omega(t-xt)-\Delta\Gamma_\omega(t)\sim
     \left\{
\begin{array}{ll}
t^{-s-1}, & \hbox{$\omega\neq 0$;}\\
t^{-s}, & \hbox{$\omega=0$,} 
\end{array}
\right.
\end{equation}
uniformly in $x$.
So $\Delta\Gamma_{\omega_1}(t,\tau)\Delta\Gamma_{\omega_2}(t,t-\tau)$
reaches zero at $t\to\infty$. Similarly, we find 
\begin{equation}
  \begin{split}
&\int_0^{t/2}\Delta\Gamma_{\omega_1}(t,\tau)\Delta\Gamma_{\omega_2}(t,t-\tau)e^{-i(\omega_1+\omega_2+\omega_3)\tau}d\tau \\
&=
     \left\{
\begin{array}{ll}
t^{-2s-c}, & 
\hbox{$c\geq 0$, $\vert\omega_{1}\vert+\vert\omega_{2}\vert+\vert\omega_{3}\vert> 0$;}\\
t^{-2s+1}, & \hbox{$\omega_1=\omega_2=\omega_3=0$.}
\end{array}
\right.
\end{split}
\end{equation}
This  integral reaches zero at $t\to\infty$ and any $s>0$, confirming proposition~\ref{Eq:OverlapIntegral}, except for $s\leq 1/2$ when all three frequencies are zero, but this is irrelevant for the TCL4 generator.

\section{\label{Sec:inaccuracy}INACCURACY OF THE BLOCH-REDFIELD MASTER EQUATION}

\begin{figure}[h]
\includegraphics[width=0.45\textwidth]{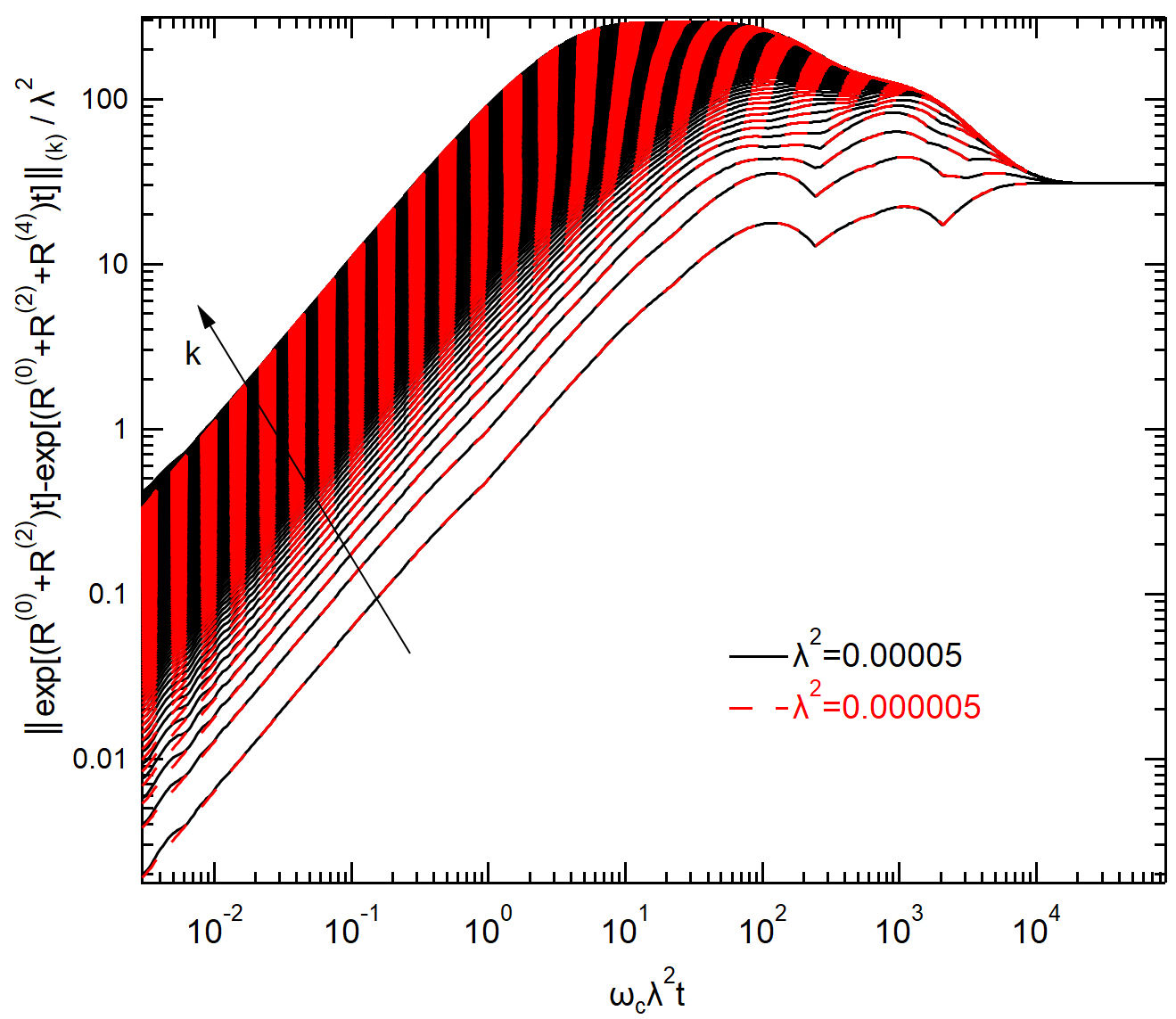}
\caption{\label{Fig1:norm} Rescaled Ky Fan norms of the difference between the TCL2 and TCL4 reduced state propagators versus rescaled time. 
$k=1,2,\,.\,.\,.\,,1225$, bottom to top. $N=35$, $T=0$ K, $\omega_c=10$, and $s=1$.}
\end{figure}

By comparing the TCL2 and TCL4 dynamics, it is possible to determine the accuracy of the TCL2 dynamics, as follows. 
The reduced dynamics is approximated by a semigroup with the reduced state propagator $\exp{(\mathcal{R}t})$,
where $\mathcal{R}$ is the asymptotic generator of the TCL master equation.
In Fig.~\ref{Fig1:norm}, the
Ky Fan $k$-norms~\cite{bhatia97} defined as
\begin{equation}
    \Vert X\Vert_{(k)}=\sum_{j=1}^k s_j(X),
\end{equation}
where $X$ is a superoperator and $s_j(X)$ are the singular values of $X$ in descending order and characterize the inaccuracy of the TCL2-dynamics. 
For $k = 1$ and $k = N^2$, the operator and trace norms are, respectively, the Ky Fan $k$-norm. 
The norms are determined on the difference between TCL2 and TCL4 state propagators.
After rescaling by $\lambda^2$, they are displayed on the vertical axis. Excellent scaling can be seen in the results, which were computed for two alternative values of $\lambda^2$.

We see that all norms collapse on the operator norm at very long time intervals. The system relaxes to the asymptotic state with a single nonzero singular value. The linear scaling of the norm with $\lambda^2$ indicates once more that the asymptotic TCL2's state error is $O(\lambda^2)$.

The significant dispersion in the Ky Fan $k$-norms over shorter timescales shows that there may be a variety of deviations between the TCL2 and TCL4 dynamics. The crucial issue is that the dynamics' inaccuracy endures throughout time exceeding the relaxation time $\sim 1/(\lambda^2\omega_c)$.  This demonstrates that the Bloch-Redfield master equation cannot accurately capture the dynamics above the relaxation time. The Davies master equation has the same problem~\cite{davies1974}.

\section{RAYLEIGH-SCHR{\"O}DINGER PERTURBATION THEORY FOR FOURTH-ORDER COHERENCES\label{Appendix:TIPT}}
Here we will derive the Rayleigh-Schr{\"o}dinger perturbation theory reduced density results seen in Sec.~IV E. As mentioned in the text, we are considering the ground state of $H_T = H_0 + \lambda V$ to be a perturbation of the ground state of $H_0 = H_S + H_B$ by the interaction $H_I = \lambda V$, where $\lambda$ is a small dimensionless coupling constant.  We will expand the ground state $|\tilde 0\rangle $ and energy $\tilde \epsilon_0$ of $H_T$  in a power series of $\lambda$,
\begin{eqnarray}
\label{Eq:TIPT_vector}
   |\tilde{0}\rangle &=& \sum_{k=0}^\infty \lambda^k |0_k\rangle,\\
   \label{Eq:TIPT_energy}
   \tilde \epsilon_0 &=& \sum_{k=0}^\infty \lambda^k \epsilon_k.
\end{eqnarray}
To zeroth order in $\lambda$, the ground state is that of the free Hamiltonian, $|0_0\rangle = |0\rangle$ and  $\epsilon_0= \tilde\epsilon_0$. Since the system and bath are isolated for $H_0$, we can express $|0\rangle = |E_1\rangle \otimes |\omega_0\rangle $ and then $\epsilon_0 = E_1 + \omega_0$, where $H_S |E_1\rangle = E_1 |E_1\rangle$ and $H_B|\omega_0\rangle = \omega_0|\omega_0 \rangle$. We will assume $E_1 =\omega_0 = 0$.

We will require both our unperturbed and perturbed state to be normalized, that is, $\langle0|0\rangle = \langle\tilde 0|\tilde 0\rangle = 1$, which gives the condition that at each order $k\ge1$ we have
\begin{equation}
\label{Eq:TIPTNormalization}
  \sum_{q=0}^k  \langle 0_q| 0_{k-q}\rangle =0.
\end{equation}
We will set $\langle0|0_k \rangle= \langle 0_k|0\rangle$ at each order. Inserting \eqref{Eq:TIPT_vector} and \eqref{Eq:TIPT_energy} into the eigenequation 
$H_T|\tilde 0  \rangle =  \tilde\epsilon_0 |\tilde 0  \rangle $, and iteratively for the ground-state corrections up to fourth order in $\lambda$, we have 
\begin{eqnarray}
    \label{Eq:TIPT_S1}
    |0_1\rangle &=&\frac{V_{k0}}{d_{0k}}|k\rangle\\
    \label{Eq:TIPT_S2}
    |0_2\rangle &=&\frac{V_{kl}V_{l0}}{d_{k0}d_{l0}}|k\rangle- \frac{1}{2} \frac{|V_{0k}|^2 }{d_{k0}^2}|0\rangle\\
    |0_3\rangle &=& \frac{V_{kl}V_{lm}V_{m0}}{d_{l0}d_{k0}d_{m0}}|k\rangle\notag\\\label{Eq:TIPT_S3}
    &+& \frac{V_{k0}|V_{l0}|^2}{d_{k0}d_{l0}}(\frac{1}{d_{k0}} + \frac{1}{2d_{l0}})|k\rangle\\
    |0_4\rangle &=&
    \frac{|V_{kl}|^2|V_{m0}|^2}{d_{k0}d_{m0}d_{kl}^2}|k\rangle \notag\\
    &+& \frac{V_{kl}V_{l0}|V_{m0}|^2}{d_{k0}d_{m0}d_{l0}}(\frac{1}{d_{0m}}+\frac{1}{d_{0k}})|k\rangle\notag\\
    &-& \frac{V_{kl}V_{lm}V_{mn}V_{n0}}{d_{k0}d_{l0}d_{0m}d_{0n}}|k\rangle \notag\\
    &-& \frac{1}{2}\frac{V_{0k}V_{kl}V_{lm}V_{m0}}{d_{k0}d_{l0}d_{m0}}(\frac{1}{d_{l0}}+\frac{1}{d_{m0}})|0\rangle\notag\\\label{Eq:TIPT_S4}
    &-& \frac{1}{2}\frac{|V_{0k}|^2}{d_{k0}d_{l0}^2}(\frac{3}{4} \frac{|V_{0l}|^2}{d_{k0}} - \frac{V_{ml}V_{lk}}{d_{m0}})|0\rangle
\end{eqnarray}
where we have used notation $V_{nm} = \langle n| V|m\rangle$, $d_{nm} = \epsilon_n - \epsilon_m$, and each term has a an implicit summation over the eigenbasis of $H_0$ for each index, excluding the states that would cause $d_{nn} = 0$ factors.  We have excluded terms from the perturbative expansion that include $V_nn$ terms, as well as terms with $V_{kl}$ where the bath components of the eigenstates would necessarily have the same or an even number difference of bosons, since we have $\langle F\rangle= 0$, since our bath coupling operator is linear in bosonic operators. Expanding then the density matrix in orders of $\lambda$, we find
\begin{eqnarray}
        \tilde \rho &=& |\tilde 0\rangle \langle\tilde 0 |  = \frac{1}{2}|0\rangle \langle 0| + \lambda|0_1\rangle \langle 0| + \lambda^2|0_2\rangle \langle 0_0| \notag\\&+& \frac{\lambda^2}{2}|0_1\rangle \langle 0_1|)+\lambda^3|0_2\rangle \langle 0_1| + \lambda^3|0_3\rangle \langle 0_0|\notag\\
&+&\frac{\lambda^4}{2}|0_2\rangle \langle 0_2|+\lambda^4|0_3\rangle \langle 0_1| \notag\\\label{Eq:rho_total_expansion}
    &+& \lambda^4|0_4\rangle \langle 0_0| + h.c. + O(\lambda^5). 
\end{eqnarray}
When calculating $\tilde \rho_S = \text{Tr}_B(\tilde \rho)$, we see that odd order terms go to zero, again because $\langle F\rangle= 0$. 
Thus we have, 
\begin{equation}
\label{Eq:TiPT_reduced_general}
\begin{split}
    \tilde \rho_S =&\frac{1}{2}|E_1\rangle \langle E_1| + \lambda^2\text{Tr}_B(|0_2\rangle\langle 0_0|+ \frac{1}{2}|0_1\rangle\langle 0_1|) \\
    +& \lambda^4\text{Tr}_B(|0_4\rangle \langle 0_0|+|0_3\rangle \langle 0_1|+\frac{1}{2}|0_2\rangle \langle 0_2|)\\ +& h.c. 
    + O(\lambda^6).
    \end{split}
\end{equation}

When evaluating these terms, we encounter expressions such as 
\begin{multline}
\label{Eq:TiPTSD_rel1}
    \sum_{\omega_\alpha \neq 0} \frac{g_{\alpha}^2}{w + \omega_\alpha} = \int_{-\infty}^{\infty}d\omega \sum_{\omega_\alpha \neq 0} \frac{g_{\alpha}^2\delta(\omega - \omega_\alpha)}{w + \omega_\alpha}\\
    = -\int_{-\infty}^{\infty}d\omega \sum_{\omega_\alpha \neq 0} \frac{ g_{\alpha}^2\delta(\omega - \omega_\alpha)}{-w - \omega_\alpha} = -\tilde{\mathcal{S}}_{-w}
\end{multline}
for $-w \neq \omega_\alpha$, which will always be the case due to our limitation of only nonzero $d_{mn}$.

Similarly, we will encounter, 
\begin{eqnarray}
\label{Eq:TiPTSD_rel2}
    \sum_{\omega_\alpha \neq 0} \frac{g_{\alpha}^2}{(w + \omega_\alpha)^2}  = -\partial \tilde{\mathcal{S}}_{-w}
\end{eqnarray}
Looking at just at the second-order terms from \eqref{Eq:TiPT_reduced_general}, tracing over the bath degrees of freedom we find  
\begin{widetext}
\begin{equation}\label{Eq:TIPT_2ndOrder}\begin{split}
\tilde{\rho}_S^{2}=& \lambda^2\text{Tr}_B(|0_2\rangle\langle 0_0| + |0_0\rangle\langle 0_2|+ \frac{1}{2}|0_1\rangle\langle 0_1|) = \sum_{\substack{n=2\\ k=1}}^{N}\sum_{\alpha}\frac{A_{nk}A_{k1}g_\alpha^2}{E_{n}(E_{k}+ \omega_\alpha)} |E_{n}\rangle \langle E_1| \\&+ \frac{1}{2}\sum_{n,m=1}^{N}\sum_{\alpha}\frac{A_{n1}A_{1m}g_\alpha^2}{(E_n + \omega_\alpha)(E_m+\omega_\alpha)}|E_{n}\rangle \langle E_{m}|
- \frac{1}{2}\sum_{n =1}^N\sum_{\alpha} \frac{|A_{1n}|^2g_\alpha^2}{(E_{n} + \omega_\alpha)^2} |E_1 \rangle\langle E_1| + h.c.
\end{split}
\end{equation}
\end{widetext}
Note that when using the Eq.~6 definition of the bath coupling operator $F$ when tracing over the bath (all Fock space eigenstates), we find
\begin{equation}
\label{Eqn:TiPT_BathInnerProds}
    \sum_{\omega_k \neq 0}|F_{0k}|^2 = \langle 0 | F |\omega_k \rangle \langle \omega_k | F|0 \rangle  = \sum_{\alpha} g_{\alpha}^2
\end{equation}
That is, for these terms the only nonzero contribution is from the single-particle states, and we can express these products in terms of the form factors. 

It should be noted that the last term in the \eqref{Eq:TIPT_2ndOrder} results from the normalization of $|\tilde 0\rangle$ giving a second-order condition that $\langle 0| 0_2\rangle = -\frac{1}{2} \langle 0_1 |0_1\rangle$. We will see that this term cancels a possible divergence in the populations. If we instead required the perturbations be orthogonal to the isolated state, $\langle 0| 0_k\rangle = 0, \forall k>0$, then at second order in alpha we would have a $\frac{\partial S(-E_{1})}{\partial E_1}$ diverging term, in the case when $a=b=1$.

We will now relate our second-order term to our continuum limit spectral densities, using \eqref{Eq:TiPTSD_rel1} and \eqref{Eq:TiPTSD_rel2}, we have 
\begin{widetext}
\begin{equation}
\begin{split}
\tilde{\rho}_S^{2} 
= -\sum_{\substack{a=2\\ b=2}}^{N}\frac{\mathcal{S}_{{1b}}}{\omega_{a1}} (A_{ab}A_{b1}|E_{a}\rangle \langle E_1| +A_{1b}A_{ba}|E_{1}\rangle \langle E_{a}|)
\\+ \sum_{\substack{a,b=2\\ a\not = b}}^{N}A_{a1}A_{1b}\frac{\mathcal{S}_{{1a}}-\mathcal{S}_{{1b}}}{\omega_{ab}}|E_{a}\rangle \langle E_{b}|
+ \sum_{a=2}^{N}|A_{1a}|^2\mathcal\partial{S}_{{1a}}(|E_{a}\rangle \langle E_{a}|-|E_1 \rangle\langle E_1|).
\end{split}
\end{equation}
leading to Eqs.~78 and~79 in the main text.  
Using this same approach for the $O(\lambda^4)$ terms, we find 

\begin{eqnarray}
\tilde{\rho}^{4}&=& |n\rangle \langle m|[\frac{(V_{nl}V_{l0}V_{km}V_{0k}-V_{nl}V_{lk}V_{0m}V_{k0}-V_{n0}V_{kl}V_{lm}V_{0k})}{d_{l0}d_{k0}d_{m0}d_{n0}}
    - \frac{V_{n0}V_{0m}|V_{k0}|^2}{d_{n0}d_{m0}d_{k0}}(\frac{1}{d_{n0}} + \frac{1}{d_{k0}} + \frac{1}{d_{m0}})]
\\ &+&|0\rangle\langle0|[\frac{1}{2}\frac{|V_{0k}|^2 (V_{kl}V_{lm}+V_{ml}V_{lk})}{d_{m0}d_{k0}d_{l0}^2} - \frac{1}{2}\frac{|V_{0k}|^2 |V_{0l}|^2}{ d_{k0}^2d_{l0}^2} 
    - \frac{1}{2}\frac{V_{k0}V_{lk}V_{ml}V_{0m}}{d_{m0}d_{l0}d_{k0}}(\frac{1}{d_{m0}}+\frac{1}{d_{k0}} +\frac{2}{d_{l0}})
      ]\\   &+&|n\rangle \langle 0 |[\frac{|V_{nl}|^2|V_{k0}|^2}{d_{n0}d_{k0}d_{nl}^2}-\frac{V_{nl}V_{lm}V_{mk}V_{k0}}{d_{n0}d_{l0}d_{m0}d_{k0}} 
    - \frac{V_{nl}V_{l0}|V_{k0}|^2}{d_{n0}d_{k0}d_{l0}}(\frac{3}{2d_{k0}}+\frac{1}{d_{n0}})]\\ 
    &+&|0\rangle\langle m| [\frac{|V_{ml}|^2|V_{k0}|^2}{d_{m0}d_{k0}d_{ml}^2}
    -\frac{V_{lm}V_{kl}V_{nk}V_{0n}}{d_{m0}d_{l0}d_{k0}d_{n0}}
    - \frac{|V_{k0}|^2V_{lm}V_{0l}}{d_{m0}d_{k0}d_{l0}}(\frac{3}{2d_{k0}}+\frac{1}{d_{m0}})].
\end{eqnarray}
Tracing over the bath degrees of freedom, we will use the notation introduced in  \eqref{Eqn:TiPT_BathInnerProds} as well as a new definition $D_{ka} = E_k + \omega_a$, assigning indices to the system and bath part of the total energy eigenvalues, i.e., $|\epsilon_i\rangle = |E_k\rangle \otimes |\omega_a\rangle$, to find
\begin{eqnarray}
\label{Eqn:rho_4_allTerms}
    \tilde{\rho}_S^{4} &=& A_{nl}A_{l1}A_{km}A_{1k}
\frac{|F_{0a}|^2|F_{b0}|^2}{E_m E_n D_{la}D_{kb}}|E_n\rangle \langle E_m|  
+A_{nl}A_{l1}A_{km}A_{1k}
\frac{F_{ca}F_{a0}F_{bc}F_{0b}}{D_{nc}D_{ma}D_{kb} D_{la}}|E_n\rangle \langle E_m| \\
&-&(A_{nl}A_{lk}A_{1m}A_{k1}
+A_{n1}A_{kl}A_{lm}A_{1k}
)\frac{|F_{0a}|^2 |F_{0b}|^2}{E_{l}D_{na}D_{ma}D_{kb}}|E_n\rangle \langle E_m| 
\notag\\
&-&(A_{nl}A_{lk}A_{1m}A_{k1}
\frac{F_{ac}F_{cb}F_{0a}F_{b0}}{D_{na}D_{ma}D_{kb}D_{lc}}+A_{n1}A_{kl}A_{lm}A_{1k}
\frac{F_{bc}F_{ca}F_{a0}F_{0b}}{D_{na}D_{ma}D_{kb}D_{lc}})|E_n\rangle \langle E_m| 
    \notag\\ &-&A_{n1}A_{1m}|A_{k1}|^2
    \frac{|F_{a0}|^2|F_{b0}|^2}{D_{kb} D_{na}D_{ma}}(\frac{1}{D_{na}} + \frac{1}{D_{kb}} + \frac{1}{D_{ma}})|E_n\rangle \langle E_m|\notag\\
    &+ & \frac{|A_{1k}|^2|F_{0b}|^2}{ D_{ma}D_{kb}}[\frac{\text{Re}(A_{kl}A_{lm}F_{bc}F_{ca})}{D_{lc}^2} +\frac{\text{Re}(A_{kl}A_{lm}F_{b0}F_{0a})}{E_l}]|E_1\rangle\langle E_1| 
\notag\\
    &-& \frac{1}{2}A_{k1}A_{lk}A_{ml}A_{1m}\frac{|F_{0a}|^2 |F_{0b}|^2}{E_lD_{ma}D_{kb}}(\frac{1}{D_{ma}}+\frac{1}{D_{kb}} + \frac{2}{E_l})|E_1\rangle\langle E_1|- \frac{1}{2}|A_{1k}|^2 |A_{1l}|^2\frac{|F_{0a}|^2 |F_{0b}|^2}{D_{kb}^2D_{la}^2}|E_1\rangle\langle E_1|
  \notag\\
    &-& \frac{1}{2}A_{k1}A_{lk}A_{ml}A_{1m}\frac{F_{b0}F_{ac}F_{cb}F_{0a}}{D_{ma}D_{kb}D_{lc}}(\frac{1}{D_{ma}}+\frac{1}{D_{kb}} + \frac{2}{D_{lc}})|E_1\rangle\langle E_1|
     \notag\\
&+&|A_{ml}|^2|A_{k1}|^2\frac{|F_{0a}|^2|F_{b0}|^2}{E_m D_{kb}(w_{ml} - \omega_a)^2}|E_1\rangle\langle E_m|-A_{lm}A_{kl}A_{nk}A_{1n}\frac{|F_{0a}|^2|F_{0b}|^2}{E_m E_k D_{la}D_{nb}}|E_1\rangle\langle E_m|\notag\\&-& |A_{k1}|^2A_{lm}A_{1l}\frac{|F_{0b}|^2|F_{0a}|^2}{E_m D_{kb}D_{la}}(\frac{3}{2 D_{kb}}+\frac{1}{E_m})|E_1\rangle\langle E_m|
-A_{lm}A_{kl}A_{nk}A_{1n}\frac{F_{a0}F_{bc}F_{ca}F_{0b}}{E_m D_{la}D_{kc}D_{nb}}|E_1\rangle\langle E_m| \notag\\
&+& (|E_{n} \rangle \langle E_1| \text{ terms}) \notag
\end{eqnarray}

\end{widetext}
where the $|E_{n} \rangle \langle E_1|$ terms are the Hermitian conjugate of the $|E_1 \rangle \langle E_m|$ expressions, with the additional change of $m\leftrightarrow n$. There are implicit summations for each index, and it should be noted that each bath energy index (e.g., $a,b,c$), represents a sum over the occupied states only, as the nonzero terms from an index being the vacuum state have been separated off.   
To evaluate these terms, plugging the bosonic operators in as in  \eqref{Eqn:TiPT_BathInnerProds}, we find
\begin{widetext}
\begin{eqnarray}
        \sum_{a,b,c} \frac{F_{a0}F_{ca}F_{bc}F_{b0}}{D_{la}D_{kb}D_{ic}} &=&\sum_{a,b}\frac{g_a^2 g_b^2}{D_{la}(E_i + \omega_a + \omega_b)} \left(\frac{1}{D_{ka}} + \frac{1}{D_{kb}} \right)\notag\\
        &=&\sum_{a,b}\left\{\frac{g_b^2}{(E_{i} + \omega_a + \omega_b)} \left[\frac{g_a^2}{D_{ka}D_{la}} - \frac{g_a^2}{(\omega_{ik} + \omega_a) D_{la}} \right] +  \frac{g_a^2}{(\omega_{ik} +\omega_a)D_{la}}\frac{g_b^2}{D_{kb}}\right\}.\label{Eqn:BathPartialFractions}
\end{eqnarray}

We consider the case $\tilde \rho_{S,nm}^{4}$ with $n\neq m$ and $n,m = 2,\ldots, N$, applying ~\eqref{Eqn:BathPartialFractions}, and after some algebra we find

    \begin{eqnarray}
\tilde{\rho}^{4}_{S,nm} &=&A_{nl}A_{l1}A_{km}A_{1k}
\frac{g_a^2 g_b^2}{E_m E_n D_{la}D_{kb}}
 \nonumber\\&-&A_{n1}A_{1m}|A_{k1}|^2
    \left[\frac{g_b^2}{\omega_{nm}D_{kb}}\left(\frac{g_a^2 }{D_{ma}^2} - \frac{g_a^2}{D_{na}^2}\right) +
    \frac{ g_b^2}{\omega_{nm}D_{kb}^2} \left(\frac{g_a^2}{D_{ma}} - \frac{g_a^2}{D_{na}}\right)\right]\nonumber\\
&+&(A_{nl}A_{lk}A_{1m}A_{k1}
+A_{n1}A_{kl}A_{lm}A_{1k})
\frac{1}{\omega_{nm}} \frac{g_b^2}{D_{kb}}\nonumber\\&\times&\left[\frac{g_a^2 }{D_{na}(\omega_{lk} +\omega_a)} - \frac{g_a^2 }{D_{ma}(\omega_{lk} +\omega_a)}-\frac{g_a^2}{E_lD_{na}} + \frac{g_a^2}{E_lD_{ma}}\right] 
\nonumber\\
&+ &A_{nl}A_{l1}A_{km}A_{1k}
\frac{1}{\omega_{nm}}\frac{g_b^2}{D_{kb}}\left[\frac{g_a^2}{D_{la}(\omega_{mk} +\omega_a)}-  \frac{g_a^2}{D_{la}(\omega_{nk} +\omega_a)}\right]
\nonumber\\
&+&A_{nl}A_{l1}A_{km}A_{1k}
\frac{1}{\omega_{nm}}\frac{g_b^2}{E_{m} + \omega_a + \omega_b}\left[\frac{g_a^2 }{D_{la}D_{ka}} -
\frac{g_a^2 }{D_{la}(\omega_{mk} + \omega_a)}\right]\nonumber\\
&-&A_{nl}A_{l1}A_{km}A_{1k}
\frac{1}{\omega_{nm}}\frac{g_b^2}{E_{n} + \omega_a + \omega_b}\left[\frac{g_a^2}{D_{la}D_{ka}}- \frac{g_a^2 }{D_{la}(\omega_{nk} + \omega_a)} \right] 
\nonumber\\
&+&(A_{nl}A_{lk}A_{1m}A_{k1}
+A_{n1}A_{kl}A_{lm}A_{1k})
\frac{1}{\omega_{nm}}\frac{g_b^2}{E_{l} + \omega_a + \omega_b}  \left(\frac{g_a^2 }{D_{na}D_{ka}} - \frac{g_a^2 }{D_{ma}D_{ka}}\right)
\nonumber\\
&-&(A_{nl}A_{lk}A_{1m}A_{k1}
+A_{n1}A_{kl}A_{lm}A_{1k})
\frac{1}{\omega_{nm}}\frac{g_b^2}{E_{l} + \omega_a + \omega_b}\left[\frac{g_a^2 }{D_{na}(\omega_{lk} + \omega_a)} - \frac{g_a^2 }{D_{ma}(\omega_{lk} + \omega_a)}\right],
\end{eqnarray}
\end{widetext}
with summations over all the indices (excluding $n,m$) still implicit. We then map from the discrete frequencies to a continuous distribution using Eqs.~\eqref{Eq:TiPTSD_rel1} and~\eqref{Eq:TiPTSD_rel2}
and the general mapping \begin{equation}
     \sum_{a} g_a^2 f(\omega_a) \mapsto \frac{1}{\pi} \int_{-\infty}^\infty d\omega\tilde{\mathcal{J}}_{\omega} f(\omega),
\end{equation}
with $f(\omega_a)$ a function describing the dependence of $\omega_a$ within each term. This gives Eq.~79. 
Evaluating the other coherence terms in Eq.~\eqref{Eqn:rho_4_allTerms}, $\tilde \rho_{S,1m}^{4}$ with $m = 2,\ldots, N$, we find

\begin{widetext}
    \begin{eqnarray}
        \tilde \rho_{S,1m}^{4} &=&A_{1l}A_{l1}A_{km}A_{1k}\frac{g_a^2 }{(\omega_{1k} +\omega_a)D_{la}}\left[\frac{g_b^2}{E_m D_{kb}}-\frac{g_b^2}{E_m(\omega_a + \omega_b)} \right]
 \notag   \\&-& A_{1l}A_{l1}A_{km}A_{1k}\frac{g_a^2 }{(\omega_{mk} +\omega_a)D_{la}}  \left[\frac{g_b^2}{E_m D_{kb}}-\frac{g_b^2}{E_m(E_{m} + \omega_a + \omega_b)} \right] \notag\\
&-& A_{pm}A_{lp}A_{kl}A_{1k}\frac{g_a^2 }{(\omega_{lk} +\omega_a)D_{pa}}\left[\frac{g_b^2}{E_m D_{kb}}-\frac{g_b^2}{E_m(E_{l} + \omega_a + \omega_b)}\right] 
 \notag\\
&+&\left(A_{1l}A_{lk}A_{1m}A_{k1}+A_{11}A_{kl}A_{lm}A_{1k}\right)\frac{g_a^2 }{(\omega_{lk} +\omega_a)D_{ma}}
\left[\frac{g_b^2}{E_{m}D_{kb}}-\frac{g_b^2}{E_{m}(E_{l} + \omega_a + \omega_b)}\right]
 \notag\\
&-&\left(A_{1l}A_{lk}A_{1m}A_{k1}+A_{11}A_{kl}A_{lm}A_{1k}\right)\frac{g_a^2 }{(\omega_{lk} +\omega_a)D_{1a}}
\left[\frac{g_b^2}{E_{m}D_{kb}}-\frac{g_b^2}{E_{m}(E_{l} + \omega_a + \omega_b)}\right] 
 \notag\\
&+&\left(A_{1l}A_{lk}A_{1m}A_{k1}+A_{11}A_{kl}A_{lm}A_{1k}\right)
\frac{g_b^2}{E_{m}(E_{l} + \omega_a + \omega_b)}   \left(\frac{g_a^2 }{D_{ma}D_{ka}} - \frac{g_a^2 }{D_{1a}D_{ka}} \right)  \notag\\&+&A_{1l}A_{l1}A_{km}A_{1k}\frac{g_a^2 }{D_{ka}D_{la}}\left[\frac{g_b^2}{E_m(\omega_a + \omega_b)}  -\frac{g_b^2}{E_m(E_{m} + \omega_a + \omega_b)} 
    \right] \notag\\
&-&A_{pm}A_{lp}A_{kl}A_{1k}\frac{g_a^2 }{D_{ka}D_{pa}}\left[\frac{g_b^2}{E_m(E_{l} + \omega_a + \omega_b)}\right]-  \frac{3}{2}A_{1l}A_{lm}|A_{k1}|^2\frac{g_a^2}{D_{ka}^2 }\frac{g_b^2}{E_m D_{lb}}
 \notag\\ &-&A_{11}A_{1m}|A_{k1}|^2\frac{ g_a^2}{D_{ka}^2 }\left(\frac{g_b^2}{E_m D_{1b}} - \frac{g_b^2}{E_m D_{mb}} \right)
+|A_{ml}|^2|A_{k1}|^2\frac{g_a^2 }{(\omega_{ml} - \omega_a)^2}\frac{ g_b^2}{E_m D_{kb}} \notag\\&-& A_{11}A_{1m}|A_{k1}|^2\frac{g_a^2 }{D_{1a}^2}\frac{g_b^2}{E_m D_{kb}}
-(A_{1m}A_{mk}A_{1l}A_{k1}
+A_{11}A_{km}A_{ml}A_{1k}
) \frac{g_a^2 }{D_{1a}D_{la}}\frac{ g_b^2}{E_{m}D_{kb}} 
 \notag\\
&-&\left(A_{km}A_{pk}A_{lp}A_{1l}\frac{1}{E_p }+ A_{1l}A_{lm}|A_{k1}|^2\frac{1}{E_m}\right)\frac{g_b^2}{E_m D_{lb}}\frac{ g_a^2}{D_{ka}}+ A_{11}A_{1m}|A_{k1}|^2\frac{g_a^2 }{D_{ma}^2}  \frac{g_b^2}{E_m D_{kb}},
    \end{eqnarray}
where the summations over open ystem indices are still implicit. 
\end{widetext}

\input{resub2.bbl}
\end{document}

%% file: resub2.bbl
%

%% file: Resub2.bbl
\begin{thebibliography}{98}%
\makeatletter
\providecommand \@ifxundefined [1]{%
 \@ifx{#1\undefined}
}%
\providecommand \@ifnum [1]{%
 \ifnum #1\expandafter \@firstoftwo
 \else \expandafter \@secondoftwo
 \fi
}%
\providecommand \@ifx [1]{%
 \ifx #1\expandafter \@firstoftwo
 \else \expandafter \@secondoftwo
 \fi
}%
\providecommand \natexlab [1]{#1}%
\providecommand \enquote  [1]{``#1''}%
\providecommand \bibnamefont  [1]{#1}%
\providecommand \bibfnamefont [1]{#1}%
\providecommand \citenamefont [1]{#1}%
\providecommand \href@noop [0]{\@secondoftwo}%
\providecommand \href [0]{\begingroup \@sanitize@url \@href}%
\providecommand \@href[1]{\@@startlink{#1}\@@href}%
\providecommand \@@href[1]{\endgroup#1\@@endlink}%
\providecommand \@sanitize@url [0]{\catcode `\\12\catcode `\$12\catcode `\&12\catcode `\#12\catcode `\^12\catcode `\_12\catcode `\%12\relax}%
\providecommand \@@startlink[1]{}%
\providecommand \@@endlink[0]{}%
\providecommand \url  [0]{\begingroup\@sanitize@url \@url }%
\providecommand \@url [1]{\endgroup\@href {#1}{\urlprefix }}%
\providecommand \urlprefix  [0]{URL }%
\providecommand \Eprint [0]{\href }%
\providecommand \doibase [0]{https://doi.org/}%
\providecommand \selectlanguage [0]{\@gobble}%
\providecommand \bibinfo  [0]{\@secondoftwo}%
\providecommand \bibfield  [0]{\@secondoftwo}%
\providecommand \translation [1]{[#1]}%
\providecommand \BibitemOpen [0]{}%
\providecommand \bibitemStop [0]{}%
\providecommand \bibitemNoStop [0]{.\EOS\space}%
\providecommand \EOS [0]{\spacefactor3000\relax}%
\providecommand \BibitemShut  [1]{\csname bibitem#1\endcsname}%
\let\auto@bib@innerbib\@empty
\bibitem [{\citenamefont {Fröhlich}\ and\ \citenamefont {Schubnel}(2016)}]{frohlich2016preparation}%
  \BibitemOpen
  \bibfield  {author} {\bibinfo {author} {\bibfnamefont {J.}~\bibnamefont {Fröhlich}}\ and\ \bibinfo {author} {\bibfnamefont {B.}~\bibnamefont {Schubnel}},\ }\bibfield  {title} {\bibinfo {title} {{The preparation of states in quantum mechanics}},\ }\href {https://doi.org/10.1063/1.4940696} {\bibfield  {journal} {\bibinfo  {journal} {Journal of Mathematical Physics}\ }\textbf {\bibinfo {volume} {57}},\ \bibinfo {pages} {042101} (\bibinfo {year} {2016})}\BibitemShut {NoStop}%
\bibitem [{\citenamefont {Zurek}(2009)}]{zurek2009quantum}%
  \BibitemOpen
  \bibfield  {author} {\bibinfo {author} {\bibfnamefont {W.~H.}\ \bibnamefont {Zurek}},\ }\bibfield  {title} {\bibinfo {title} {Quantum darwinism},\ }\href@noop {} {\bibfield  {journal} {\bibinfo  {journal} {Nature physics}\ }\textbf {\bibinfo {volume} {5}},\ \bibinfo {pages} {181} (\bibinfo {year} {2009})}\BibitemShut {NoStop}%
\bibitem [{\citenamefont {Shandera}\ \emph {et~al.}(2018)\citenamefont {Shandera}, \citenamefont {Agarwal},\ and\ \citenamefont {Kamal}}]{shandera2018open}%
  \BibitemOpen
  \bibfield  {author} {\bibinfo {author} {\bibfnamefont {S.}~\bibnamefont {Shandera}}, \bibinfo {author} {\bibfnamefont {N.}~\bibnamefont {Agarwal}},\ and\ \bibinfo {author} {\bibfnamefont {A.}~\bibnamefont {Kamal}},\ }\bibfield  {title} {\bibinfo {title} {Open quantum cosmological system},\ }\href@noop {} {\bibfield  {journal} {\bibinfo  {journal} {Physical Review D}\ }\textbf {\bibinfo {volume} {98}},\ \bibinfo {pages} {083535} (\bibinfo {year} {2018})}\BibitemShut {NoStop}%
\bibitem [{\citenamefont {Engel}\ \emph {et~al.}(2007)\citenamefont {Engel}, \citenamefont {Calhoun}, \citenamefont {Read}, \citenamefont {Ahn}, \citenamefont {Man{\v{c}}al}, \citenamefont {Cheng}, \citenamefont {Blankenship},\ and\ \citenamefont {Fleming}}]{engel2007evidence}%
  \BibitemOpen
  \bibfield  {author} {\bibinfo {author} {\bibfnamefont {G.~S.}\ \bibnamefont {Engel}}, \bibinfo {author} {\bibfnamefont {T.~R.}\ \bibnamefont {Calhoun}}, \bibinfo {author} {\bibfnamefont {E.~L.}\ \bibnamefont {Read}}, \bibinfo {author} {\bibfnamefont {T.-K.}\ \bibnamefont {Ahn}}, \bibinfo {author} {\bibfnamefont {T.}~\bibnamefont {Man{\v{c}}al}}, \bibinfo {author} {\bibfnamefont {Y.-C.}\ \bibnamefont {Cheng}}, \bibinfo {author} {\bibfnamefont {R.~E.}\ \bibnamefont {Blankenship}},\ and\ \bibinfo {author} {\bibfnamefont {G.~R.}\ \bibnamefont {Fleming}},\ }\bibfield  {title} {\bibinfo {title} {Evidence for wavelike energy transfer through quantum coherence in photosynthetic systems},\ }\href@noop {} {\bibfield  {journal} {\bibinfo  {journal} {Nature}\ }\textbf {\bibinfo {volume} {446}},\ \bibinfo {pages} {782} (\bibinfo {year} {2007})}\BibitemShut {NoStop}%
\bibitem [{\citenamefont {Collini}\ and\ \citenamefont {Scholes}(2009)}]{collini2009coherent}%
  \BibitemOpen
  \bibfield  {author} {\bibinfo {author} {\bibfnamefont {E.}~\bibnamefont {Collini}}\ and\ \bibinfo {author} {\bibfnamefont {G.~D.}\ \bibnamefont {Scholes}},\ }\bibfield  {title} {\bibinfo {title} {Coherent intrachain energy migration in a conjugated polymer at room temperature},\ }\href@noop {} {\bibfield  {journal} {\bibinfo  {journal} {science}\ }\textbf {\bibinfo {volume} {323}},\ \bibinfo {pages} {369} (\bibinfo {year} {2009})}\BibitemShut {NoStop}%
\bibitem [{\citenamefont {Carleo}\ \emph {et~al.}(2012)\citenamefont {Carleo}, \citenamefont {Becca}, \citenamefont {Schir{\'o}},\ and\ \citenamefont {Fabrizio}}]{carleo2012localization}%
  \BibitemOpen
  \bibfield  {author} {\bibinfo {author} {\bibfnamefont {G.}~\bibnamefont {Carleo}}, \bibinfo {author} {\bibfnamefont {F.}~\bibnamefont {Becca}}, \bibinfo {author} {\bibfnamefont {M.}~\bibnamefont {Schir{\'o}}},\ and\ \bibinfo {author} {\bibfnamefont {M.}~\bibnamefont {Fabrizio}},\ }\bibfield  {title} {\bibinfo {title} {Localization and glassy dynamics of many-body quantum systems},\ }\href@noop {} {\bibfield  {journal} {\bibinfo  {journal} {Scientific reports}\ }\textbf {\bibinfo {volume} {2}},\ \bibinfo {pages} {243} (\bibinfo {year} {2012})}\BibitemShut {NoStop}%
\bibitem [{\citenamefont {Feynman}\ and\ \citenamefont {Vernon}(1963)}]{FEYNMAN1963118}%
  \BibitemOpen
  \bibfield  {author} {\bibinfo {author} {\bibfnamefont {R.}~\bibnamefont {Feynman}}\ and\ \bibinfo {author} {\bibfnamefont {F.}~\bibnamefont {Vernon}},\ }\bibfield  {title} {\bibinfo {title} {The theory of a general quantum system interacting with a linear dissipative system},\ }\href {https://doi.org/https://doi.org/10.1016/0003-4916(63)90068-X} {\bibfield  {journal} {\bibinfo  {journal} {Annals of Physics}\ }\textbf {\bibinfo {volume} {24}},\ \bibinfo {pages} {118} (\bibinfo {year} {1963})}\BibitemShut {NoStop}%
\bibitem [{\citenamefont {Meyer}\ \emph {et~al.}(1990)\citenamefont {Meyer}, \citenamefont {Manthe},\ and\ \citenamefont {Cederbaum}}]{meyer1990multi}%
  \BibitemOpen
  \bibfield  {author} {\bibinfo {author} {\bibfnamefont {H.-D.}\ \bibnamefont {Meyer}}, \bibinfo {author} {\bibfnamefont {U.}~\bibnamefont {Manthe}},\ and\ \bibinfo {author} {\bibfnamefont {L.~S.}\ \bibnamefont {Cederbaum}},\ }\bibfield  {title} {\bibinfo {title} {The multi-configurational time-dependent hartree approach},\ }\href@noop {} {\bibfield  {journal} {\bibinfo  {journal} {Chemical Physics Letters}\ }\textbf {\bibinfo {volume} {165}},\ \bibinfo {pages} {73} (\bibinfo {year} {1990})}\BibitemShut {NoStop}%
\bibitem [{\citenamefont {Strathearn}\ \emph {et~al.}(2018)\citenamefont {Strathearn}, \citenamefont {Kirton}, \citenamefont {Kilda}, \citenamefont {Keeling},\ and\ \citenamefont {Lovett}}]{Strathearn2018}%
  \BibitemOpen
  \bibfield  {author} {\bibinfo {author} {\bibfnamefont {A.}~\bibnamefont {Strathearn}}, \bibinfo {author} {\bibfnamefont {P.}~\bibnamefont {Kirton}}, \bibinfo {author} {\bibfnamefont {D.}~\bibnamefont {Kilda}}, \bibinfo {author} {\bibfnamefont {J.}~\bibnamefont {Keeling}},\ and\ \bibinfo {author} {\bibfnamefont {B.~W.}\ \bibnamefont {Lovett}},\ }\bibfield  {title} {\bibinfo {title} {{Efficient non-Markovian quantum dynamics using time-evolving matrix product operators}},\ }\href {https://doi.org/10.1038/s41467-018-05617-3} {\bibfield  {journal} {\bibinfo  {journal} {Nat. Commun.}\ }\textbf {\bibinfo {volume} {9}},\ \bibinfo {pages} {3322} (\bibinfo {year} {2018})}\BibitemShut {NoStop}%
\bibitem [{\citenamefont {Makri}(2020)}]{makri2020small}%
  \BibitemOpen
  \bibfield  {author} {\bibinfo {author} {\bibfnamefont {N.}~\bibnamefont {Makri}},\ }\bibfield  {title} {\bibinfo {title} {Small matrix disentanglement of the path integral: overcoming the exponential tensor scaling with memory length},\ }\href@noop {} {\bibfield  {journal} {\bibinfo  {journal} {The Journal of Chemical Physics}\ }\textbf {\bibinfo {volume} {152}} (\bibinfo {year} {2020})}\BibitemShut {NoStop}%
\bibitem [{\citenamefont {Tanimura}\ and\ \citenamefont {Kubo}(1989)}]{tanimura1989time}%
  \BibitemOpen
  \bibfield  {author} {\bibinfo {author} {\bibfnamefont {Y.}~\bibnamefont {Tanimura}}\ and\ \bibinfo {author} {\bibfnamefont {R.}~\bibnamefont {Kubo}},\ }\bibfield  {title} {\bibinfo {title} {Time evolution of a quantum system in contact with a nearly gaussian-markoffian noise bath},\ }\href@noop {} {\bibfield  {journal} {\bibinfo  {journal} {Journal of the Physical Society of Japan}\ }\textbf {\bibinfo {volume} {58}},\ \bibinfo {pages} {101} (\bibinfo {year} {1989})}\BibitemShut {NoStop}%
\bibitem [{\citenamefont {Breuer}\ and\ \citenamefont {Petruccione}(2007)}]{BreuerHeinz-Peter1961-2007TToO}%
  \BibitemOpen
  \bibfield  {author} {\bibinfo {author} {\bibfnamefont {H.-P.}\ \bibnamefont {Breuer}}\ and\ \bibinfo {author} {\bibfnamefont {F.}~\bibnamefont {Petruccione}},\ }\href {https://oxford.universitypressscholarship.com/view/10.1093/acprof:oso/9780199213900.001.0001/acprof-9780199213900} {\emph {\bibinfo {title} {The theory of open quantum systems}}}\ (\bibinfo  {publisher} {Oxford University Press},\ \bibinfo {address} {Oxford},\ \bibinfo {year} {2007})\BibitemShut {NoStop}%
\bibitem [{\citenamefont {Bloch}(1946)}]{bloch1946nuclear}%
  \BibitemOpen
  \bibfield  {author} {\bibinfo {author} {\bibfnamefont {F.}~\bibnamefont {Bloch}},\ }\bibfield  {title} {\bibinfo {title} {Nuclear induction},\ }\href@noop {} {\bibfield  {journal} {\bibinfo  {journal} {Physical review}\ }\textbf {\bibinfo {volume} {70}},\ \bibinfo {pages} {460} (\bibinfo {year} {1946})}\BibitemShut {NoStop}%
\bibitem [{\citenamefont {REDFIELD}(1965)}]{REDFIELD19651}%
  \BibitemOpen
  \bibfield  {author} {\bibinfo {author} {\bibfnamefont {A.}~\bibnamefont {REDFIELD}},\ }\bibfield  {title} {\bibinfo {title} {The theory of relaxation processes* *this work was started while the author was at harvard university, and was then partially supported by joint services contract n5ori-76, project order i.},\ }in\ \href {https://doi.org/https://doi.org/10.1016/B978-1-4832-3114-3.50007-6} {\emph {\bibinfo {booktitle} {Advances in Magnetic Resonance}}},\ \bibinfo {series} {Advances in Magnetic and Optical Resonance}, Vol.~\bibinfo {volume} {1},\ \bibinfo {editor} {edited by\ \bibinfo {editor} {\bibfnamefont {J.~S.}\ \bibnamefont {Waugh}}}\ (\bibinfo  {publisher} {Academic Press},\ \bibinfo {year} {1965})\ pp.\ \bibinfo {pages} {1 -- 32}\BibitemShut {NoStop}%
\bibitem [{\citenamefont {Liu}\ \emph {et~al.}(2018)\citenamefont {Liu}, \citenamefont {Yan}, \citenamefont {Xu}, \citenamefont {Song},\ and\ \citenamefont {Shi}}]{liu2018exact}%
  \BibitemOpen
  \bibfield  {author} {\bibinfo {author} {\bibfnamefont {Y.-y.}\ \bibnamefont {Liu}}, \bibinfo {author} {\bibfnamefont {Y.-m.}\ \bibnamefont {Yan}}, \bibinfo {author} {\bibfnamefont {M.}~\bibnamefont {Xu}}, \bibinfo {author} {\bibfnamefont {K.}~\bibnamefont {Song}},\ and\ \bibinfo {author} {\bibfnamefont {Q.}~\bibnamefont {Shi}},\ }\bibfield  {title} {\bibinfo {title} {Exact generator and its high order expansions in time-convolutionless generalized master equation: Applications to spin-boson model and excitation energy transfer},\ }\href@noop {} {\bibfield  {journal} {\bibinfo  {journal} {Chinese Journal of Chemical Physics}\ }\textbf {\bibinfo {volume} {31}},\ \bibinfo {pages} {575} (\bibinfo {year} {2018})}\BibitemShut {NoStop}%
\bibitem [{\citenamefont {Becker}\ \emph {et~al.}(2022)\citenamefont {Becker}, \citenamefont {Schnell},\ and\ \citenamefont {Thingna}}]{becker2022canonically}%
  \BibitemOpen
  \bibfield  {author} {\bibinfo {author} {\bibfnamefont {T.}~\bibnamefont {Becker}}, \bibinfo {author} {\bibfnamefont {A.}~\bibnamefont {Schnell}},\ and\ \bibinfo {author} {\bibfnamefont {J.}~\bibnamefont {Thingna}},\ }\bibfield  {title} {\bibinfo {title} {Canonically consistent quantum master equation},\ }\href@noop {} {\bibfield  {journal} {\bibinfo  {journal} {Physical Review Letters}\ }\textbf {\bibinfo {volume} {129}},\ \bibinfo {pages} {200403} (\bibinfo {year} {2022})}\BibitemShut {NoStop}%
\bibitem [{\citenamefont {Fleming}\ and\ \citenamefont {Cummings}(2011)}]{Fleming}%
  \BibitemOpen
  \bibfield  {author} {\bibinfo {author} {\bibfnamefont {C.~H.}\ \bibnamefont {Fleming}}\ and\ \bibinfo {author} {\bibfnamefont {N.~I.}\ \bibnamefont {Cummings}},\ }\bibfield  {title} {\bibinfo {title} {Accuracy of perturbative master equations},\ }\href {https://doi.org/10.1103/PhysRevE.83.031117} {\bibfield  {journal} {\bibinfo  {journal} {Phys. Rev. E}\ }\textbf {\bibinfo {volume} {83}},\ \bibinfo {pages} {031117} (\bibinfo {year} {2011})}\BibitemShut {NoStop}%
\bibitem [{\citenamefont {Tupkary}\ \emph {et~al.}(2022)\citenamefont {Tupkary}, \citenamefont {Dhar}, \citenamefont {Kulkarni},\ and\ \citenamefont {Purkayastha}}]{tupkary2021fundamental}%
  \BibitemOpen
  \bibfield  {author} {\bibinfo {author} {\bibfnamefont {D.}~\bibnamefont {Tupkary}}, \bibinfo {author} {\bibfnamefont {A.}~\bibnamefont {Dhar}}, \bibinfo {author} {\bibfnamefont {M.}~\bibnamefont {Kulkarni}},\ and\ \bibinfo {author} {\bibfnamefont {A.}~\bibnamefont {Purkayastha}},\ }\bibfield  {title} {\bibinfo {title} {Fundamental limitations in lindblad descriptions of systems weakly coupled to baths},\ }\href {https://doi.org/10.1103/PhysRevA.105.032208} {\bibfield  {journal} {\bibinfo  {journal} {Phys. Rev. A}\ }\textbf {\bibinfo {volume} {105}},\ \bibinfo {pages} {032208} (\bibinfo {year} {2022})}\BibitemShut {NoStop}%
\bibitem [{\citenamefont {Davies}(1974)}]{davies1974}%
  \BibitemOpen
  \bibfield  {author} {\bibinfo {author} {\bibfnamefont {E.~B.}\ \bibnamefont {Davies}},\ }\bibfield  {title} {\bibinfo {title} {Markovian master equations},\ }\href {https://projecteuclid.org:443/euclid.cmp/1103860160} {\bibfield  {journal} {\bibinfo  {journal} {Comm. Math. Phys.}\ }\textbf {\bibinfo {volume} {39}},\ \bibinfo {pages} {91} (\bibinfo {year} {1974})}\BibitemShut {NoStop}%
\bibitem [{\citenamefont {Schaller}\ and\ \citenamefont {Brandes}(2008)}]{Schaller}%
  \BibitemOpen
  \bibfield  {author} {\bibinfo {author} {\bibfnamefont {G.}~\bibnamefont {Schaller}}\ and\ \bibinfo {author} {\bibfnamefont {T.}~\bibnamefont {Brandes}},\ }\bibfield  {title} {\bibinfo {title} {Preservation of positivity by dynamical coarse graining},\ }\href {https://doi.org/10.1103/PhysRevA.78.022106} {\bibfield  {journal} {\bibinfo  {journal} {Phys. Rev. A}\ }\textbf {\bibinfo {volume} {78}},\ \bibinfo {pages} {022106} (\bibinfo {year} {2008})}\BibitemShut {NoStop}%
\bibitem [{\citenamefont {Benatti}\ \emph {et~al.}(2010)\citenamefont {Benatti}, \citenamefont {Floreanini},\ and\ \citenamefont {Marzolino}}]{Benatti}%
  \BibitemOpen
  \bibfield  {author} {\bibinfo {author} {\bibfnamefont {F.}~\bibnamefont {Benatti}}, \bibinfo {author} {\bibfnamefont {R.}~\bibnamefont {Floreanini}},\ and\ \bibinfo {author} {\bibfnamefont {U.}~\bibnamefont {Marzolino}},\ }\bibfield  {title} {\bibinfo {title} {Entangling two unequal atoms through a common bath},\ }\href {https://doi.org/10.1103/PhysRevA.81.012105} {\bibfield  {journal} {\bibinfo  {journal} {Phys. Rev. A}\ }\textbf {\bibinfo {volume} {81}},\ \bibinfo {pages} {012105} (\bibinfo {year} {2010})}\BibitemShut {NoStop}%
\bibitem [{\citenamefont {Benatti}\ \emph {et~al.}(2009)\citenamefont {Benatti}, \citenamefont {Floreanini},\ and\ \citenamefont {Marzolino}}]{Benatti2}%
  \BibitemOpen
  \bibfield  {author} {\bibinfo {author} {\bibfnamefont {F.}~\bibnamefont {Benatti}}, \bibinfo {author} {\bibfnamefont {R.}~\bibnamefont {Floreanini}},\ and\ \bibinfo {author} {\bibfnamefont {U.}~\bibnamefont {Marzolino}},\ }\bibfield  {title} {\bibinfo {title} {Environment-induced entanglement in a refined weak-coupling limit},\ }\href {https://doi.org/10.1209/0295-5075/88/20011} {\bibfield  {journal} {\bibinfo  {journal} {{EPL} (Europhysics Letters)}\ }\textbf {\bibinfo {volume} {88}},\ \bibinfo {pages} {20011} (\bibinfo {year} {2009})}\BibitemShut {NoStop}%
\bibitem [{\citenamefont {Majenz}\ \emph {et~al.}(2013)\citenamefont {Majenz}, \citenamefont {Albash}, \citenamefont {Breuer},\ and\ \citenamefont {Lidar}}]{Majenz}%
  \BibitemOpen
  \bibfield  {author} {\bibinfo {author} {\bibfnamefont {C.}~\bibnamefont {Majenz}}, \bibinfo {author} {\bibfnamefont {T.}~\bibnamefont {Albash}}, \bibinfo {author} {\bibfnamefont {H.-P.}\ \bibnamefont {Breuer}},\ and\ \bibinfo {author} {\bibfnamefont {D.~A.}\ \bibnamefont {Lidar}},\ }\bibfield  {title} {\bibinfo {title} {Coarse graining can beat the rotating-wave approximation in quantum markovian master equations},\ }\href {https://doi.org/10.1103/PhysRevA.88.012103} {\bibfield  {journal} {\bibinfo  {journal} {Phys. Rev. A}\ }\textbf {\bibinfo {volume} {88}},\ \bibinfo {pages} {012103} (\bibinfo {year} {2013})}\BibitemShut {NoStop}%
\bibitem [{\citenamefont {Cresser}\ and\ \citenamefont {Facer}(2017)}]{Cresser}%
  \BibitemOpen
  \bibfield  {author} {\bibinfo {author} {\bibfnamefont {J.~D.}\ \bibnamefont {Cresser}}\ and\ \bibinfo {author} {\bibfnamefont {C.}~\bibnamefont {Facer}},\ }\href@noop {} {\bibinfo {title} {Coarse-graining in the derivation of markovian master equations and its significance in quantum thermodynamics}} (\bibinfo {year} {2017}),\ \Eprint {https://arxiv.org/abs/1710.09939} {arXiv:1710.09939 [quant-ph]} \BibitemShut {NoStop}%
\bibitem [{\citenamefont {Farina}\ and\ \citenamefont {Giovannetti}(2019)}]{Giovannetti}%
  \BibitemOpen
  \bibfield  {author} {\bibinfo {author} {\bibfnamefont {D.}~\bibnamefont {Farina}}\ and\ \bibinfo {author} {\bibfnamefont {V.}~\bibnamefont {Giovannetti}},\ }\bibfield  {title} {\bibinfo {title} {Open-quantum-system dynamics: Recovering positivity of the redfield equation via the partial secular approximation},\ }\href {https://doi.org/10.1103/PhysRevA.100.012107} {\bibfield  {journal} {\bibinfo  {journal} {Phys. Rev. A}\ }\textbf {\bibinfo {volume} {100}},\ \bibinfo {pages} {012107} (\bibinfo {year} {2019})}\BibitemShut {NoStop}%
\bibitem [{\citenamefont {Hartmann}\ and\ \citenamefont {Strunz}(2020)}]{Hartmann}%
  \BibitemOpen
  \bibfield  {author} {\bibinfo {author} {\bibfnamefont {R.}~\bibnamefont {Hartmann}}\ and\ \bibinfo {author} {\bibfnamefont {W.~T.}\ \bibnamefont {Strunz}},\ }\bibfield  {title} {\bibinfo {title} {Accuracy assessment of perturbative master equations: Embracing nonpositivity},\ }\href {https://doi.org/10.1103/PhysRevA.101.012103} {\bibfield  {journal} {\bibinfo  {journal} {Phys. Rev. A}\ }\textbf {\bibinfo {volume} {101}},\ \bibinfo {pages} {012103} (\bibinfo {year} {2020})}\BibitemShut {NoStop}%
\bibitem [{\citenamefont {Mozgunov}\ and\ \citenamefont {Lidar}(2020)}]{mozgunov}%
  \BibitemOpen
  \bibfield  {author} {\bibinfo {author} {\bibfnamefont {E.}~\bibnamefont {Mozgunov}}\ and\ \bibinfo {author} {\bibfnamefont {D.}~\bibnamefont {Lidar}},\ }\bibfield  {title} {\bibinfo {title} {Completely positive master equation for arbitrary driving and small level spacing},\ }\href {https://doi.org/https://doi.org/10.22331/q-2020-02-06-227} {\ \textbf {\bibinfo {volume} {4}},\ \bibinfo {pages} {227} (\bibinfo {year} {2020})},\ \Eprint {https://arxiv.org/abs/1908.01095} {1908.01095 [Quantum]} \BibitemShut {NoStop}%
\bibitem [{\citenamefont {Vogt}\ \emph {et~al.}(2013)\citenamefont {Vogt}, \citenamefont {Jeske},\ and\ \citenamefont {Cole}}]{Vogt}%
  \BibitemOpen
  \bibfield  {author} {\bibinfo {author} {\bibfnamefont {N.}~\bibnamefont {Vogt}}, \bibinfo {author} {\bibfnamefont {J.}~\bibnamefont {Jeske}},\ and\ \bibinfo {author} {\bibfnamefont {J.~H.}\ \bibnamefont {Cole}},\ }\bibfield  {title} {\bibinfo {title} {Stochastic bloch-redfield theory: Quantum jumps in a solid-state environment},\ }\href {https://doi.org/10.1103/PhysRevB.88.174514} {\bibfield  {journal} {\bibinfo  {journal} {Phys. Rev. B}\ }\textbf {\bibinfo {volume} {88}},\ \bibinfo {pages} {174514} (\bibinfo {year} {2013})}\BibitemShut {NoStop}%
\bibitem [{\citenamefont {Tscherbul}\ and\ \citenamefont {Brumer}(2015)}]{Tscherbul}%
  \BibitemOpen
  \bibfield  {author} {\bibinfo {author} {\bibfnamefont {T.~V.}\ \bibnamefont {Tscherbul}}\ and\ \bibinfo {author} {\bibfnamefont {P.}~\bibnamefont {Brumer}},\ }\bibfield  {title} {\bibinfo {title} {Partial secular bloch-redfield master equation for incoherent excitation of multilevel quantum systems},\ }\href {https://doi.org/10.1063/1.4908130} {\bibfield  {journal} {\bibinfo  {journal} {The Journal of Chemical Physics}\ }\textbf {\bibinfo {volume} {142}},\ \bibinfo {pages} {104107} (\bibinfo {year} {2015})},\ \Eprint {https://arxiv.org/abs/https://doi.org/10.1063/1.4908130} {https://doi.org/10.1063/1.4908130} \BibitemShut {NoStop}%
\bibitem [{\citenamefont {Trushechkin}(2021{\natexlab{a}})}]{Trushechkin}%
  \BibitemOpen
  \bibfield  {author} {\bibinfo {author} {\bibfnamefont {A.}~\bibnamefont {Trushechkin}},\ }\bibfield  {title} {\bibinfo {title} {Unified gorini-kossakowski-lindblad-sudarshan quantum master equation beyond the secular approximation},\ }\href {https://doi.org/10.1103/PhysRevA.103.062226} {\bibfield  {journal} {\bibinfo  {journal} {Phys. Rev. A}\ }\textbf {\bibinfo {volume} {103}},\ \bibinfo {pages} {062226} (\bibinfo {year} {2021}{\natexlab{a}})}\BibitemShut {NoStop}%
\bibitem [{\citenamefont {Gerry}\ and\ \citenamefont {Segal}(2023)}]{gerry2023full}%
  \BibitemOpen
  \bibfield  {author} {\bibinfo {author} {\bibfnamefont {M.}~\bibnamefont {Gerry}}\ and\ \bibinfo {author} {\bibfnamefont {D.}~\bibnamefont {Segal}},\ }\bibfield  {title} {\bibinfo {title} {Full counting statistics and coherences: Fluctuation symmetry in heat transport with the unified quantum master equation},\ }\href@noop {} {\bibfield  {journal} {\bibinfo  {journal} {Physical Review E}\ }\textbf {\bibinfo {volume} {107}},\ \bibinfo {pages} {054115} (\bibinfo {year} {2023})}\BibitemShut {NoStop}%
\bibitem [{\citenamefont {Palmieri}\ \emph {et~al.}(2009)\citenamefont {Palmieri}, \citenamefont {Abramavicius},\ and\ \citenamefont {Mukamel}}]{Benoit}%
  \BibitemOpen
  \bibfield  {author} {\bibinfo {author} {\bibfnamefont {B.}~\bibnamefont {Palmieri}}, \bibinfo {author} {\bibfnamefont {D.}~\bibnamefont {Abramavicius}},\ and\ \bibinfo {author} {\bibfnamefont {S.}~\bibnamefont {Mukamel}},\ }\bibfield  {title} {\bibinfo {title} {Lindblad equations for strongly coupled populations and coherences in photosynthetic complexes},\ }\href {https://doi.org/10.1063/1.3142485} {\bibfield  {journal} {\bibinfo  {journal} {The Journal of Chemical Physics}\ }\textbf {\bibinfo {volume} {130}},\ \bibinfo {pages} {204512} (\bibinfo {year} {2009})},\ \Eprint {https://arxiv.org/abs/https://doi.org/10.1063/1.3142485} {https://doi.org/10.1063/1.3142485} \BibitemShut {NoStop}%
\bibitem [{\citenamefont {Kir\ifmmode~\check{s}\else \v{s}\fi{}anskas}\ \emph {et~al.}(2018)\citenamefont {Kir\ifmmode~\check{s}\else \v{s}\fi{}anskas}, \citenamefont {Francki\'e},\ and\ \citenamefont {Wacker}}]{perlind}%
  \BibitemOpen
  \bibfield  {author} {\bibinfo {author} {\bibfnamefont {G.}~\bibnamefont {Kir\ifmmode~\check{s}\else \v{s}\fi{}anskas}}, \bibinfo {author} {\bibfnamefont {M.}~\bibnamefont {Francki\'e}},\ and\ \bibinfo {author} {\bibfnamefont {A.}~\bibnamefont {Wacker}},\ }\bibfield  {title} {\bibinfo {title} {Phenomenological position and energy resolving lindblad approach to quantum kinetics},\ }\href {https://doi.org/10.1103/PhysRevB.97.035432} {\bibfield  {journal} {\bibinfo  {journal} {Phys. Rev. B}\ }\textbf {\bibinfo {volume} {97}},\ \bibinfo {pages} {035432} (\bibinfo {year} {2018})}\BibitemShut {NoStop}%
\bibitem [{\citenamefont {Ptaszy\ifmmode~\acute{n}\else \'{n}\fi{}ski}\ and\ \citenamefont {Esposito}(2019)}]{Massimiliano}%
  \BibitemOpen
  \bibfield  {author} {\bibinfo {author} {\bibfnamefont {K.}~\bibnamefont {Ptaszy\ifmmode~\acute{n}\else \'{n}\fi{}ski}}\ and\ \bibinfo {author} {\bibfnamefont {M.}~\bibnamefont {Esposito}},\ }\bibfield  {title} {\bibinfo {title} {Thermodynamics of quantum information flows},\ }\href {https://doi.org/10.1103/PhysRevLett.122.150603} {\bibfield  {journal} {\bibinfo  {journal} {Phys. Rev. Lett.}\ }\textbf {\bibinfo {volume} {122}},\ \bibinfo {pages} {150603} (\bibinfo {year} {2019})}\BibitemShut {NoStop}%
\bibitem [{\citenamefont {Kleinherbers}\ \emph {et~al.}(2020)\citenamefont {Kleinherbers}, \citenamefont {Szpak}, \citenamefont {K\"onig},\ and\ \citenamefont {Sch\"utzhold}}]{Kleinherbers}%
  \BibitemOpen
  \bibfield  {author} {\bibinfo {author} {\bibfnamefont {E.}~\bibnamefont {Kleinherbers}}, \bibinfo {author} {\bibfnamefont {N.}~\bibnamefont {Szpak}}, \bibinfo {author} {\bibfnamefont {J.}~\bibnamefont {K\"onig}},\ and\ \bibinfo {author} {\bibfnamefont {R.}~\bibnamefont {Sch\"utzhold}},\ }\bibfield  {title} {\bibinfo {title} {Relaxation dynamics in a hubbard dimer coupled to fermionic baths: Phenomenological description and its microscopic foundation},\ }\href {https://doi.org/10.1103/PhysRevB.101.125131} {\bibfield  {journal} {\bibinfo  {journal} {Phys. Rev. B}\ }\textbf {\bibinfo {volume} {101}},\ \bibinfo {pages} {125131} (\bibinfo {year} {2020})}\BibitemShut {NoStop}%
\bibitem [{\citenamefont {Davidović}(2020)}]{Davidovic2020}%
  \BibitemOpen
  \bibfield  {author} {\bibinfo {author} {\bibfnamefont {D.}~\bibnamefont {Davidović}},\ }\bibfield  {title} {\bibinfo {title} {Completely positive, simple, and possibly highly accurate approximation of the redfield equation},\ }\href {https://doi.org/10.22331/q-2020-09-21-326} {\bibfield  {journal} {\bibinfo  {journal} {Quantum}\ }\textbf {\bibinfo {volume} {4}},\ \bibinfo {pages} {326} (\bibinfo {year} {2020})}\BibitemShut {NoStop}%
\bibitem [{\citenamefont {Nathan}\ and\ \citenamefont {Rudner}(2020)}]{Nathan}%
  \BibitemOpen
  \bibfield  {author} {\bibinfo {author} {\bibfnamefont {F.}~\bibnamefont {Nathan}}\ and\ \bibinfo {author} {\bibfnamefont {M.~S.}\ \bibnamefont {Rudner}},\ }\bibfield  {title} {\bibinfo {title} {Universal lindblad equation for open quantum systems},\ }\href {https://doi.org/10.1103/PhysRevB.102.115109} {\bibfield  {journal} {\bibinfo  {journal} {Phys. Rev. B}\ }\textbf {\bibinfo {volume} {102}},\ \bibinfo {pages} {115109} (\bibinfo {year} {2020})}\BibitemShut {NoStop}%
\bibitem [{\citenamefont {Merkli}(2020)}]{merkli2020quantum}%
  \BibitemOpen
  \bibfield  {author} {\bibinfo {author} {\bibfnamefont {M.}~\bibnamefont {Merkli}},\ }\bibfield  {title} {\bibinfo {title} {Quantum markovian master equations: Resonance theory shows validity for all time scales},\ }\href@noop {} {\bibfield  {journal} {\bibinfo  {journal} {Annals of Physics}\ }\textbf {\bibinfo {volume} {412}},\ \bibinfo {pages} {167996} (\bibinfo {year} {2020})}\BibitemShut {NoStop}%
\bibitem [{\citenamefont {Merkli}(2022{\natexlab{a}})}]{merkli2022dynamics}%
  \BibitemOpen
  \bibfield  {author} {\bibinfo {author} {\bibfnamefont {M.}~\bibnamefont {Merkli}},\ }\bibfield  {title} {\bibinfo {title} {Dynamics of open quantum systems i, oscillation and decay},\ }\href@noop {} {\bibfield  {journal} {\bibinfo  {journal} {Quantum}\ }\textbf {\bibinfo {volume} {6}},\ \bibinfo {pages} {615} (\bibinfo {year} {2022}{\natexlab{a}})}\BibitemShut {NoStop}%
\bibitem [{\citenamefont {Merkli}(2022{\natexlab{b}})}]{merkli2022dynamics2}%
  \BibitemOpen
  \bibfield  {author} {\bibinfo {author} {\bibfnamefont {M.}~\bibnamefont {Merkli}},\ }\bibfield  {title} {\bibinfo {title} {Dynamics of open quantum systems ii, markovian approximation},\ }\href@noop {} {\bibfield  {journal} {\bibinfo  {journal} {Quantum}\ }\textbf {\bibinfo {volume} {6}},\ \bibinfo {pages} {616} (\bibinfo {year} {2022}{\natexlab{b}})}\BibitemShut {NoStop}%
\bibitem [{\citenamefont {Albash}\ \emph {et~al.}(2012)\citenamefont {Albash}, \citenamefont {Boixo}, \citenamefont {Lidar},\ and\ \citenamefont {Zanardi}}]{Albash_2012}%
  \BibitemOpen
  \bibfield  {author} {\bibinfo {author} {\bibfnamefont {T.}~\bibnamefont {Albash}}, \bibinfo {author} {\bibfnamefont {S.}~\bibnamefont {Boixo}}, \bibinfo {author} {\bibfnamefont {D.~A.}\ \bibnamefont {Lidar}},\ and\ \bibinfo {author} {\bibfnamefont {P.}~\bibnamefont {Zanardi}},\ }\bibfield  {title} {\bibinfo {title} {Quantum adiabatic markovian master equations},\ }\href {https://doi.org/10.1088/1367-2630/14/12/123016} {\bibfield  {journal} {\bibinfo  {journal} {New Journal of Physics}\ }\textbf {\bibinfo {volume} {14}},\ \bibinfo {pages} {123016} (\bibinfo {year} {2012})}\BibitemShut {NoStop}%
\bibitem [{\citenamefont {Becker}\ \emph {et~al.}(2021)\citenamefont {Becker}, \citenamefont {Wu},\ and\ \citenamefont {Eckardt}}]{Becker}%
  \BibitemOpen
  \bibfield  {author} {\bibinfo {author} {\bibfnamefont {T.}~\bibnamefont {Becker}}, \bibinfo {author} {\bibfnamefont {L.-N.}\ \bibnamefont {Wu}},\ and\ \bibinfo {author} {\bibfnamefont {A.}~\bibnamefont {Eckardt}},\ }\bibfield  {title} {\bibinfo {title} {Lindbladian approximation beyond ultraweak coupling},\ }\href {https://doi.org/10.1103/PhysRevE.104.014110} {\bibfield  {journal} {\bibinfo  {journal} {Phys. Rev. E}\ }\textbf {\bibinfo {volume} {104}},\ \bibinfo {pages} {014110} (\bibinfo {year} {2021})}\BibitemShut {NoStop}%
\bibitem [{\citenamefont {Rivas}(2017)}]{Rivas}%
  \BibitemOpen
  \bibfield  {author} {\bibinfo {author} {\bibfnamefont {A.}~\bibnamefont {Rivas}},\ }\bibfield  {title} {\bibinfo {title} {Refined weak-coupling limit: Coherence, entanglement, and non-markovianity},\ }\href {https://doi.org/10.1103/PhysRevA.95.042104} {\bibfield  {journal} {\bibinfo  {journal} {Phys. Rev. A}\ }\textbf {\bibinfo {volume} {95}},\ \bibinfo {pages} {042104} (\bibinfo {year} {2017})}\BibitemShut {NoStop}%
\bibitem [{\citenamefont {Winczewski}\ \emph {et~al.}(2021)\citenamefont {Winczewski}, \citenamefont {Mandarino}, \citenamefont {Horodecki},\ and\ \citenamefont {Alicki}}]{winczewski}%
  \BibitemOpen
  \bibfield  {author} {\bibinfo {author} {\bibfnamefont {M.}~\bibnamefont {Winczewski}}, \bibinfo {author} {\bibfnamefont {A.}~\bibnamefont {Mandarino}}, \bibinfo {author} {\bibfnamefont {M.}~\bibnamefont {Horodecki}},\ and\ \bibinfo {author} {\bibfnamefont {R.}~\bibnamefont {Alicki}},\ }\href@noop {} {\bibinfo {title} {Bypassing the intermediate times dilemma for open quantum system}} (\bibinfo {year} {2021}),\ \Eprint {https://arxiv.org/abs/2106.05776} {arXiv:2106.05776 [quant-ph]} \BibitemShut {NoStop}%
\bibitem [{\citenamefont {D'Abbruzzo}\ \emph {et~al.}(2023)\citenamefont {D'Abbruzzo}, \citenamefont {Cavina},\ and\ \citenamefont {Giovannetti}}]{DAbbruzzo}%
  \BibitemOpen
  \bibfield  {author} {\bibinfo {author} {\bibfnamefont {A.}~\bibnamefont {D'Abbruzzo}}, \bibinfo {author} {\bibfnamefont {V.}~\bibnamefont {Cavina}},\ and\ \bibinfo {author} {\bibfnamefont {V.}~\bibnamefont {Giovannetti}},\ }\bibfield  {title} {\bibinfo {title} {A time-dependent regularization of the redfield equation},\ }\href@noop {} {\bibfield  {journal} {\bibinfo  {journal} {SciPost Physics}\ }\textbf {\bibinfo {volume} {15}},\ \bibinfo {pages} {117} (\bibinfo {year} {2023})}\BibitemShut {NoStop}%
\bibitem [{\citenamefont {Potts}\ \emph {et~al.}(2021)\citenamefont {Potts}, \citenamefont {Kalaee},\ and\ \citenamefont {Wacker}}]{potts2021thermodynamically}%
  \BibitemOpen
  \bibfield  {author} {\bibinfo {author} {\bibfnamefont {P.~P.}\ \bibnamefont {Potts}}, \bibinfo {author} {\bibfnamefont {A.~A.~S.}\ \bibnamefont {Kalaee}},\ and\ \bibinfo {author} {\bibfnamefont {A.}~\bibnamefont {Wacker}},\ }\bibfield  {title} {\bibinfo {title} {A thermodynamically consistent markovian master equation beyond the secular approximation},\ }\href@noop {} {\bibfield  {journal} {\bibinfo  {journal} {New Journal of Physics}\ }\textbf {\bibinfo {volume} {23}},\ \bibinfo {pages} {123013} (\bibinfo {year} {2021})}\BibitemShut {NoStop}%
\bibitem [{\citenamefont {Tupkary}\ \emph {et~al.}(2023)\citenamefont {Tupkary}, \citenamefont {Dhar}, \citenamefont {Kulkarni},\ and\ \citenamefont {Purkayastha}}]{tupkary2023searching}%
  \BibitemOpen
  \bibfield  {author} {\bibinfo {author} {\bibfnamefont {D.}~\bibnamefont {Tupkary}}, \bibinfo {author} {\bibfnamefont {A.}~\bibnamefont {Dhar}}, \bibinfo {author} {\bibfnamefont {M.}~\bibnamefont {Kulkarni}},\ and\ \bibinfo {author} {\bibfnamefont {A.}~\bibnamefont {Purkayastha}},\ }\bibfield  {title} {\bibinfo {title} {Searching for lindbladians obeying local conservation laws and showing thermalization},\ }\href@noop {} {\bibfield  {journal} {\bibinfo  {journal} {arXiv preprint arXiv:2301.02146}\ } (\bibinfo {year} {2023})}\BibitemShut {NoStop}%
\bibitem [{\citenamefont {Uchiyama}(2023)}]{uchiyama2023dynamics}%
  \BibitemOpen
  \bibfield  {author} {\bibinfo {author} {\bibfnamefont {C.}~\bibnamefont {Uchiyama}},\ }\bibfield  {title} {\bibinfo {title} {Dynamics of a quantum interacting system-extended global approach beyond the born-markov and secular approximation},\ }\href@noop {} {\bibfield  {journal} {\bibinfo  {journal} {arXiv preprint arXiv:2303.02926}\ } (\bibinfo {year} {2023})}\BibitemShut {NoStop}%
\bibitem [{\citenamefont {Winczewski}\ and\ \citenamefont {Alicki}(2021)}]{winczewski2021renormalization}%
  \BibitemOpen
  \bibfield  {author} {\bibinfo {author} {\bibfnamefont {M.}~\bibnamefont {Winczewski}}\ and\ \bibinfo {author} {\bibfnamefont {R.}~\bibnamefont {Alicki}},\ }\bibfield  {title} {\bibinfo {title} {Renormalization in the theory of open quantum systems via the self-consistency condition},\ }\href@noop {} {\bibfield  {journal} {\bibinfo  {journal} {arXiv preprint arXiv:2112.11962}\ } (\bibinfo {year} {2021})}\BibitemShut {NoStop}%
\bibitem [{\citenamefont {Tokuyama}\ and\ \citenamefont {Mori}(1976)}]{tokuyama1976statistical}%
  \BibitemOpen
  \bibfield  {author} {\bibinfo {author} {\bibfnamefont {M.}~\bibnamefont {Tokuyama}}\ and\ \bibinfo {author} {\bibfnamefont {H.}~\bibnamefont {Mori}},\ }\bibfield  {title} {\bibinfo {title} {Statistical-mechanical theory of the boltzmann equation and fluctuations in $\mu$ space},\ }\href@noop {} {\bibfield  {journal} {\bibinfo  {journal} {Progress of Theoretical Physics}\ }\textbf {\bibinfo {volume} {56}},\ \bibinfo {pages} {1073} (\bibinfo {year} {1976})}\BibitemShut {NoStop}%
\bibitem [{\citenamefont {Yoon}\ \emph {et~al.}(1975)\citenamefont {Yoon}, \citenamefont {Deutch},\ and\ \citenamefont {Freed}}]{Yoon}%
  \BibitemOpen
  \bibfield  {author} {\bibinfo {author} {\bibfnamefont {B.}~\bibnamefont {Yoon}}, \bibinfo {author} {\bibfnamefont {J.}~\bibnamefont {Deutch}},\ and\ \bibinfo {author} {\bibfnamefont {J.~H.}\ \bibnamefont {Freed}},\ }\bibfield  {title} {\bibinfo {title} {A comparison of generalized cumulant and projection operator methods in spin-relaxation theory},\ }\href@noop {} {\bibfield  {journal} {\bibinfo  {journal} {The Journal of Chemical Physics}\ }\textbf {\bibinfo {volume} {62}},\ \bibinfo {pages} {4687} (\bibinfo {year} {1975})}\BibitemShut {NoStop}%
\bibitem [{\citenamefont {Mukamel}\ \emph {et~al.}(1978)\citenamefont {Mukamel}, \citenamefont {Oppenheim},\ and\ \citenamefont {Ross}}]{Mukamel}%
  \BibitemOpen
  \bibfield  {author} {\bibinfo {author} {\bibfnamefont {S.}~\bibnamefont {Mukamel}}, \bibinfo {author} {\bibfnamefont {I.}~\bibnamefont {Oppenheim}},\ and\ \bibinfo {author} {\bibfnamefont {J.}~\bibnamefont {Ross}},\ }\bibfield  {title} {\bibinfo {title} {Statistical reduction for strongly driven simple quantum systems},\ }\href {https://doi.org/10.1103/PhysRevA.17.1988} {\bibfield  {journal} {\bibinfo  {journal} {Phys. Rev. A}\ }\textbf {\bibinfo {volume} {17}},\ \bibinfo {pages} {1988} (\bibinfo {year} {1978})}\BibitemShut {NoStop}%
\bibitem [{\citenamefont {Shibata}\ \emph {et~al.}(1977)\citenamefont {Shibata}, \citenamefont {Takahashi},\ and\ \citenamefont {Hashitsume}}]{shibata1977generalized}%
  \BibitemOpen
  \bibfield  {author} {\bibinfo {author} {\bibfnamefont {F.}~\bibnamefont {Shibata}}, \bibinfo {author} {\bibfnamefont {Y.}~\bibnamefont {Takahashi}},\ and\ \bibinfo {author} {\bibfnamefont {N.}~\bibnamefont {Hashitsume}},\ }\bibfield  {title} {\bibinfo {title} {A generalized stochastic liouville equation. non-markovian versus memoryless master equations},\ }\href@noop {} {\bibfield  {journal} {\bibinfo  {journal} {Journal of Statistical Physics}\ }\textbf {\bibinfo {volume} {17}},\ \bibinfo {pages} {171} (\bibinfo {year} {1977})}\BibitemShut {NoStop}%
\bibitem [{\citenamefont {Shibata}\ and\ \citenamefont {Arimitsu}(1980)}]{shibata1980expansion}%
  \BibitemOpen
  \bibfield  {author} {\bibinfo {author} {\bibfnamefont {F.}~\bibnamefont {Shibata}}\ and\ \bibinfo {author} {\bibfnamefont {T.}~\bibnamefont {Arimitsu}},\ }\bibfield  {title} {\bibinfo {title} {Expansion formulas in nonequilibrium statistical mechanics},\ }\href@noop {} {\bibfield  {journal} {\bibinfo  {journal} {Journal of the Physical Society of Japan}\ }\textbf {\bibinfo {volume} {49}},\ \bibinfo {pages} {891} (\bibinfo {year} {1980})}\BibitemShut {NoStop}%
\bibitem [{\citenamefont {Laird}\ \emph {et~al.}(1991)\citenamefont {Laird}, \citenamefont {Budimir},\ and\ \citenamefont {Skinner}}]{laird1991quantum}%
  \BibitemOpen
  \bibfield  {author} {\bibinfo {author} {\bibfnamefont {B.~B.}\ \bibnamefont {Laird}}, \bibinfo {author} {\bibfnamefont {J.}~\bibnamefont {Budimir}},\ and\ \bibinfo {author} {\bibfnamefont {J.~L.}\ \bibnamefont {Skinner}},\ }\bibfield  {title} {\bibinfo {title} {Quantum-mechanical derivation of the bloch equations: Beyond the weak-coupling limit},\ }\href@noop {} {\bibfield  {journal} {\bibinfo  {journal} {The Journal of chemical physics}\ }\textbf {\bibinfo {volume} {94}},\ \bibinfo {pages} {4391} (\bibinfo {year} {1991})}\BibitemShut {NoStop}%
\bibitem [{\citenamefont {Reichman}\ \emph {et~al.}(1997)\citenamefont {Reichman}, \citenamefont {Brown},\ and\ \citenamefont {Neu}}]{Reichman}%
  \BibitemOpen
  \bibfield  {author} {\bibinfo {author} {\bibfnamefont {D.~R.}\ \bibnamefont {Reichman}}, \bibinfo {author} {\bibfnamefont {F.~L.~H.}\ \bibnamefont {Brown}},\ and\ \bibinfo {author} {\bibfnamefont {P.}~\bibnamefont {Neu}},\ }\bibfield  {title} {\bibinfo {title} {Cumulant expansions and the spin-boson problem},\ }\href {https://doi.org/10.1103/PhysRevE.55.2328} {\bibfield  {journal} {\bibinfo  {journal} {Phys. Rev. E}\ }\textbf {\bibinfo {volume} {55}},\ \bibinfo {pages} {2328} (\bibinfo {year} {1997})}\BibitemShut {NoStop}%
\bibitem [{\citenamefont {Breuer}\ \emph {et~al.}(1999)\citenamefont {Breuer}, \citenamefont {Kappler},\ and\ \citenamefont {Petruccione}}]{Breuer_1999}%
  \BibitemOpen
  \bibfield  {author} {\bibinfo {author} {\bibfnamefont {H.-P.}\ \bibnamefont {Breuer}}, \bibinfo {author} {\bibfnamefont {B.}~\bibnamefont {Kappler}},\ and\ \bibinfo {author} {\bibfnamefont {F.}~\bibnamefont {Petruccione}},\ }\bibfield  {title} {\bibinfo {title} {Stochastic wave-function method for non-markovian quantum master equations},\ }\href {https://doi.org/10.1103/physreva.59.1633} {\bibfield  {journal} {\bibinfo  {journal} {Physical Review A}\ }\textbf {\bibinfo {volume} {59}},\ \bibinfo {pages} {1633–1643} (\bibinfo {year} {1999})}\BibitemShut {NoStop}%
\bibitem [{\citenamefont {Jang}\ \emph {et~al.}(2002)\citenamefont {Jang}, \citenamefont {Cao},\ and\ \citenamefont {Silbey}}]{JangS}%
  \BibitemOpen
  \bibfield  {author} {\bibinfo {author} {\bibfnamefont {S.}~\bibnamefont {Jang}}, \bibinfo {author} {\bibfnamefont {J.}~\bibnamefont {Cao}},\ and\ \bibinfo {author} {\bibfnamefont {R.~J.}\ \bibnamefont {Silbey}},\ }\bibfield  {title} {\bibinfo {title} {Fourth-order quantum master equation and its markovian bath limit},\ }\href {https://doi.org/10.1063/1.1445105} {\bibfield  {journal} {\bibinfo  {journal} {The Journal of Chemical Physics}\ }\textbf {\bibinfo {volume} {116}},\ \bibinfo {pages} {2705} (\bibinfo {year} {2002})},\ \Eprint {https://arxiv.org/abs/https://doi.org/10.1063/1.1445105} {https://doi.org/10.1063/1.1445105} \BibitemShut {NoStop}%
\bibitem [{\citenamefont {Trushechkin}(2019)}]{trushechkin2019higher}%
  \BibitemOpen
  \bibfield  {author} {\bibinfo {author} {\bibfnamefont {A.}~\bibnamefont {Trushechkin}},\ }\bibfield  {title} {\bibinfo {title} {Higher-order corrections to the redfield equation with respect to the system-bath coupling based on the hierarchical equations of motion},\ }\href@noop {} {\bibfield  {journal} {\bibinfo  {journal} {Lobachevskii Journal of Mathematics}\ }\textbf {\bibinfo {volume} {40}},\ \bibinfo {pages} {1606} (\bibinfo {year} {2019})}\BibitemShut {NoStop}%
\bibitem [{\citenamefont {Trushechkin}(2021{\natexlab{b}})}]{trushechkin2021derivation}%
  \BibitemOpen
  \bibfield  {author} {\bibinfo {author} {\bibfnamefont {A.~S.}\ \bibnamefont {Trushechkin}},\ }\bibfield  {title} {\bibinfo {title} {Derivation of the redfield quantum master equation and corrections to it by the bogoliubov method},\ }\href@noop {} {\bibfield  {journal} {\bibinfo  {journal} {Proceedings of the Steklov Institute of Mathematics}\ }\textbf {\bibinfo {volume} {313}},\ \bibinfo {pages} {246} (\bibinfo {year} {2021}{\natexlab{b}})}\BibitemShut {NoStop}%
\bibitem [{\citenamefont {Nestmann}\ and\ \citenamefont {Timm}(2019)}]{nestmann2019timeconvolutionless}%
  \BibitemOpen
  \bibfield  {author} {\bibinfo {author} {\bibfnamefont {K.}~\bibnamefont {Nestmann}}\ and\ \bibinfo {author} {\bibfnamefont {C.}~\bibnamefont {Timm}},\ }\href@noop {} {\bibinfo {title} {Time-convolutionless master equation: Perturbative expansions to arbitrary order and application to quantum dots}} (\bibinfo {year} {2019}),\ \Eprint {https://arxiv.org/abs/1903.05132} {arXiv:1903.05132 [cond-mat.mes-hall]} \BibitemShut {NoStop}%
\bibitem [{\citenamefont {Karasev}\ and\ \citenamefont {Teretenkov}(2023)}]{karasev2023timeconvolutionless}%
  \BibitemOpen
  \bibfield  {author} {\bibinfo {author} {\bibfnamefont {A.~Y.}\ \bibnamefont {Karasev}}\ and\ \bibinfo {author} {\bibfnamefont {A.~E.}\ \bibnamefont {Teretenkov}},\ }\href@noop {} {\bibinfo {title} {Time-convolutionless master equations for composite open quantum systems}} (\bibinfo {year} {2023}),\ \Eprint {https://arxiv.org/abs/2304.08627} {arXiv:2304.08627 [quant-ph]} \BibitemShut {NoStop}%
\bibitem [{\citenamefont {De~Roeck}\ and\ \citenamefont {Kupiainen}(2013)}]{de2013approach}%
  \BibitemOpen
  \bibfield  {author} {\bibinfo {author} {\bibfnamefont {W.}~\bibnamefont {De~Roeck}}\ and\ \bibinfo {author} {\bibfnamefont {A.}~\bibnamefont {Kupiainen}},\ }\bibfield  {title} {\bibinfo {title} {Approach to ground state and time-independent photon bound for massless spin-boson models},\ }in\ \href@noop {} {\emph {\bibinfo {booktitle} {Annales Henri Poincar{\'e}}}},\ Vol.~\bibinfo {volume} {14}\ (\bibinfo {organization} {Springer},\ \bibinfo {year} {2013})\ pp.\ \bibinfo {pages} {253--311}\BibitemShut {NoStop}%
\bibitem [{\citenamefont {Derezi{\'n}ski}\ and\ \citenamefont {G{\'e}rard}(2004)}]{derezinski2004scattering}%
  \BibitemOpen
  \bibfield  {author} {\bibinfo {author} {\bibfnamefont {J.}~\bibnamefont {Derezi{\'n}ski}}\ and\ \bibinfo {author} {\bibfnamefont {C.}~\bibnamefont {G{\'e}rard}},\ }\bibfield  {title} {\bibinfo {title} {Scattering theory of infrared divergent pauli-fierz hamiltonians},\ }in\ \href@noop {} {\emph {\bibinfo {booktitle} {Annales Henri Poincar{\'e}}}},\ Vol.~\bibinfo {volume} {5}\ (\bibinfo {organization} {Springer},\ \bibinfo {year} {2004})\ pp.\ \bibinfo {pages} {523--577}\BibitemShut {NoStop}%
\bibitem [{\citenamefont {Araki}\ and\ \citenamefont {Woods}(1963)}]{araki1963representations}%
  \BibitemOpen
  \bibfield  {author} {\bibinfo {author} {\bibfnamefont {H.}~\bibnamefont {Araki}}\ and\ \bibinfo {author} {\bibfnamefont {E.}~\bibnamefont {Woods}},\ }\bibfield  {title} {\bibinfo {title} {Representations of the canonical commutation relations describing a nonrelativistic infinite free bose gas},\ }\href@noop {} {\bibfield  {journal} {\bibinfo  {journal} {Journal of Mathematical Physics}\ }\textbf {\bibinfo {volume} {4}},\ \bibinfo {pages} {637} (\bibinfo {year} {1963})}\BibitemShut {NoStop}%
\bibitem [{\citenamefont {Merkli}(2006)}]{merkli2006ideal}%
  \BibitemOpen
  \bibfield  {author} {\bibinfo {author} {\bibfnamefont {M.}~\bibnamefont {Merkli}},\ }\bibfield  {title} {\bibinfo {title} {The ideal quantum gas},\ }in\ \href@noop {} {\emph {\bibinfo {booktitle} {Open Quantum Systems I: The Hamiltonian Approach}}}\ (\bibinfo  {publisher} {Springer},\ \bibinfo {year} {2006})\ pp.\ \bibinfo {pages} {183--233}\BibitemShut {NoStop}%
\bibitem [{\citenamefont {Spohn}(1989)}]{spohn1989ground}%
  \BibitemOpen
  \bibfield  {author} {\bibinfo {author} {\bibfnamefont {H.}~\bibnamefont {Spohn}},\ }\bibfield  {title} {\bibinfo {title} {Ground state (s) of the spin-boson hamiltonian},\ }\href@noop {} {\bibfield  {journal} {\bibinfo  {journal} {Communications in mathematical physics}\ }\textbf {\bibinfo {volume} {123}},\ \bibinfo {pages} {277} (\bibinfo {year} {1989})}\BibitemShut {NoStop}%
\bibitem [{\citenamefont {Strominger}(2018)}]{strominger2018lectures}%
  \BibitemOpen
  \bibfield  {author} {\bibinfo {author} {\bibfnamefont {A.}~\bibnamefont {Strominger}},\ }\href@noop {} {\bibinfo {title} {Lectures on the infrared structure of gravity and gauge theory}} (\bibinfo {year} {2018}),\ \Eprint {https://arxiv.org/abs/1703.05448} {arXiv:1703.05448 [hep-th]} \BibitemShut {NoStop}%
\bibitem [{\citenamefont {Jak{\v{s}}i{\'c}}\ and\ \citenamefont {Pillet}(1996)}]{jakvsic1996model}%
  \BibitemOpen
  \bibfield  {author} {\bibinfo {author} {\bibfnamefont {V.}~\bibnamefont {Jak{\v{s}}i{\'c}}}\ and\ \bibinfo {author} {\bibfnamefont {C.-A.}\ \bibnamefont {Pillet}},\ }\bibfield  {title} {\bibinfo {title} {On a model for quantum friction iii. ergodic properties of the spin-boson system},\ }\href@noop {} {\bibfield  {journal} {\bibinfo  {journal} {Communications in Mathematical Physics}\ }\textbf {\bibinfo {volume} {178}},\ \bibinfo {pages} {627} (\bibinfo {year} {1996})}\BibitemShut {NoStop}%
\bibitem [{\citenamefont {Bach}\ \emph {et~al.}(2000)\citenamefont {Bach}, \citenamefont {Fr{\"o}hlich},\ and\ \citenamefont {Sigal}}]{bach2000return}%
  \BibitemOpen
  \bibfield  {author} {\bibinfo {author} {\bibfnamefont {V.}~\bibnamefont {Bach}}, \bibinfo {author} {\bibfnamefont {J.}~\bibnamefont {Fr{\"o}hlich}},\ and\ \bibinfo {author} {\bibfnamefont {I.~M.}\ \bibnamefont {Sigal}},\ }\bibfield  {title} {\bibinfo {title} {Return to equilibrium},\ }\href@noop {} {\bibfield  {journal} {\bibinfo  {journal} {Journal of Mathematical Physics}\ }\textbf {\bibinfo {volume} {41}},\ \bibinfo {pages} {3985} (\bibinfo {year} {2000})}\BibitemShut {NoStop}%
\bibitem [{\citenamefont {Weiss}(2012)}]{ulrich}%
  \BibitemOpen
  \bibfield  {author} {\bibinfo {author} {\bibfnamefont {U.}~\bibnamefont {Weiss}},\ }\href {https://doi.org/10.1142/8334} {\emph {\bibinfo {title} {Quantum Dissipative Systems}}},\ \bibinfo {edition} {4th}\ ed.\ (\bibinfo  {publisher} {WORLD SCIENTIFIC},\ \bibinfo {year} {2012})\ \Eprint {https://arxiv.org/abs/https://www.worldscientific.com/doi/pdf/10.1142/8334} {https://www.worldscientific.com/doi/pdf/10.1142/8334} \BibitemShut {NoStop}%
\bibitem [{\citenamefont {Nemati}\ \emph {et~al.}(2022)\citenamefont {Nemati}, \citenamefont {Henkel},\ and\ \citenamefont {Anders}}]{nemati2022coupling}%
  \BibitemOpen
  \bibfield  {author} {\bibinfo {author} {\bibfnamefont {S.}~\bibnamefont {Nemati}}, \bibinfo {author} {\bibfnamefont {C.}~\bibnamefont {Henkel}},\ and\ \bibinfo {author} {\bibfnamefont {J.}~\bibnamefont {Anders}},\ }\bibfield  {title} {\bibinfo {title} {Coupling function from bath density of states},\ }\href@noop {} {\bibfield  {journal} {\bibinfo  {journal} {Europhysics Letters}\ }\textbf {\bibinfo {volume} {139}},\ \bibinfo {pages} {36002} (\bibinfo {year} {2022})}\BibitemShut {NoStop}%
\bibitem [{\citenamefont {Faupin}\ and\ \citenamefont {Sigal}(2014)}]{faupin2014rayleigh}%
  \BibitemOpen
  \bibfield  {author} {\bibinfo {author} {\bibfnamefont {J.}~\bibnamefont {Faupin}}\ and\ \bibinfo {author} {\bibfnamefont {I.~M.}\ \bibnamefont {Sigal}},\ }\bibfield  {title} {\bibinfo {title} {On rayleigh scattering in non-relativistic quantum electrodynamics},\ }\href@noop {} {\bibfield  {journal} {\bibinfo  {journal} {Communications in Mathematical Physics}\ }\textbf {\bibinfo {volume} {328}},\ \bibinfo {pages} {1199} (\bibinfo {year} {2014})}\BibitemShut {NoStop}%
\bibitem [{\citenamefont {De~Roeck}\ \emph {et~al.}(2015)\citenamefont {De~Roeck}, \citenamefont {Griesemer},\ and\ \citenamefont {Kupiainen}}]{de2015asymptotic}%
  \BibitemOpen
  \bibfield  {author} {\bibinfo {author} {\bibfnamefont {W.}~\bibnamefont {De~Roeck}}, \bibinfo {author} {\bibfnamefont {M.}~\bibnamefont {Griesemer}},\ and\ \bibinfo {author} {\bibfnamefont {A.}~\bibnamefont {Kupiainen}},\ }\bibfield  {title} {\bibinfo {title} {Asymptotic completeness for the massless spin-boson model},\ }\href@noop {} {\bibfield  {journal} {\bibinfo  {journal} {Advances in Mathematics}\ }\textbf {\bibinfo {volume} {268}},\ \bibinfo {pages} {62} (\bibinfo {year} {2015})}\BibitemShut {NoStop}%
\bibitem [{\citenamefont {Arai}\ and\ \citenamefont {Hirokawa}(1997)}]{arai1997existence}%
  \BibitemOpen
  \bibfield  {author} {\bibinfo {author} {\bibfnamefont {A.}~\bibnamefont {Arai}}\ and\ \bibinfo {author} {\bibfnamefont {M.}~\bibnamefont {Hirokawa}},\ }\bibfield  {title} {\bibinfo {title} {On the existence and uniqueness of ground states of a generalized spin-boson model},\ }\href@noop {} {\bibfield  {journal} {\bibinfo  {journal} {journal of functional analysis}\ }\textbf {\bibinfo {volume} {151}},\ \bibinfo {pages} {455} (\bibinfo {year} {1997})}\BibitemShut {NoStop}%
\bibitem [{\citenamefont {Bach}\ \emph {et~al.}(1999)\citenamefont {Bach}, \citenamefont {Fr{\"o}hlich},\ and\ \citenamefont {Sigal}}]{bach1999spectral}%
  \BibitemOpen
  \bibfield  {author} {\bibinfo {author} {\bibfnamefont {V.}~\bibnamefont {Bach}}, \bibinfo {author} {\bibfnamefont {J.}~\bibnamefont {Fr{\"o}hlich}},\ and\ \bibinfo {author} {\bibfnamefont {I.~M.}\ \bibnamefont {Sigal}},\ }\bibfield  {title} {\bibinfo {title} {Spectral analysis for systems of atoms and molecules coupled to the quantized radiation field},\ }\href@noop {} {\bibfield  {journal} {\bibinfo  {journal} {Communications in Mathematical Physics}\ }\textbf {\bibinfo {volume} {207}},\ \bibinfo {pages} {249} (\bibinfo {year} {1999})}\BibitemShut {NoStop}%
\bibitem [{\citenamefont {Gerard}(2000)}]{gerard2000existence}%
  \BibitemOpen
  \bibfield  {author} {\bibinfo {author} {\bibfnamefont {C.}~\bibnamefont {Gerard}},\ }\bibfield  {title} {\bibinfo {title} {On the existence of ground states for massless pauli-fierz hamiltonians},\ }\href@noop {} {\bibfield  {journal} {\bibinfo  {journal} {Annales Henri Poincar}\ }\textbf {\bibinfo {volume} {1}},\ \bibinfo {pages} {443} (\bibinfo {year} {2000})}\BibitemShut {NoStop}%
\bibitem [{\citenamefont {Griesemer}\ \emph {et~al.}(2001)\citenamefont {Griesemer}, \citenamefont {Lieb},\ and\ \citenamefont {Loss}}]{griesemer2001ground}%
  \BibitemOpen
  \bibfield  {author} {\bibinfo {author} {\bibfnamefont {M.}~\bibnamefont {Griesemer}}, \bibinfo {author} {\bibfnamefont {E.~H.}\ \bibnamefont {Lieb}},\ and\ \bibinfo {author} {\bibfnamefont {M.}~\bibnamefont {Loss}},\ }\bibfield  {title} {\bibinfo {title} {Ground states in non-relativistic quantum electrodynamics},\ }\href@noop {} {\bibfield  {journal} {\bibinfo  {journal} {Inventiones mathematicae}\ }\textbf {\bibinfo {volume} {145}},\ \bibinfo {pages} {557} (\bibinfo {year} {2001})}\BibitemShut {NoStop}%
\bibitem [{\citenamefont {Hasler}\ and\ \citenamefont {Herbst}(2011)}]{hasler2011ground}%
  \BibitemOpen
  \bibfield  {author} {\bibinfo {author} {\bibfnamefont {D.}~\bibnamefont {Hasler}}\ and\ \bibinfo {author} {\bibfnamefont {I.}~\bibnamefont {Herbst}},\ }\bibfield  {title} {\bibinfo {title} {Ground states in the spin boson model},\ }in\ \href@noop {} {\emph {\bibinfo {booktitle} {Annales Henri Poincar{\'e}}}},\ Vol.~\bibinfo {volume} {12}\ (\bibinfo {organization} {Springer},\ \bibinfo {year} {2011})\ pp.\ \bibinfo {pages} {621--677}\BibitemShut {NoStop}%
\bibitem [{\citenamefont {Abdesselam}(2011)}]{abdesselam2011ground}%
  \BibitemOpen
  \bibfield  {author} {\bibinfo {author} {\bibfnamefont {A.}~\bibnamefont {Abdesselam}},\ }\bibfield  {title} {\bibinfo {title} {The ground state energy of the massless spin-boson model},\ }in\ \href@noop {} {\emph {\bibinfo {booktitle} {Annales Henri Poincar{\'e}}}},\ Vol.~\bibinfo {volume} {12}\ (\bibinfo {organization} {Springer},\ \bibinfo {year} {2011})\ pp.\ \bibinfo {pages} {1321--1347}\BibitemShut {NoStop}%
\bibitem [{\citenamefont {Leggett}\ \emph {et~al.}(1987)\citenamefont {Leggett}, \citenamefont {Chakravarty}, \citenamefont {Dorsey}, \citenamefont {Fisher}, \citenamefont {Garg},\ and\ \citenamefont {Zwerger}}]{Leggett}%
  \BibitemOpen
  \bibfield  {author} {\bibinfo {author} {\bibfnamefont {A.~J.}\ \bibnamefont {Leggett}}, \bibinfo {author} {\bibfnamefont {S.}~\bibnamefont {Chakravarty}}, \bibinfo {author} {\bibfnamefont {A.~T.}\ \bibnamefont {Dorsey}}, \bibinfo {author} {\bibfnamefont {M.~P.~A.}\ \bibnamefont {Fisher}}, \bibinfo {author} {\bibfnamefont {A.}~\bibnamefont {Garg}},\ and\ \bibinfo {author} {\bibfnamefont {W.}~\bibnamefont {Zwerger}},\ }\bibfield  {title} {\bibinfo {title} {Dynamics of the dissipative two-state system},\ }\href {https://doi.org/10.1103/RevModPhys.59.1} {\bibfield  {journal} {\bibinfo  {journal} {Rev. Mod. Phys.}\ }\textbf {\bibinfo {volume} {59}},\ \bibinfo {pages} {1} (\bibinfo {year} {1987})}\BibitemShut {NoStop}%
\bibitem [{\citenamefont {Cresser}\ and\ \citenamefont {Anders}(2021)}]{cresser2021weak}%
  \BibitemOpen
  \bibfield  {author} {\bibinfo {author} {\bibfnamefont {J.}~\bibnamefont {Cresser}}\ and\ \bibinfo {author} {\bibfnamefont {J.}~\bibnamefont {Anders}},\ }\bibfield  {title} {\bibinfo {title} {Weak and ultrastrong coupling limits of the quantum mean force gibbs state},\ }\href@noop {} {\bibfield  {journal} {\bibinfo  {journal} {Physical Review Letters}\ }\textbf {\bibinfo {volume} {127}},\ \bibinfo {pages} {250601} (\bibinfo {year} {2021})}\BibitemShut {NoStop}%
\bibitem [{\citenamefont {Thingna}\ \emph {et~al.}(2012)\citenamefont {Thingna}, \citenamefont {Wang},\ and\ \citenamefont {Hänggi}}]{Thingna}%
  \BibitemOpen
  \bibfield  {author} {\bibinfo {author} {\bibfnamefont {J.}~\bibnamefont {Thingna}}, \bibinfo {author} {\bibfnamefont {J.-S.}\ \bibnamefont {Wang}},\ and\ \bibinfo {author} {\bibfnamefont {P.}~\bibnamefont {Hänggi}},\ }\bibfield  {title} {\bibinfo {title} {Generalized gibbs state with modified redfield solution: Exact agreement up to second order},\ }\href {https://doi.org/10.1063/1.4718706} {\bibfield  {journal} {\bibinfo  {journal} {The Journal of Chemical Physics}\ }\textbf {\bibinfo {volume} {136}},\ \bibinfo {pages} {194110} (\bibinfo {year} {2012})},\ \Eprint {https://arxiv.org/abs/https://doi.org/10.1063/1.4718706} {https://doi.org/10.1063/1.4718706} \BibitemShut {NoStop}%
\bibitem [{\citenamefont {Lee}\ and\ \citenamefont {Yeo}(2022)}]{lee2022perturbative}%
  \BibitemOpen
  \bibfield  {author} {\bibinfo {author} {\bibfnamefont {J.~S.}\ \bibnamefont {Lee}}\ and\ \bibinfo {author} {\bibfnamefont {J.}~\bibnamefont {Yeo}},\ }\bibfield  {title} {\bibinfo {title} {Perturbative steady states of completely positive quantum master equations},\ }\href@noop {} {\bibfield  {journal} {\bibinfo  {journal} {Physical Review E}\ }\textbf {\bibinfo {volume} {106}},\ \bibinfo {pages} {054145} (\bibinfo {year} {2022})}\BibitemShut {NoStop}%
\bibitem [{\citenamefont {{\L}obejko}\ \emph {et~al.}(2022)\citenamefont {{\L}obejko}, \citenamefont {Winczewski}, \citenamefont {Su{\'a}rez}, \citenamefont {Alicki},\ and\ \citenamefont {Horodecki}}]{lobejko2022towards}%
  \BibitemOpen
  \bibfield  {author} {\bibinfo {author} {\bibfnamefont {M.}~\bibnamefont {{\L}obejko}}, \bibinfo {author} {\bibfnamefont {M.}~\bibnamefont {Winczewski}}, \bibinfo {author} {\bibfnamefont {G.}~\bibnamefont {Su{\'a}rez}}, \bibinfo {author} {\bibfnamefont {R.}~\bibnamefont {Alicki}},\ and\ \bibinfo {author} {\bibfnamefont {M.}~\bibnamefont {Horodecki}},\ }\bibfield  {title} {\bibinfo {title} {Towards reconciliation of completely positive open system dynamics with the equilibration postulate},\ }\href@noop {} {\bibfield  {journal} {\bibinfo  {journal} {arXiv preprint arXiv:2204.00643}\ } (\bibinfo {year} {2022})}\BibitemShut {NoStop}%
\bibitem [{\citenamefont {Trushechkin}\ \emph {et~al.}(2022)\citenamefont {Trushechkin}, \citenamefont {Merkli}, \citenamefont {Cresser},\ and\ \citenamefont {Anders}}]{trushechkin2022open}%
  \BibitemOpen
  \bibfield  {author} {\bibinfo {author} {\bibfnamefont {A.}~\bibnamefont {Trushechkin}}, \bibinfo {author} {\bibfnamefont {M.}~\bibnamefont {Merkli}}, \bibinfo {author} {\bibfnamefont {J.}~\bibnamefont {Cresser}},\ and\ \bibinfo {author} {\bibfnamefont {J.}~\bibnamefont {Anders}},\ }\bibfield  {title} {\bibinfo {title} {Open quantum system dynamics and the mean force gibbs state},\ }\href@noop {} {\bibfield  {journal} {\bibinfo  {journal} {AVS Quantum Science}\ }\textbf {\bibinfo {volume} {4}} (\bibinfo {year} {2022})}\BibitemShut {NoStop}%
\bibitem [{\citenamefont {Bulla}\ \emph {et~al.}(2003)\citenamefont {Bulla}, \citenamefont {Tong},\ and\ \citenamefont {Vojta}}]{bulla2003numerical}%
  \BibitemOpen
  \bibfield  {author} {\bibinfo {author} {\bibfnamefont {R.}~\bibnamefont {Bulla}}, \bibinfo {author} {\bibfnamefont {N.-H.}\ \bibnamefont {Tong}},\ and\ \bibinfo {author} {\bibfnamefont {M.}~\bibnamefont {Vojta}},\ }\bibfield  {title} {\bibinfo {title} {Numerical renormalization group for bosonic systems and application to the sub-ohmic spin-boson model},\ }\href@noop {} {\bibfield  {journal} {\bibinfo  {journal} {Physical review letters}\ }\textbf {\bibinfo {volume} {91}},\ \bibinfo {pages} {170601} (\bibinfo {year} {2003})}\BibitemShut {NoStop}%
\bibitem [{\citenamefont {Chin}\ \emph {et~al.}(2011)\citenamefont {Chin}, \citenamefont {Prior}, \citenamefont {Huelga},\ and\ \citenamefont {Plenio}}]{PhysRevLett.107.160601}%
  \BibitemOpen
  \bibfield  {author} {\bibinfo {author} {\bibfnamefont {A.~W.}\ \bibnamefont {Chin}}, \bibinfo {author} {\bibfnamefont {J.}~\bibnamefont {Prior}}, \bibinfo {author} {\bibfnamefont {S.~F.}\ \bibnamefont {Huelga}},\ and\ \bibinfo {author} {\bibfnamefont {M.~B.}\ \bibnamefont {Plenio}},\ }\bibfield  {title} {\bibinfo {title} {Generalized polaron ansatz for the ground state of the sub-ohmic spin-boson model: An analytic theory of the localization transition},\ }\href {https://doi.org/10.1103/PhysRevLett.107.160601} {\bibfield  {journal} {\bibinfo  {journal} {Phys. Rev. Lett.}\ }\textbf {\bibinfo {volume} {107}},\ \bibinfo {pages} {160601} (\bibinfo {year} {2011})}\BibitemShut {NoStop}%
\bibitem [{\citenamefont {M{\"u}ller}\ \emph {et~al.}(2015)\citenamefont {M{\"u}ller}, \citenamefont {Lisenfeld}, \citenamefont {Shnirman},\ and\ \citenamefont {Poletto}}]{muller2015interacting}%
  \BibitemOpen
  \bibfield  {author} {\bibinfo {author} {\bibfnamefont {C.}~\bibnamefont {M{\"u}ller}}, \bibinfo {author} {\bibfnamefont {J.}~\bibnamefont {Lisenfeld}}, \bibinfo {author} {\bibfnamefont {A.}~\bibnamefont {Shnirman}},\ and\ \bibinfo {author} {\bibfnamefont {S.}~\bibnamefont {Poletto}},\ }\bibfield  {title} {\bibinfo {title} {Interacting two-level defects as sources of fluctuating high-frequency noise in superconducting circuits},\ }\href@noop {} {\bibfield  {journal} {\bibinfo  {journal} {Physical Review B}\ }\textbf {\bibinfo {volume} {92}},\ \bibinfo {pages} {035442} (\bibinfo {year} {2015})}\BibitemShut {NoStop}%
\bibitem [{\citenamefont {Burnett}\ \emph {et~al.}(2019)\citenamefont {Burnett}, \citenamefont {Bengtsson}, \citenamefont {Scigliuzzo}, \citenamefont {Niepce}, \citenamefont {Kudra}, \citenamefont {Delsing},\ and\ \citenamefont {Bylander}}]{burnett2019decoherence}%
  \BibitemOpen
  \bibfield  {author} {\bibinfo {author} {\bibfnamefont {J.~J.}\ \bibnamefont {Burnett}}, \bibinfo {author} {\bibfnamefont {A.}~\bibnamefont {Bengtsson}}, \bibinfo {author} {\bibfnamefont {M.}~\bibnamefont {Scigliuzzo}}, \bibinfo {author} {\bibfnamefont {D.}~\bibnamefont {Niepce}}, \bibinfo {author} {\bibfnamefont {M.}~\bibnamefont {Kudra}}, \bibinfo {author} {\bibfnamefont {P.}~\bibnamefont {Delsing}},\ and\ \bibinfo {author} {\bibfnamefont {J.}~\bibnamefont {Bylander}},\ }\bibfield  {title} {\bibinfo {title} {Decoherence benchmarking of superconducting qubits},\ }\href@noop {} {\bibfield  {journal} {\bibinfo  {journal} {npj Quantum Information}\ }\textbf {\bibinfo {volume} {5}},\ \bibinfo {pages} {54} (\bibinfo {year} {2019})}\BibitemShut {NoStop}%
\bibitem [{\citenamefont {Jock}\ \emph {et~al.}(2022)\citenamefont {Jock}, \citenamefont {Jacobson}, \citenamefont {Rudolph}, \citenamefont {Ward}, \citenamefont {Carroll},\ and\ \citenamefont {Luhman}}]{jock2022silicon}%
  \BibitemOpen
  \bibfield  {author} {\bibinfo {author} {\bibfnamefont {R.~M.}\ \bibnamefont {Jock}}, \bibinfo {author} {\bibfnamefont {N.~T.}\ \bibnamefont {Jacobson}}, \bibinfo {author} {\bibfnamefont {M.}~\bibnamefont {Rudolph}}, \bibinfo {author} {\bibfnamefont {D.~R.}\ \bibnamefont {Ward}}, \bibinfo {author} {\bibfnamefont {M.~S.}\ \bibnamefont {Carroll}},\ and\ \bibinfo {author} {\bibfnamefont {D.~R.}\ \bibnamefont {Luhman}},\ }\bibfield  {title} {\bibinfo {title} {A silicon singlet-triplet qubit driven by spin-valley coupling},\ }\href@noop {} {\bibfield  {journal} {\bibinfo  {journal} {Nature communications}\ }\textbf {\bibinfo {volume} {13}},\ \bibinfo {pages} {641} (\bibinfo {year} {2022})}\BibitemShut {NoStop}%
\bibitem [{\citenamefont {Carroll}\ \emph {et~al.}(2022)\citenamefont {Carroll}, \citenamefont {Rosenblatt}, \citenamefont {Jurcevic}, \citenamefont {Lauer},\ and\ \citenamefont {Kandala}}]{carroll2022dynamics}%
  \BibitemOpen
  \bibfield  {author} {\bibinfo {author} {\bibfnamefont {M.}~\bibnamefont {Carroll}}, \bibinfo {author} {\bibfnamefont {S.}~\bibnamefont {Rosenblatt}}, \bibinfo {author} {\bibfnamefont {P.}~\bibnamefont {Jurcevic}}, \bibinfo {author} {\bibfnamefont {I.}~\bibnamefont {Lauer}},\ and\ \bibinfo {author} {\bibfnamefont {A.}~\bibnamefont {Kandala}},\ }\bibfield  {title} {\bibinfo {title} {Dynamics of superconducting qubit relaxation times},\ }\href@noop {} {\bibfield  {journal} {\bibinfo  {journal} {npj Quantum Information}\ }\textbf {\bibinfo {volume} {8}},\ \bibinfo {pages} {132} (\bibinfo {year} {2022})}\BibitemShut {NoStop}%
\bibitem [{\citenamefont {Shnirman}\ and\ \citenamefont {Sch{\"o}n}(2003)}]{shnirman2003dephasing}%
  \BibitemOpen
  \bibfield  {author} {\bibinfo {author} {\bibfnamefont {A.}~\bibnamefont {Shnirman}}\ and\ \bibinfo {author} {\bibfnamefont {G.}~\bibnamefont {Sch{\"o}n}},\ }\bibfield  {title} {\bibinfo {title} {Dephasing and renormalization in quantum two-level systems},\ }in\ \href@noop {} {\emph {\bibinfo {booktitle} {Quantum Noise in Mesoscopic Physics}}}\ (\bibinfo  {publisher} {Springer},\ \bibinfo {year} {2003})\ pp.\ \bibinfo {pages} {357--370}\BibitemShut {NoStop}%
\bibitem [{\citenamefont {Carney}\ \emph {et~al.}(2017)\citenamefont {Carney}, \citenamefont {Chaurette}, \citenamefont {Neuenfeld},\ and\ \citenamefont {Semenoff}}]{carney2017infrared}%
  \BibitemOpen
  \bibfield  {author} {\bibinfo {author} {\bibfnamefont {D.}~\bibnamefont {Carney}}, \bibinfo {author} {\bibfnamefont {L.}~\bibnamefont {Chaurette}}, \bibinfo {author} {\bibfnamefont {D.}~\bibnamefont {Neuenfeld}},\ and\ \bibinfo {author} {\bibfnamefont {G.~W.}\ \bibnamefont {Semenoff}},\ }\bibfield  {title} {\bibinfo {title} {Infrared quantum information},\ }\href@noop {} {\bibfield  {journal} {\bibinfo  {journal} {Physical Review Letters}\ }\textbf {\bibinfo {volume} {119}},\ \bibinfo {pages} {180502} (\bibinfo {year} {2017})}\BibitemShut {NoStop}%
\bibitem [{\citenamefont {Derezi{\'n}ski}(2003)}]{derezinski2003van}%
  \BibitemOpen
  \bibfield  {author} {\bibinfo {author} {\bibfnamefont {J.}~\bibnamefont {Derezi{\'n}ski}},\ }\bibfield  {title} {\bibinfo {title} {Van hove hamiltonians--exactly solvable models of the infrared and ultraviolet problem},\ }in\ \href@noop {} {\emph {\bibinfo {booktitle} {Annales Henri Poincar{\'e}}}},\ Vol.~\bibinfo {volume} {4}\ (\bibinfo {organization} {Springer},\ \bibinfo {year} {2003})\ pp.\ \bibinfo {pages} {713--738}\BibitemShut {NoStop}%
\bibitem [{\citenamefont {Dingle}(1973)}]{dingle1973asymptotic}%
  \BibitemOpen
  \bibfield  {author} {\bibinfo {author} {\bibfnamefont {R.~B.}\ \bibnamefont {Dingle}},\ }\href@noop {} {\emph {\bibinfo {title} {Asymptotic expansions: their derivation and interpretation}}}\ (\bibinfo  {publisher} {Academic Press},\ \bibinfo {year} {1973})\BibitemShut {NoStop}%
\bibitem [{\citenamefont {Olver}(1997)}]{olver1997asymptotics}%
  \BibitemOpen
  \bibfield  {author} {\bibinfo {author} {\bibfnamefont {F.}~\bibnamefont {Olver}},\ }\href@noop {} {\emph {\bibinfo {title} {Asymptotics and special functions}}}\ (\bibinfo  {publisher} {AK Peters/CRC Press},\ \bibinfo {year} {1997})\BibitemShut {NoStop}%
\bibitem [{\citenamefont {Bhatia}(1997)}]{bhatia97}%
  \BibitemOpen
  \bibfield  {author} {\bibinfo {author} {\bibfnamefont {R.}~\bibnamefont {Bhatia}},\ }\href@noop {} {\emph {\bibinfo {title} {Matrix Analysis}}},\ Vol.\ \bibinfo {volume} {169}\ (\bibinfo  {publisher} {Springer},\ \bibinfo {year} {1997})\BibitemShut {NoStop}%
\end{thebibliography}
